\documentclass[useAMS,usenatbib,usegraphicx]{mn2e}
\usepackage{times}
\input{psfig.sty}

\def\gs{\mathrel{\raise0.35ex\hbox{$\scriptstyle >$}\kern-0.6em
\lower0.40ex\hbox{{$\scriptstyle \sim$}}}}
\def\ls{\mathrel{\raise0.35ex\hbox{$\scriptstyle <$}\kern-0.6em
\lower0.40ex\hbox{{$\scriptstyle \sim$}}}}
\def\m@th{\mathsurround=0pt }
\def\eqalign#1{\null\,\vcenter{\openup1\jot \m@th
 \ialign{\strut\hfil$\displaystyle{##}$&$\displaystyle{{}##}$\hfil
 \crcr#1\crcr}}\,}

\title[Identification of submillimetre galaxies in the SHADES Source Catalogue]
      {The SCUBA HAlf Degree Extragalactic Survey --- III.\
       Identification of radio and mid-infrared counterparts to
       submillimetre galaxies}
\author[Ivison et al.]
       {R.\,J.\ Ivison,$^{\! 1,2}$ T.\,R.\ Greve,$^{\! 3}$ J.\,S.\ Dunlop,$^{\! 2}$  
        J.\,A.\ Peacock,$^{\! 2}$ E.\ Egami,$^{\! 4}$ Ian Smail,$^{\! 5}$ 
	E.\ Ibar,$^{\! 2}$ \and E.\ van Kampen,$^{\! 6}$ I.\ Aretxaga,$^{\! 7}$
        T.\ Babbedge,$^{\! 8}$
        A.\,D.\ Biggs,$^{\! 1}$ A.\,W.\ Blain,$^{\! 3}$ S.\,C.\ Chapman,$^{\! 9}$ \and
        D.\,L.\ Clements,$^{\! 8}$ K.\ Coppin,$^{\! 5,10}$ D.\ Farrah,$^{\! 11}$
        M.\ Halpern,$^{\! 10}$
        D.\,H.\ Hughes,$^{\! 7}$ M.\,J.\ Jarvis,$^{\! 12}$ \and
        T.\ Jenness,$^{\! 13}$ J.\,R.\ Jones,$^{\! 14}$ A.\,M.\,J.\ Mortier,$^{\! 2}$
        S.\ Oliver,$^{\! 15}$ C.\ Papovich,$^{\! 4}$
        P.\,G.\ P\'{e}rez-Gonz\'{a}lez,$^{\! 16}$ \and
        A.\ Pope,$^{\! 10}$ S.\ Rawlings,$^{\! 17}$
        G.\,H.\ Rieke,$^{\! 4}$ M.\ Rowan-Robinson,$^{\! 8}$ 
        R.\,S.\ Savage,$^{\! 15}$ D.\ Scott,$^{\! 10}$ \and M.\ Seigar,$^{\! 18}$ 
        S.\ Serjeant,$^{\! 19}$ C.\ Simpson,$^{\! 20}$ J.\,A.\ Stevens,$^{\! 12}$ 
        M.\ Vaccari,$^{\! 21,8}$ J.\ Wagg$^{7,22}$ and \and C.\,J.\ Willott$^{23}$
        \vspace*{1mm}\\
        $^1$    UK Astronomy Technology Centre, Royal Observatory, Blackford Hill,
                Edinburgh EH9 3HJ\\
        $^2$    Scottish Universities Physics Alliance,
                Institute for Astronomy, University of Edinburgh, Blackford Hill,
                Edinburgh EH9 3HJ\\
	$^3$    Astronomy Department, California Institute of
                Technology, Pasadena, CA 91125, USA\\
	$^4$    Steward Observatory, University of Arizona, 933 N.\
                Cherry Avenue, Tuscon, AZ 85721, USA\\
        $^5$    Institute for Computational Cosmology, University of Durham,
                South Road, Durham DH1 3LE\\
	$^6$    Institute for Astrophysics, University of Innsbruck, Technikerstr.\ 25,
		A-6020 Innsbruck, Austria\\
	$^7$    Instituto Nacional de Astrof\'isica, Optica y Electr\'onica,
		Apartado Postal 51 y 216, 72000 Puebla, Pue., Mexico\\
        $^8$    Astrophysics Group, Blackett Laboratory, Imperial College London,
                Prince Consort Road, London SW7 2BW\\
	$^9$	Institute of Astronomy, Madingley Road, Cambridge, CB3 0HA\\
        $^{10}$ Department of Physics \& Astronomy, University of British Columbia,
                Vancouver, BC V6T 1Z1, Canada\\
        $^{11}$ Department of Astronomy, Cornell University, 106 Space Sciences, Ithaca,
                NY 14853, USA\\
        $^{12}$ Centre for Astrophysics Research, University of Hertfordshire,
                College Lane, Hatfield AL10 9AB\\
        $^{13}$ Joint Astronomy Centre, 660 N.\ A`oh\={o}k\={u} Place, University Park,
                Hilo, HI 96720, USA\\
        $^{14}$ Center for Astrophysics and Space Astronomy, 389 UCB, Boulder,
                CO 80309, USA\\
	$^{15}$ Astronomy Centre, University of Sussex, Falmer, Brighton BN1 9QH\\
        $^{16}$ Departmento de Astrof\'{\i}sica y CC.\ de Atm\'osfera, Facultad de
                CC.\ F\'{\i}sicas, Universidad Complutense de Madrid, 28040 Madrid, Spain\\
        $^{17}$ Astrophysics, Department of Physics, Denys Wilkinson Building,
                Keble Road, Oxford OX1 3RH\\
        $^{18}$ Department of Physics and Astronomy, 4129 Frederick Reines Hall,
                University of California, Irvine, CA 92697, USA\\
        $^{19}$ Astrophysics Group, Department of Physics, The Open University,
                Milton Keynes MK7 6AA\\
	$^{20}$ Astrophysics Research Institute, Liverpool John Moores University,
		Twelve Quays House, Egerton Wharf, Birkenhead CH41 1LD\\
        $^{21}$ Department of Astronomy, University of Padova, Vicolo dell'Osservatorio 3,
                I-35122, Italy\\
        $^{22}$ National Radio Astronomy Observatory, P.O.\ Box 0, Socorro, NM 87801, USA\\
        $^{23}$ Physics Department, University of Ottawa, 150 Louis Pasteur, Pavillon
                MacDonald Hall, Ottawa, Ontario K1N 6N5, Canada
}

\date{Accepted ... ; Received ... ; in original form 2007 February 20}

\pagerange{000--000}

\begin{document}

\maketitle

\begin{abstract}
Determining an accurate position for a submillimetre (submm) galaxy
(SMG) is the crucial step that enables us to move from the basic
properties of an SMG sample -- source counts and 2-D clustering -- to
an assessment of their detailed, multi-wavelength properties, their
contribution to the history of cosmic star formation and their links
with present-day galaxy populations. In this paper, we identify robust
radio and/or infrared (IR) counterparts, and hence accurate positions,
for over two thirds of the SCUBA HAlf-Degree Extragalactic Survey
(SHADES) Source Catalogue, presenting optical, 24-$\mu$m and radio
images of each SMG. Observed trends in identification rate have given
no strong rationale for pruning the sample. Uncertainties in submm
position are found to be consistent with theoretical expectations,
with no evidence for significant additional sources of
error. Employing the submm/radio redshift indicator, via a
parameterisation appropriate for radio-identified SMGs with
spectroscopic redshifts, yields a median redshift of 2.8 for the
radio-identified subset of SHADES, somewhat higher than the median
spectroscopic redshift. We present a diagnostic colour-colour plot,
exploiting {\em Spitzer} photometry, in which we identify regions
commensurate with SMGs at very high redshift. Finally, we find that
significantly more SMGs have multiple robust counterparts than would
be expected by chance, indicative of physical associations. These
multiple systems are most common amongst the brightest SMGs and are
typically separated by 2--6\,arcsec, $\sim$15--50/sin $i$\,kpc at
$z\sim\rm 2$, consistent with early bursts seen in merger simulations.
\end{abstract}


\section{Introduction}

Observational cosmology in the submm waveband has been one of the few
fields that can claim to have beaten Moore's Law (Moore 1965), the
other notable astronomical exception being the Virgo consortium's
`Millennium Simulation' (Springel et al.\ 2005). It has benefited
enormously from the development of bolometer arrays such as SCUBA
(Holland et al.\ 1999) and MAMBO (Kreysa et al.\ 1998): the
commissioning of these groundbreaking cameras, on the James Clerk
Maxwell Telescope (JCMT) and the IRAM 30-m telescope, respectively,
yielded a thousand-fold increase in mapping speed over single-pixel
devices such as UKT14 (Duncan et al.\ 1990). A decade on, the next
generation of cameras exemplified by LABOCA (Kreysa et al.\ 2003) and
SCUBA-2 (Holland et al.\ 2003) will yield a similar 
increase in mapping speed over existing arrays.

SCUBA brought about a radical shift in our understanding of the
formation and evolution of galaxies, with the discovery that luminous,
dusty galaxies were a thousand times more abundant in the early
Universe than at the present day (Smail, Ivison \& Blain 1997; Hughes
et al.\ 1998; Barger et al.\ 1998; Eales et al.\ 1999). SCUBA was
capable of providing only approximate coordinates so it was
immediately clear that the nature of these sources would remain a
mystery until more accurate positions could be determined -- the
subject of this paper. To refine positions provided by SCUBA, we are
reliant on radio observations; the radio emission is a high-resolution
proxy for the rest-frame far-IR emission observed in the submm (Ivison
et al.\ 1998, 2000, 2002; Smail et al.\ 2000; Webb et al.\ 2003a;
Clements et al.\ 2004; Dannerbauer et al.\ 2004; Borys et al.\ 2004;
Garrett, Knudsen \& van der Werf 2005; Voss et al.\ 2006). Although
likely to be inefficient in the era of SCUBA-2, radio imaging also
enabled large samples of SMGs to be acquired by targeting optically
faint $\mu$Jy radio sources (OFRS) using SCUBA's fast PHOTOM mode
(Barger, Cowie \& Richards 1999; Chapman et al.\ 2002).

Mid-IR imaging with {\em Spitzer} has also proved useful for refining
SMG positions (Egami et al.\ 2004; Ivison et al.\ 2004; Pope et al.\
2006; Ashby et al.\ 2006), albeit with poor angular resolution and an
imprecise connection to bolometric luminosity. To be useful, such data
need to be close to the 24-$\mu$m confusion limit ($\sim$50\,$\mu$Jy),
so radio imaging is likely to remain the preferred procedure.

Radio and submm flux densities, taken together, are sensitive to
redshift (Carilli \& Yun 1999; Dunne, Clements \& Eales 2000;
Rengarajan \& Takeuchi 2001), albeit limited to $z\ls\rm 3$ by the
depth of radio imaging available currently. This approach is the
subject of paper {\sc iv} in this series (Aretxaga et al.\
2007). Early work in this vein constrained the median redshift of the
SMG population to be $z\gs\rm 2$ (Carilli \& Yun 2000; Smail et al.\
2000; Ivison et al.\ 2002).

The true triumph of the radio identification procedure, however, has
been in identifying the correct optical/IR counterparts so that their
morphologies, colours, magnitudes, etc.\ can be determined
unambiguously; more importantly, this has also allowed spectroscopists
to place their slits accurately, sometimes on apparently blank sky
when optical counterparts were too faint for existing imaging ($R_{\rm
AB}\gs\rm 26$, e.g.\ LE850.12 and SSA13.332 -- Chapman et al.\
2005). This painstaking approach was slow to pay dividends, with only
a handful of redshifts reported initially (Ivison et al.\ 1998, 2000;
Barger et al.\ 1999; Ledlow et al.\ 2002; Knudsen, van der Werf \&
Jaffe 2003; Simpson et al.\ 2004). Deeper radio observations allied
with the largest existing submm surveys and the OFRS technique
resulted eventually in the acquisition of approximately 100
spectroscopic redshifts, the majority by Chapman et al.\ (2003,
2005). This has enabled the direct detection of colossal molecular gas
reservoirs in a representative sample of SMGs (Neri et al.\ 2003;
Greve et al.\ 2005; Tacconi et al.\ 2006), following on from the
pioneering CO detections of Frayer et al.\ (1998, 1999). It allowed
Alexander et al.\ (2005a, 2005b) to suggest that the bulk of the SMG
population contains obscured, often Compton-thick, active galactic
nuclei (AGN) via the first meaningful analysis of their X-ray
properties; it permitted a rigorous test of the radio/far-IR relation
at high redshift, via observations near the peak of SMG spectral
energy distributions (SEDs) at 350\,$\mu$m (Kovacs et al.\ 2006) and,
finally, it allowed a thorough analysis of their rest-frame optical
photometric and spectroscopic properties (Smail et al.\ 2004; Swinbank
et al.\ 2004; Takata et al.\ 2006).

Until now, the most adventurous blank-field surveys have covered a few
$\times$100\,arcmin$^2$, detecting typically 40 galaxies (Scott et
al.\ 2002; Webb et al.\ 2003a; Borys et al.\ 2003; Greve et al.\
2004). The properties of these galaxies were quickly characterised
over the entire observable spectral range (Lilly et al.\ 1999; Eales
et al.\ 2000; Gear et al.\ 2000; Lutz et al.\ 2001; Fox et al.\ 2002;
Ivison et al.\ 2002; Webb et al.\ 2003a, 2003b; Waskett et al.\ 2003;
Borys et al.\ 2004; Dunlop et al.\ 2004; Pope et al.\ 2005, 2006), but
it soon became clear that some of the key remaining questions -- the
degree of clustering and the role played by AGN -- could only be
addressed by a significantly larger sample selected homogeneously from
contiguous sky.

Despite the steep slope of the submm number counts (Blain et al.\
1998, 1999), the 850-$\mu$m confusion limit -- set at around 2\,mJy by
the JCMT's 15-m primary -- dictates that we must map more sky if we
are to obtain larger samples with well-characterised positions and
flux densities. SHADES aimed to detect 200 SMGs over two
0.25-degree$^2$ fields -- the Lockman Hole (LH; 10h 52m,
+57$^{\circ}$.4) and the Subaru-{\em XMM-Newton} Deep Field (SXDF; 02h
18m, $-$5$^{\circ}$.0). See Mortier et al.\ (2005), Paper {\sc i} of
this series, for a description of its motivation and design. SCUBA was
retired in 2005 July, before SHADES could be completed, after two
years plagued by cryogenic problems. The SHADES Source Catalogue,
gleaned from 800\,arcmin$^2$ and comprising 120 SMGs in the LH and the
SXDF, is presented in Paper {\sc ii} of this series (Coppin et al.\
2006).

In this, Paper {\sc iii}, we identify radio and/or mid-IR counterparts
and hence accurate positions for the SHADES sample using 1.4-GHz radio
imaging from the National Radio Astronomy Observatory's
(NRAO\footnote{NRAO is operated by Associated Universities Inc., under
a cooperative agreement with the National Science Foundation.}) Very
Large Array (VLA) and 24-$\mu$m data from MIPS (Rieke et al.\ 2004) on
board {\em Spitzer} (Werner et al.\ 2004). This is the crucial step
that allows us to move from the basic properties of an SMG sample --
source counts and 2-dimensional clustering -- to an assessment of
their detailed properties across the entire accessible wavelength
range, their contribution to the history of cosmic star formation and
their links with present-day galaxy populations. In \S2 we describe
the data exploited in \S3 to find radio and mid-IR counterparts for
our SMG sample. We use these associations in \S4 to determine the
positional uncertainty associated with SMGs, comparing with theory
developed in Appendix~B. In \S5 we discuss SMGs with multiple, robust
counterparts and in \S6 we explore identification trends. Finally, in
\S7 and \S8 we utilise the magnitudes and colours of SMGs, now
robustly identified, to constrain their redshift distribution and to
identify outliers. We assume $\Omega_m=\rm 0.27$, $\Omega_\Lambda=\rm
0.73$, $H_0=\rm 71$\,km\,s$^{-1}$\,Mpc$^{-1}$ throughout (Spergel et
al.\ 2003).

\section{Observations}

\subsection{1.4-GHz radio imaging}

%
%
\begin{figure}
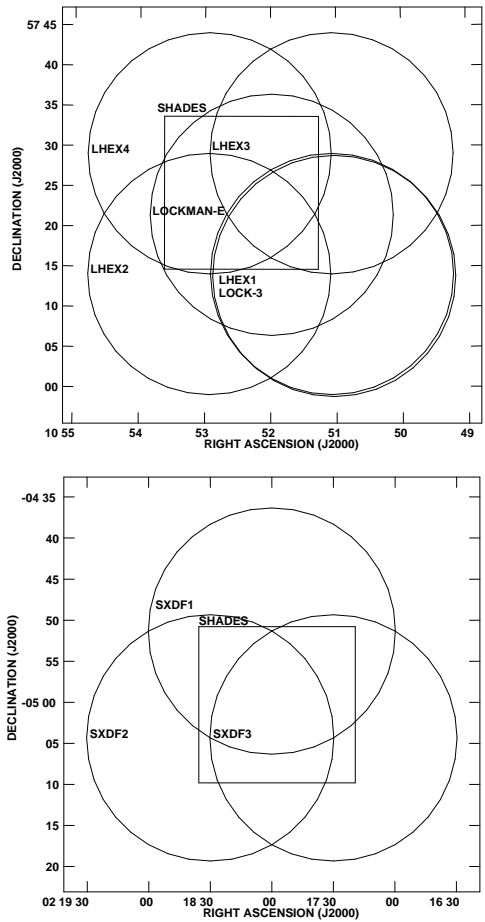

\centerline{\psfig{file=f1a.eps,angle=0,width=2.5in}}
\vspace*{3mm}
\centerline{\psfig{file=f1b.eps,angle=0,width=2.5in}}
\vspace{-0.5cm}
\noindent{\small\addtolength{\baselineskip}{-3pt}}
\caption{Individual pointings for radio mosaics in the SHADES fields,
together with an indication of the submm coverage. The diameter of the
circles is the {\sc fwhm} of the VLA's primary beam at 1.4\,GHz.}
\label{illustration}
\end{figure}

Wide-field radio images were obtained using the VLA. The LH
data used here, comprising 75\,hr of integration on a field designated
{\sc lockman-e}, were described in detail by Ivison et al.\
(2002). The data have since been re-analysed by Biggs \& Ivison
(2006), using the 37-piece mosaicing technique described by Owen et
al.\ (2005), together with additional self-calibration. The resulting
image covers most of the primary beam, out to a radius of
23\,arcmin. Once combined, the noise level is unusually uniform,
4.2\,$\mu$Jy\,beam$^{-1}$ r.m.s.\ in the centre of the field, with a
1.3-arcsec synthesised beam ({\sc fwhm}).

We also utilise a new low-resolution map, made by tapering our {\sc
lockman-e} data to give a 4.2-arcsec synthesised beam and then
mosaicing with B-configuration data taken for several nearby
pointings: a new field, 11\,arcmin to the south west, designated {\sc
lock-3}, plus archival data for fields designated {\sc lhex1, lhex2,
lhex3} and {\sc lhex4}, where {\sc lhex4} comprises 31\,hr of
integration, 11\,arcmin to the north east of {\sc lockman-e}.
Fig.~\ref{illustration} illustrates the mosaic of pointings. These
data, together with matched-resolution 610-MHz imaging from the Giant
Metre-wave Telescope in Pune, India, are described by Ibar et al.\
(in preparation).

We obtained new 1.4-GHz data for the SXDF, again using the VLA, during
2003. Many of these data were affected by interference and by a
prolonged failure of the correlator, but the equivalent of around
60\,hr of normal integration were salvaged. These A-configuration data
were combined with the B- and C-configuration data described by
Simpson et al.\ (2006) resulting in a 9:3:1 ratio of recorded A:B:C
visibilities, evenly distributed in three pointings separated by
15\,arcmin (see Fig.~\ref{illustration}). Each pointing was imaged as
a 37-piece mosaic, as with the LH. The final image was
knitted together and corrected for the response of the primary beam
using the {\sc aips} task, {\sc flatn}. The resulting noise level is
around 6.3\,$\mu$Jy\,beam$^{-1}$ in the best regions of the map,
though as high as 8.4\,$\mu$Jy\,beam$^{-1}$ near bright, complex radio
emitters, with a synthesised beam measuring around 1.7\,arcsec ({\sc
  fwhm}). As with the LH, we also utilise a low-resolution map,
tapering our entire dataset to give a 4.2-arcsec synthesised beam.

%
%
\setcounter{table}{0}
\begin{table*}
\scriptsize
\caption{Radio properties of SMGs in the Lockman Hole SHADES Source Catalogue.}
\vspace{0.2cm}
\begin{center}
\begin{tabular}{lcccccccccl}
Nickname&\multicolumn{2}{c}{Position at 850$\mu$m}&$S_{\rm 850\mu m}$&SNR$^a$&\multicolumn{2}{c}{Position at 1.4\,GHz}&$S_{\rm 1.4GHz}^b$&Submm--radio&$P^c$&Notes\\
    &$\alpha_{\rm J2000}$&$\delta_{\rm J2000}$&($S+, S-$)&&$\alpha_{\rm J2000}$&$\delta_{\rm J2000}$&&separation$^c$&&\\
    &h m s&$^{\circ}\ '\ ''$&/mJy&&h m s&$^{\circ}\ '\ ''$&/$\mu$Jy&/arcsec&&\\
&&&&&&&&&&\\
LOCK850.01& 10 52 01.417 &+57 24 43.04 & 8.8 (1.0, 1.0) &8.54&10 52 01.249 &+57 24 45.76 & 78.9\,$\pm$\,4.7       &3.04&{\bf 0.011}&$z$=2.148\\
LOCK850.02& 10 52 57.316 &+57 21 05.79 &13.4 (2.1, 2.1) &6.83&10 52 57.014 &+57 21 08.31 & 40.7\,$\pm$\,5.6       &3.51&{\bf 0.026}&\\
          &              &             &                &    &10 52 57.084 &+57 21 02.82 & 52.4\,$\pm$\,5.6       &3.51&{\bf 0.020}&\\
LOCK850.03& 10 52 38.247 &+57 24 36.54 &10.9 (1.8, 1.9) &6.39&10 52 38.401 &+57 24 39.50 & 35.0\,$\pm$\,5.2       &3.21&{\bf 0.027}&\\
          &              &             &                &    &10 52 38.299 &+57 24 35.76 & 25.8\,$\pm$\,4.9       &0.89&{\bf 0.005}&$z$=3.036\\
LOCK850.04& 10 52 04.171 &+57 26 58.85 &10.6 (1.7, 1.8) &6.42&10 52 03.691 &+57 27 07.06 & 47.0\,$\pm$\,5.7       &(9.08)&(0.104)&$z$=1.48\\
          &              &             &                &    &10 52 04.079 &+57 26 58.52 & 32.0\,$\pm$\,5.1       &0.81&{\bf 0.004}&\\
          &              &             &                &    &10 52 04.226 &+57 26 55.46 & 73.0\,$\pm$\,5.0       &3.42&{\bf 0.014}&\\
LOCK850.05& 10 53 02.615 &+57 18 26.95 & 8.1 (2.0, 2.1) &4.90&---          &---          &5$\sigma$\,$<$\,22      &--- &---  &24$\mu$m id\\
LOCK850.06& 10 52 04.131 &+57 25 26.34 & 6.8 (1.3, 1.3) &5.83&10 52 04.013 &+57 25 24.20 &{\em 15.0\,$\pm$\,4.8}  &2.34&{\bf 0.038}&\\
          &              &             &                &    &10 52 03.549 &+57 25 17.38 & 22.2\,$\pm$\,4.6       &(10.1)&(0.176)&\\
LOCK850.07& 10 53 01.403 &+57 25 54.24 & 8.5 (1.8, 1.9) &5.30&10 53 00.956 &+57 25 52.06 & 42.6\,$\pm$\,5.8       &4.22&{\bf 0.032}&\\
LOCK850.08& 10 51 53.862 &+57 18 39.75 & 5.4 (1.1, 1.2) &5.24&---          &---          &5$\sigma$\,$<$\,22      &--- &---  &24$\mu$m id\\
LOCK850.09& 10 52 16.088 &+57 25 04.11 & 5.9 (1.6, 1.6) &4.67&10 52 15.636 &+57 25 04.26 & 52.6\,$\pm$\,4.7       &3.65&{\bf 0.021}&\\
LOCK850.10& 10 52 48.607 &+57 32 58.58 & 9.1 (2.7, 2.9) &4.53&10 52 48.992 &+57 32 56.26 & 25.5\,$\pm$\,6.3      &3.87 &{\bf 0.048}&\\
LOCK850.11& 10 51 29.531 &+57 24 05.21 & 6.2 (1.7, 1.8) &4.53&10 51 29.824 &+57 24 15.19 &{\em 19.0\,$\pm$\,5.4}  &(10.3)&(0.181)&Confused at 24$\mu$m\\
LOCK850.12& 10 52 27.612 &+57 25 13.08 & 6.1 (1.7, 1.7) &4.58&10 52 27.579 &+57 25 12.46 & 44.3\,$\pm$\,5.1       &0.67&{\bf 0.002}&$z$=2.14?\\
          &              &             &                &    &10 52 28.793 &+57 25 16.01 & 19.2\,$\pm$\,4.5       &(9.98)&(0.180)&\\
LOCK850.13& 10 51 32.333 &+57 31 34.76 & 5.6 (2.3, 2.9) &3.89&---          &---          &5$\sigma$\,$<$\,28      &--- &---  &\\
LOCK850.14& 10 52 30.110 &+57 22 15.55 & 7.2 (1.8, 1.9) &4.84&10 52 28.995 &+57 22 22.42 & 25.3\,$\pm$\,4.2       &(11.3)&(0.178)&\\
          &              &             &                &    &10 52 30.717 &+57 22 09.56 & 37.4\,$\pm$\,4.2       &7.74&0.068&$z$=2.611\\
LOCK850.15& 10 53 19.200 &+57 21 10.64 &13.2 (4.3, 5.0) &4.51&10 53 19.025 &+57 21 09.47 & 43.9\,$\pm$\,7.8       &1.84&{\bf 0.009}&\\
          &              &             &                &    &10 53 19.271 &+57 21 08.45 & 61.5\,$\pm$\,7.6       &2.26&{\bf 0.009}&\\
          &              &             &                &    &10 53 19.067 &+57 21 16.28 &{\em 22.6\,$\pm$\,7.1}  &5.74&0.071&\\
LOCK850.16& 10 51 51.453 &+57 26 37.00 & 5.8 (1.8, 1.9) &4.32&10 51 51.690 &+57 26 36.09 & 106\,$\pm$\,6          &2.12&{\bf 0.004}&$z$=1.147\\
          &              &             &                &    &10 51 50.113 &+57 26 35.73 & 115\,$\pm$\,6          &(10.9)&(0.059)&\\
LOCK850.17& 10 51 58.250 &+57 18 00.81 & 4.7 (1.3, 1.3) &4.49&10 51 58.018 &+57 18 00.27 & 92.3\,$\pm$\,4.5       &1.96&{\bf 0.004}&$z$=2.239\\
LOCK850.18& 10 52 27.693 &+57 22 17.75 & 6.0 (1.9, 2.1) &4.27&10 52 27.778 &+57 22 18.18 & 29.4\,$\pm$\,4.4       &0.81&{\bf 0.004}&$z$=1.956\\
          &              &             &                &    &10 52 28.995 &+57 22 22.42 & 25.3\,$\pm$\,4.1       &(11.5)&(0.178)&\\
LOCK850.19& 10 52 35.709 &+57 31 19.05 & 5.1 (2.0, 2.4) &3.92&---          &---          &5$\sigma$\,$<$\,27      &--- &---  &24$\mu$m id\\
LOCK850.21& 10 52 56.858 &+57 30 38.05 & 4.1 (2.0, 2.5) &3.62&---          &---          &5$\sigma$\,$<$\,30      &--- &---  &24$\mu$m id\\
LOCK850.22& 10 51 37.551 &+57 33 23.32 & 7.5 (3.2, 4.2) &4.00&---          &---          &5$\sigma$\,$<$\,30      &--- &---  &24$\mu$m id\\
LOCK850.23& 10 52 13.737 &+57 31 54.11 & 4.3 (1.9, 2.4) &3.71&---          &---          &5$\sigma$\,$<$\,25      &--- &---  &\\
LOCK850.24& 10 52 00.227 &+57 20 38.05 & 2.7 (1.2, 1.2) &3.60&10 52 00.445 &+57 20 40.16 & 28.5\,$\pm$\,4.8       &2.75&{\bf 0.026}&\\
LOCK850.26& 10 52 40.950 &+57 23 12.01 & 5.8 (2.4, 2.9) &3.93&10 52 40.726 &+57 23 15.18 & 14.5\,$\pm$\,5.5       &3.65&0.064&\\
          &              &             &                &    &10 52 40.698 &+57 23 09.96 & 31.4\,$\pm$\,5.2       &2.89&{\bf 0.026}&\\
          &              &             &                &    &10 52 41.453 &+57 23 20.65 & 1,050\,$\pm$\,50       &(9.55)&(0.004)&\\
LOCK850.27& 10 52 03.574 &+57 18 13.46 & 5.0 (1.3, 1.3) &4.63&10 52 04.579 &+57 18 06.11 & 20.0\,$\pm$\,4.5       &(11.0)&(0.181)&\\
LOCK850.28& 10 52 57.001 &+57 31 07.14 & 6.4 (1.7, 1.8) &4.67&10 52 57.667 &+57 30 58.71 & 63.0\,$\pm$\,8.2       &(9.99)&(0.091)&Candidate id\\
LOCK850.29& 10 51 30.923 &+57 20 35.95 & 6.7 (2.0, 2.2) &4.39&10 51 31.305 &+57 20 40.28 & 23.7\,$\pm$\,4.9       &5.32&0.066&Radio+24$\mu$m id\\
LOCK850.30& 10 52 07.786 &+57 19 06.59 & 4.7 (1.5, 1.6) &4.19&10 52 07.490 &+57 19 04.01 & 245\,$\pm$\,13         &3.52&{\bf 0.004}&$z$=2.689\\
          &              &             &                &    &10 52 08.054 &+57 19 02.58 & 20.0\,$\pm$\,4.2       &4.56&0.064&\\
LOCK850.31& 10 52 16.055 &+57 16 21.11 & 6.0 (1.8, 2.0) &4.34&10 52 15.989 &+57 16 19.34 & 43.0\,$\pm$\,4.7       &1.85&{\bf 0.010}&\\
LOCK850.33& 10 51 55.975 &+57 23 11.76 & 3.8 (1.0, 1.1) &4.45&10 51 55.470 &+57 23 12.77 & 51.0\,$\pm$\,4.3       &4.21&{\bf 0.027}&$z$=2.686\\
LOCK850.34& 10 52 13.502 &+57 33 28.14 &14.0 (3.1, 3.2) &5.37&10 52 13.584 &+57 33 20.81 &{\em 28.7\,$\pm$\,8.7}  &7.36&0.075&Radio+24$\mu$m id\\
          &              &             &                &    &10 52 14.202 &+57 33 28.30 &58.4\,$\pm$\,8.5        &5.63&{\bf 0.035}&\\
LOCK850.35& 10 52 46.915 &+57 20 56.25 & 6.1 (2.2, 2.4) &4.12&10 52 46.655 &+57 20 52.54 &{\em 17.4\,$\pm$\,5.0}  &4.27&0.065&\\
LOCK850.36& 10 52 09.335 &+57 18 06.78 & 6.3 (1.7, 1.8) &4.55&---          &---          &5$\sigma$\,$<$\,20      &--- &---  &\\
LOCK850.37& 10 51 24.130 &+57 23 34.86 & 7.5 (2.9, 3.5) &4.10&10 51 24.595 &+57 23 31.08 &{\em 14.8\,$\pm$\,5.4}  &5.33&0.078&24$\mu$m id\\
          &              &             &                &    &10 51 24.342 &+57 23 36.18 &41.8\,$\pm$\,8.7        &2.16&{\bf 0.013}&\\
LOCK850.38& 10 53 07.104 &+57 24 31.39 & 4.3 (2.2, 2.7) &3.63&10 53 07.253 &+57 24 30.82 &{\em 24.4\,$\pm$\,6.7}  &1.33&{\bf 0.011}&\\
          &              &             &                &    &10 53 06.568 &+57 24 32.65 &{\em 13.8\,$\pm$\,6.5}  &4.51&0.075&\\
          &              &             &                &    &10 53 06.933 &+57 24 27.27 &{\em 20.9\,$\pm$\,6.2}  &4.35&0.059&\\
LOCK850.39& 10 52 24.851 &+57 16 09.80 & 6.5 (2.2, 2.5) &4.20&10 52 25.643 &+57 16 07.65 &5$\sigma$\,$<$\,20      &--- &---  &\\
LOCK850.40& 10 52 02.014 &+57 19 15.80 & 3.0 (1.1, 1.2) &3.79&10 52 01.721 &+57 19 17.00 & 16.2\,$\pm$\,4.3       &2.66&{\bf 0.042}&\\
          &              &             &                &    &10 52 02.070 &+57 19 23.13 & 18.0\,$\pm$\,4.9       &7.34&0.075&\\
LOCK850.41& 10 51 59.861 &+57 24 23.60 & 3.8 (0.9, 1.0) &4.54&10 52 00.248 &+57 24 21.69 & 43.6\,$\pm$\,4.7       &3.66&{\bf 0.026}&$z$=0.689\\
          &              &             &                &    &10 51 59.760 &+57 24 24.94 & 22.1\,$\pm$\,4.8       &1.57&{\bf 0.015}&\\
LOCK850.43& 10 52 57.169 &+57 23 51.81 & 4.9 (2.1, 2.6) &3.80&10 52 56.561 &+57 23 52.80 & 25.4\,$\pm$\,5.4       &5.01&0.060&24$\mu$m id\\
          &              &             &                &    &10 52 56.655 &+57 23 54.13 & 19.4\,$\pm$\,5.5       &4.76&0.067&Train wreck?\\
          &              &             &                &    &10 52 56.576 &+57 23 58.62 & 40.8\,$\pm$\,5.9       &(8.33)&(0.105)&24$\mu$m id\\
LOCK850.47& 10 52 35.629 &+57 25 14.04 & 3.5 (1.7, 2.1) &3.54&10 52 35.138 &+57 25 16.04 &5$\sigma$\,$<$\,22      &--- &---  &\\
LOCK850.48& 10 52 56.239 &+57 32 45.82 & 5.4 (2.1, 2.5) &3.94&10 52 55.181 &+57 32 45.38 & 43.7\,$\pm$\,10.0      &(8.53)&(0.103)&24$\mu$m id\\
LOCK850.52& 10 52 45.531 &+57 31 21.94 & 3.9 (2.2, 2.7) &3.52&10 52 45.808 &+57 31 19.86 & 38.7\,$\pm$\,8.0       &3.05&{\bf 0.023}&\\
LOCK850.53& 10 52 40.488 &+57 19 28.42 & 4.4 (2.3, 2.9) &3.62&---          &---          &5$\sigma$\,$<$\,21      &--- &---  &24$\mu$m id\\
LOCK850.60& 10 51 43.583 &+57 24 45.97 & 3.1 (1.7, 2.0) &3.40&10 51 43.488 &+57 24 35.90 & 22.3\,$\pm$\,4.9       &(10.1)&(0.176)&\\
LOCK850.63& 10 51 53.906 &+57 25 05.07 & 3.6 (1.2, 1.3) &4.00&10 51 54.261 &+57 25 02.55 & 22.6\,$\pm$\,4.8       &3.82&{\bf 0.049}&\\
LOCK850.64& 10 52 51.808 &+57 32 42.23 & 5.8 (2.5, 3.2) &3.87&10 52 52.231 &+57 32 32.39 & 45.5\,$\pm$\,7.4       &(10.4)&(0.124)&\\
          &              &             &                &    &10 52 53.121 &+57 32 40.22 & 31.7\,$\pm$\,7.4       &(10.8)&(0.159)&\\
LOCK850.66& 10 51 38.687 &+57 20 17.24 & 4.2 (1.9, 2.2) &3.74&---          &---          &5$\sigma$\,$<$\,21      &--- &---  &\\
LOCK850.67& 10 52 08.998 &+57 23 55.13 & 2.5 (1.5, 1.5) &3.30&---          &---          &5$\sigma$\,$<$\,21      &--- &---  &\\
LOCK850.70& 10 51 48.516 &+57 30 46.69 & 3.8 (2.2, 2.5) &3.52&10 51 47.894 &+57 30 44.37 &{\em 21.9\,$\pm$\,7.2}  &5.52&0.070&24$\mu$m id\\
&&&&&&&&&\\													   
\end{tabular}
\end{center}

\noindent
$a$) Raw signal-to-noise ratio (SNR), before deboosting.\\
$b$) Integrated flux densities; for tentative detections, these are given in {\em italics}.\\
$c$) Possible counterparts with 8.0--12.5-arcsec offsets are listed in parentheses
for completeness. Reliable identifications ($P\le\rm 0.05$) are listed in {\bf bold}.

\label{radio-lh}
\end{table*}

%
%
\setcounter{table}{0}
\begin{table*}
\scriptsize
\caption{Cont...}
\vspace{0.2cm}
\begin{center}
\begin{tabular}{lcccccccccl}
Nickname&\multicolumn{2}{c}{Position at 850$\mu$m}&$S_{\rm 850\mu m}$&SNR&\multicolumn{2}{c}{Position at 1.4\,GHz}&$S_{\rm 1.4GHz}$&Submm--radio&$P$&Notes\\
    &$\alpha_{\rm J2000}$&$\delta_{\rm J2000}$&($S+, S-$)&&$\alpha_{\rm J2000}$&$\delta_{\rm J2000}$&&separation&&\\
    &h m s&$^{\circ}\ '\ ''$&/mJy&&h m s&$^{\circ}\ '\ ''$&/$\mu$Jy&/arcsec&&\\
&&&&&&&&&&\\
LOCK850.71& 10 52 18.618 &+57 19 03.79 & 3.9 (1.8, 2.0) &3.69&10 52 19.086 &+57 18 57.87 & 95.8\,$\pm$\,4.6       &7.03&{\bf 0.030}&\\
LOCK850.73& 10 51 41.660 &+57 22 17.63 & 3.5 (1.9, 2.3) &3.48&10 51 41.705 &+57 22 20.10 & 26.7\,$\pm$\,4.6       &2.50&{\bf 0.025}&\\ 
          &              &             &                &    &10 51 41.992 &+57 22 17.52 & 27.3\,$\pm$\,4.8       &2.69&{\bf 0.027}&\\
LOCK850.75& 10 53 15.927 &+57 26 45.47 & 4.4 (2.2, 2.6) &3.68&10 53 15.439 &+57 26 37.42 & 27.1\,$\pm$\,7.8       &(8.96)&(0.150)&Radio+24$\mu$m id\\
LOCK850.76& 10 51 48.516 &+57 28 38.69 & 4.7 (2.5, 3.1) &3.66&10 51 49.101 &+57 28 40.28 & 48.0\,$\pm$\,6.0       &4.98&{\bf 0.036}&\\
LOCK850.77& 10 51 57.004 &+57 22 10.07 & 3.2 (1.2, 1.3) &3.84&10 51 57.153 &+57 22 09.58 & {\em 15.5\,$\pm$\,4.4} &1.30&{\bf 0.017}&\\
          &              &             &                &    &10 51 57.665 &+57 22 12.35 & 39.5\,$\pm$\,7.8       &5.81&{\bf 0.050}&\\
LOCK850.78& 10 51 45.333 &+57 17 38.68 & 4.5 (2.2, 2.7) &3.70&---          &---          &5$\sigma$\,$<$\,23      &--- &---  &\\
LOCK850.79& 10 51 52.104 &+57 21 27.38 & 3.1 (1.3, 1.5) &3.65&10 51 52.594 &+57 21 24.43 & 22.4\,$\pm$\,4.5       &4.94&0.064&24$\mu$m id\\
          &              &             &                &    &10 51 51.198 &+57 21 27.29 & 26.3\,$\pm$\,4.6       &7.33&0.077&Plausible id\\
LOCK850.81& 10 52 31.989 &+57 18 00.40 & 5.3 (1.9, 2.3) &4.01&10 52 31.523 &+57 17 51.67 & 55.2\,$\pm$\,5.3       &(9.51)&(0.096)&\\
LOCK850.83& 10 53 07.939 &+57 28 39.14 & 3.1 (2.0, 2.1) &3.37&---          &---          &5$\sigma$\,$<$\,28      &--- &---  &24$\mu$m id\\
LOCK850.87& 10 51 53.302 &+57 17 33.38 & 3.4 (1.5, 1.7) &3.64&10 51 53.365 &+57 17 30.05 & 84.5\,$\pm$\,5.3       &3.37&{\bf 0.012}&\\
LOCK850.100&10 51 39.056 &+57 15 09.81 &11.2 (4.2, 5.3) &4.30&10 51 38.877 &+57 15 03.90 &{\em 19.8\,$\pm$\,6.3}  &6.09&0.077&Radio+24$\mu$m id\\
\end{tabular}
\end{center}
\end{table*}

%
%
\setcounter{table}{1}
\begin{table*}
\scriptsize
\caption{Radio properties of SMGs in the SXDF SHADES Source Catalogue.}
\vspace{0.2cm}
\begin{center}
\begin{tabular}{lcccccccccl}
Nickname&\multicolumn{2}{c}{Position at 850$\mu$m}&$S_{\rm 850\mu m}$&SNR$^a$&\multicolumn{2}{c}{Position at 1.4\,GHz}&$S_{\rm 1.4GHz}^b$&Submm--radio&$P$&Notes\\
    &$\alpha_{\rm J2000}$&$\delta_{\rm J2000}$&($S+, S-$)&&$\alpha_{\rm J2000}$&$\delta_{\rm J2000}$&&separation&&\\
    &h m s&$^{\circ}\ '\ ''$&/mJy&&h m s&$^{\circ}\ '\ ''$&/$\mu$Jy&/arcsec&&\\
&&&&&&&&&&\\
SXDF850.01&02 17 30.531&$-$04 59 36.96&10.4 (1.5, 1.4)&7.35&02 17 30.629&$-$04 59 36.70&54.3\,$\pm$\,9.7       &1.49&{\bf 0.005}&\\
SXDF850.02&02 18 03.509&$-$04 55 27.24&10.1 (1.6, 1.6)&6.62&02 18 03.556&$-$04 55 27.55&66.2\,$\pm$\,10.9      &0.77&{\bf 0.001}&\\
SXDF850.03&02 17 42.144&$-$04 56 28.22& 8.8 (1.5, 1.6)&5.95&02 17 42.128&$-$04 56 27.67&77.2\,$\pm$\,9.3       &0.60&{\bf 0.001}&\\
SXDF850.04&02 17 38.621&$-$05 03 37.47& 4.4 (1.7, 2.0)&3.88&02 17 38.680&$-$05 03 39.46&185\,$\pm$\,12         &2.18&{\bf 0.002}&\\
SXDF850.05&02 18 02.876&$-$05 00 32.75& 8.4 (1.7, 1.9)&5.35&02 18 02.858&$-$05 00 30.91&574\,$\pm$\,10         &1.86&{\bf 0.001}&\\
SXDF850.06&02 17 29.769&$-$05 03 26.81& 8.2 (2.2, 2.2)&4.72&02 17 30.224&$-$05 03 25.37&66.6\,$\pm$\,12.7      &6.95&{\bf 0.034}&\\
          &            &              &               &    &02 17 29.926&$-$05 03 22.01&47.4\,$\pm$\,10.8      &5.34&{\bf 0.033}&\\
          &            &              &               &    &02 17 29.753&$-$05 03 18.50&92.9\,$\pm$\,9.6       &(8.31)&(0.044)&\\
SXDF850.07&02 17 38.921&$-$05 05 23.72& 7.1 (1.5, 1.6)&5.16&02 17 38.878&$-$05 05 28.03&41.2\,$\pm$\,11.3      &4.36&{\bf 0.029}&\\
SXDF850.08&02 17 44.432&$-$04 55 54.72& 6.0 (1.8, 1.9)&4.39&02 17 44.137&$-$04 55 48.72&52.0\,$\pm$\,9.5       &7.45&{\bf 0.042}&\\
SXDF850.09&02 17 56.422&$-$04 58 06.74& 6.4 (2.0, 2.1)&4.35&02 17 55.772&$-$04 58 14.31&{\em 46.0\,$\pm$\,10.5}&(12.3)&(0.110)&\\
SXDF850.10&02 18 25.248&$-$04 55 57.21& 7.7 (2.6, 3.1)&4.24&02 18 24.975&$-$04 56 02.85&149\,$\pm$\,12         &6.97&{\bf 0.017}&\\
          &            &              &               &    &02 18 25.797&$-$04 55 51.31&47.4\,$\pm$\,10.4      &(10.1)&(0.094)&\\
SXDF850.11&02 17 25.117&$-$04 59 37.44& 4.5 (1.9, 2.2)&3.81&02 17 25.101&$-$04 59 33.77&56.8\,$\pm$\,10.0      &3.68&{\bf 0.018}&\\
SXDF850.12&02 17 59.369&$-$05 05 03.74& 5.7 (1.7, 1.8)&4.34&02 17 59.294&$-$05 05 04.04&{\em 42.0\,$\pm$\,10.8}&1.16&{\bf 0.004}&\\
SXDF850.14&02 18 19.256&$-$05 02 44.21& 4.8 (1.9, 2.1)&3.93&02 18 18.748&$-$05 02 49.25&{\em 30.4\,$\pm$\,11.4}&(9.11)&(0.109)&\\
          &            &              &               &    &02 18 19.018&$-$05 02 48.90&{\em 40.0\,$\pm$\,11.1}&5.89&{\bf 0.040}&\\
SXDF850.15&02 18 15.699&$-$04 54 05.22& 6.2 (1.6, 1.6)&4.76&---         &---           &5$\sigma$\,$<$\,37     &--- &---  &\\
SXDF850.16&02 18 13.887&$-$04 57 41.74& 4.8 (1.7, 1.8)&4.10&02 18 13.805&$-$04 57 43.22&36.5\,$\pm$\,8.8       &1.92&{\bf 0.011}&\\
SXDF850.17&02 17 54.980&$-$04 53 02.83& 7.6 (1.7, 1.7)&5.25&---         &---           &5$\sigma$\,$<$\,39     &--- &---  &\\
SXDF850.18&02 17 57.790&$-$05 00 29.75& 6.4 (2.0, 2.2)&4.30&02 17 57.591&$-$05 00 33.69&{\em 40.8\,$\pm$\,9.0} &4.94&{\bf 0.034}&\\
SXDF850.19&02 18 28.149&$-$04 58 39.21& 4.3 (1.8, 2.1)&3.79&02 18 27.782&$-$04 58 37.17&95.9\,$\pm$\,10.1      &5.86&{\bf 0.020}&\\
SXDF850.20&02 17 44.182&$-$05 02 15.97& 4.4 (2.0, 2.2)&3.78&---         &---           &5$\sigma$\,$<$\,34     &--- &---  &\\
SXDF850.21&02 17 42.803&$-$05 04 27.71& 5.2 (2.0, 2.2)&3.99&02 17 42.499&$-$05 04 24.50&690\,$\pm$\,50         &5.56&{\bf 0.002}&\\
SXDF850.22&02 18 00.379&$-$05 07 41.50& 6.2 (2.3, 2.6)&4.08&---         &---           &5$\sigma$\,$<$\,36     &--- &---  &\\
SXDF850.23&02 17 42.526&$-$05 05 45.47& 5.2 (1.7, 2.0)&4.12&02 17 42.455&$-$05 05 45.88&71.3\,$\pm$\,10.1      &1.14&{\bf 0.002}&\\
SXDF850.24&02 17 34.578&$-$05 04 37.71& 5.1 (2.0, 2.3)&3.93&02 17 34.696&$-$05 04 39.18&35.3\,$\pm$\,10.3      &2.30&{\bf 0.014}&\\
          &            &              &               &    &02 17 34.749&$-$05 04 30.47&42.3\,$\pm$\,12.0      &7.68&{\bf 0.047}&\\
SXDF850.25&02 18 12.120&$-$05 05 55.74& 4.0 (2.1, 2.5)&3.58&---         &---           &5$\sigma$\,$<$\,38     &--- &---  &\\
SXDF850.27&02 18 07.861&$-$05 01 48.49& 5.6 (2.0, 2.3)&4.08&02 18 07.934&$-$05 01 45.38&316\,$\pm$\,12         &3.30&{\bf 0.002}&\\
SXDF850.28&02 18 07.043&$-$04 59 15.50& 4.8 (2.2, 2.7)&3.76&02 18 06.920&$-$04 59 12.72&96.7\,$\pm$\,10.4      &3.34&{\bf 0.009}&\\
          &            &              &               &    &02 18 06.831&$-$04 59 17.52&96.2\,$\pm$\,9.6       &3.76&{\bf 0.011}&\\
          &            &              &               &    &02 18 06.419&$-$04 59 20.05&57.9\,$\pm$\,9.0       &(10.4)&(0.085)&\\
SXDF850.29&02 18 16.468&$-$04 55 11.82& 5.3 (1.8, 1.9)&4.15&02 18 16.484&$-$04 55 08.66&245\,$\pm$\,9          &3.17&{\bf 0.003}&\\
SXDF850.30&02 17 40.305&$-$05 01 16.22& 5.7 (2.0, 2.2)&4.14&02 17 40.020&$-$05 01 15.32&29.3\,$\pm$\,11.3      &4.35&{\bf 0.037}&\\
SXDF850.31&02 17 36.301&$-$04 55 57.46& 6.0 (1.7, 2.0)&4.37&02 17 35.856&$-$04 55 55.10&55.9\,$\pm$\,11.8      &7.07&{\bf 0.039}&\\
SXDF850.32&02 17 22.888&$-$05 00 38.10& 6.0 (2.4, 3.0)&3.96&---         &---           &5$\sigma$\,$<$\,40     &--- &-- - &\\
SXDF850.35&02 18 00.888&$-$04 53 11.24& 5.3 (1.8, 2.1)&4.06&02 18 00.867&$-$04 53 05.71&45.1\,$\pm$\,11.3      &5.54&{\bf 0.035}&\\
SXDF850.36&02 18 32.272&$-$04 59 47.21& 5.4 (1.8, 1.9)&4.20&---         &---           &5$\sigma$\,$<$\,38     &--- &---  &\\
SXDF850.37&02 17 24.445&$-$04 58 39.93& 4.5 (2.2, 2.6)&3.71&02 17 24.569&$-$04 58 41.29&{\em 40.9\,$\pm$\,9.2} &2.30&{\bf 0.013}&\\
SXDF850.38&02 18 25.427&$-$04 57 14.71& 3.8 (2.3, 2.7)&3.49&02 18 25.176&$-$04 57 19.70&{\em 49.8\,$\pm$\,18.2}&6.25&{\bf 0.037}&\\
SXDF850.39&02 17 50.595&$-$04 55 40.16& 4.0 (1.7, 2.1)&3.69&---         &---           &5$\sigma$\,$<$\,37     &--- &---  &\\
SXDF850.40&02 17 29.669&$-$05 00 59.21& 3.6 (1.5, 1.6)&3.78&02 17 29.625&$-$05 00 58.57&{\em 40.3\,$\pm$\,9.5} &0.92&{\bf 0.003}&\\
SXDF850.45&02 18 29.328&$-$05 05 40.71&21.9 (6.2, 6.8)&4.92&---         &---           &5$\sigma$\,$<$\,40     &--- &---  &\\
SXDF850.47&02 17 33.887&$-$04 58 57.71& 3.0 (1.6, 1.9)&3.39&02 17 34.363&$-$04 58 57.23&175\,$\pm$\,11         &7.15&{\bf 0.015}&\\
          &            &              &               &    &02 17 34.400&$-$04 58 59.76&43.1\,$\pm$\,10.1      &7.95&{\bf 0.048}&\\
          &            &              &               &    &02 17 33.616&$-$04 58 58.21&64.2\,$\pm$\,13.2      &4.09&{\bf 0.018}&\\
SXDF850.48&02 17 24.621&$-$04 57 17.68& 7.6 (2.5, 2.9)&4.28&---         &---           &5$\sigma$\,$<$\,39     &--- &---  &\\
SXDF850.49&02 18 20.259&$-$04 56 48.47& 3.3 (2.0, 2.2)&3.43&---         &---           &5$\sigma$\,$<$\,35     &--- &---  &\\
SXDF850.50&02 18 02.858&$-$04 56 45.49& 5.3 (2.0, 2.5)&3.93&02 18 02.827&$-$04 56 47.80&{\em 38.8\,$\pm$\,12.7}&2.36&{\bf 0.014}&\\
SXDF850.52&02 18 04.896&$-$05 04 53.74& 3.2 (1.8, 2.1)&3.41&02 18 05.118&$-$05 04 52.12&89.3\,$\pm$\,11.1      &3.69&{\bf 0.011}&\\
          &            &              &               &    &02 18 04.972&$-$05 05 01.02&88.8\,$\pm$\,10.3      &7.37&{\bf 0.029}&\\
SXDF850.55&02 17 52.190&$-$05 04 46.50& 3.9 (2.2, 2.7)&3.52&02 17 51.865&$-$05 04 46.96&{\em 41.7\,$\pm$\,13.8}&4.88&{\bf 0.033}&\\
SXDF850.56&02 17 50.679&$-$05 06 31.82& 3.6 (2.2, 2.5)&3.47&---         &---           &5$\sigma$\,$<$\,40     &--- &---  &\\
SXDF850.63&02 17 45.802&$-$04 57 50.49& 4.1 (1.7, 2.1)&3.73&---         &---           &5$\sigma$\,$<$\,38     &--- &---  &\\
SXDF850.65&02 18 07.935&$-$05 04 03.24& 4.3 (1.9, 2.4)&3.70&---         &---           &5$\sigma$\,$<$\,27     &--- &---  &\\
SXDF850.69&02 17 51.395&$-$05 02 50.82& 3.6 (2.1, 2.4)&3.49&---         &---           &5$\sigma$\,$<$\,38     &--- &---  &61.4\,$\pm$\,10.4\,$\mu$Jy, 13.0$''$ to SSW\\
SXDF850.70&02 18 11.199&$-$05 02 47.16& 4.0 (1.9, 2.3)&3.64&---         &---           &5$\sigma$\,$<$\,29     &--- &---  &\\
SXDF850.71&02 18 21.235&$-$04 59 03.22& 4.1 (1.9, 2.4)&3.66&---         &---           &5$\sigma$\,$<$\,35     &--- &---  &24$\mu$m id\\
SXDF850.74&02 17 58.732&$-$04 54 28.83& 3.3 (1.8, 2.1)&3.45&02 17 58.729&$-$04 54 33.41&38.9\,$\pm$\,12.7      &4.58&{\bf 0.032}&\\
SXDF850.76&02 17 55.781&$-$05 06 21.82& 4.4 (2.0, 2.4)&3.73&02 17 56.308&$-$05 06 24.91&84.2\,$\pm$\,13.1      &(8.46)&(0.049)&\\
SXDF850.77&02 17 36.432&$-$05 04 32.15& 3.0 (2.0, 2.1)&3.35&02 17 35.951&$-$05 04 25.97&43.8\,$\pm$\,10.7      &(9.48)&(0.093)&\\
          &            &              &               &    &02 17 36.175&$-$05 04 33.26&34.0\,$\pm$\,9.9       &4.00&{\bf 0.047}&\\
SXDF850.86&02 18 17.184&$-$05 04 04.70& 3.6 (1.9, 2.2)&3.54&---         &---           &5$\sigma$\,$<$\,37     &--- &---  &\\
SXDF850.88&02 18 00.994&$-$05 04 48.49& 4.5 (2.1, 2.5)&3.74&02 18 01.494&$-$05 04 43.74&40.8\,$\pm$\,9.3       &(8.85)&(0.091)&54.5\,$\pm$\,10.8\,$\mu$Jy, 12.6$''$ to ESE\\
SXDF850.91&02 17 34.808&$-$04 57 23.93& 3.5 (2.1, 2.5)&3.43&---         &---           &5$\sigma$\,$<$\,35     &--- &---  &\\
SXDF850.93&02 17 33.082&$-$04 58 13.48& 3.1 (2.0, 2.1)&3.36&---         &---           &5$\sigma$\,$<$\,28     &--- &---  &\\
SXDF850.94&02 17 40.079&$-$04 58 17.73& 4.1 (1.8, 2.1)&3.75&---         &---           &5$\sigma$\,$<$\,41     &--- &---  &\\
SXDF850.95&02 17 41.715&$-$04 58 33.70& 3.4 (1.9, 2.2)&3.47&---         &---           &5$\sigma$\,$<$\,35     &--- &---  &\\
SXDF850.96&02 18 00.000&$-$05 02 12.75& 4.7 (2.1, 2.5)&3.79&02 18 00.238&$-$05 02 16.83&{\em 37.5\,$\pm$\,8.4} &5.41&{\bf 0.039}&85.3\,$\pm$\,9.9\,$\mu$Jy, 12.7$''$ to NNW\\
SXDF850.119&02 17 56.345&$-$04 52 55.24&4.5 (2.1, 2.5)&3.73&02 17 56.205&$-$04 53 03.36&71.9\,$\pm$\,8.7       &(8.39)&(0.056)&\\
           &            &              &              &    &02 17 56.005&$-$04 52 51.96&{\em 38.0\,$\pm$\,9.7} &6.06&{\bf 0.043}&\\
\end{tabular}
\end{center}

\noindent
$a$) Raw SNR, before deboosting.\\
$b$) Flux densities for tentative detections are given in {\em italics}.\\
$c$) Possible counterparts with 8.0--12.5-arcsec offsets are listed in parentheses
for completeness. Reliable identifications ($P\le\rm 0.05$) are listed in {\bf bold}.

\label{radio-sxdf}
\end{table*}

%
%
\setcounter{table}{2}
\begin{table}
\scriptsize
\caption{Mid-IR properties of SMGs in the LH SHADES Source Catalogue.}
\vspace{0.2cm}
\begin{center}
\begin{tabular}{lccrccc}
Nickname&\multicolumn{2}{c}{Position at 24\,$\mu$m}&ID$^a$&$S_{24\mu\rm m}$&Off &$P^b$\\
    &$\alpha_{\rm J2000}$&$\delta_{\rm J2000}$&       &/$\mu$Jy        &-set &   \\
    &h m s&$^{\circ}\ '\ ''$                  &       &                &$''$ &   \\
&&&&&&\\
LOCK850.01& 10 52 01.30 &+57 24 46.1 & 1934&217\,$\pm$\,16          & 3.20  &{\bf 0.024} \\
LOCK850.02& 10 52 57.07 &+57 21 02.9 &19460&545\,$\pm$\,31          & 3.51  &{\bf 0.010} \\
LOCK850.03& 10 52 38.66 &+57 24 43.7 &17451&73.7\,$\pm$\,21.3       & 7.90  &0.196  \\
          & 10 52 38.31 &+57 24 39.5 &20054&183\,$\pm$\,33          & 3.00  &{\bf 0.026} \\
          & 10 52 38.31 &+57 24 34.8 &20603&175\,$\pm$\,23          & 1.81  &{\bf 0.012} \\
LOCK850.04& 10 52 04.21 &+57 26 55.6 &15970&261\,$\pm$\,73          & 3.27  &{\bf 0.020} \\
          & 10 52 04.04 &+57 26 58.3 &15971&179\,$\pm$\,68          & 1.19  &{\bf 0.006} \\
          & 10 52 03.67 &+57 27 07.0 & 3707&1,104\,$\pm$\,33        &(9.10) &(0.026)\\
LOCK850.05& 10 53 02.86 &+57 18 23.9 &11921&58.6\,$\pm$\,15.1       & 3.64  &0.107  \\
LOCK850.06& 10 52 04.12 &+57 25 25.8 &11922&75.1\,$\pm$\,12.7       & 0.55  &{\bf 0.005} \\
          & 10 52 03.51 &+57 25 17.1 &17409&379\,$\pm$\,18          &(10.5) &(0.107)\\
          & 10 52 05.19 &+57 25 22.9 &20753&540\,$\pm$\,48          &(9.22) &(0.060)\\
LOCK850.07& 10 53 00.97 &+57 25 52.2 & 5670&341\,$\pm$\,21          & 4.05  &{\bf 0.021} \\
LOCK850.08& 10 51 53.69 &+57 18 34.9 & 1811&481\,$\pm$\,25          & 5.05  &{\bf 0.021} \\
LOCK850.09& 10 52 15.73 &+57 25 01.7 &13577&159\,$\pm$\,73          & 3.76  &{\bf 0.043} \\
          & 10 52 15.65 &+57 25 04.5 &13578&466\,$\pm$\,74          & 3.56  &{\bf 0.012} \\
LOCK850.10& 10 52 47.39 &+57 32 57.9 &16088&65.9\,$\pm$\,11.1       &(9.82) &(0.429)\\
          & 10 52 48.27 &+57 32 51.0 &17604&79.6\,$\pm$\,10.8       &(8.05) &(0.313)\\
LOCK850.11& 10 51 29.16 &+57 24 06.8 & 8740&112\,$\pm$\,57          & 3.39  &0.053  \\
          & 10 51 29.39 &+57 24 10.3 & 8741&177\,$\pm$\,51          & 5.22  &0.063  \\
          & 10 51 29.81 &+57 24 16.3 & 8742&111\,$\pm$\,17          &(11.3) &(0.349)\\
LOCK850.12& 10 52 27.60 &+57 25 12.4 & 3757&263\,$\pm$\,19          & 0.69  &{\bf 0.001} \\
LOCK850.13& 10 51 31.45 &+57 31 29.1 &11931&240\,$\pm$\,17          &(9.09) &(0.137)\\
          & 10 51 31.77 &+57 31 41.2 &11932&172\,$\pm$\,14          & 7.88  &0.110  \\
LOCK850.14& 10 52 30.72 &+57 22 09.4 & 5560&188\,$\pm$\,16          & 7.88  &0.102  \\
          & 10 52 29.06 &+57 22 21.8 & 5563&103\,$\pm$\,13          &(10.5) &(0.343)\\
LOCK850.15& 10 53 19.26 &+57 21 08.3 & 3834&353\,$\pm$\,20          & 2.39  &{\bf 0.009} \\
          & 10 53 18.99 &+57 21 15.6 & 3836&70.4\,$\pm$\,12.1       & 5.24  &0.141  \\
LOCK850.16& 10 51 51.67 &+57 26 36.0 & 3626&314\,$\pm$\,24          & 2.02  &{\bf 0.008} \\
LOCK850.17& 10 51 58.48 &+57 18 01.2 &13387&64.2\,$\pm$\,26.1       & 1.90  &{\bf 0.040} \\
          & 10 51 57.96 &+57 17 59.9 &17315&239\,$\pm$\,18          & 2.52  &{\bf 0.015} \\
LOCK850.18& 10 52 29.06 &+57 22 21.8 & 5563&103\,$\pm$\,13          &(11.8) &(0.381)\\
LOCK850.19& 10 52 36.09 &+57 31 19.6 &13661&118\,$\pm$\,15          & 3.12  &{\bf 0.045} \\
          & 10 52 35.52 &+57 31 11.7 &17536&242\,$\pm$\,19          & 7.51  &0.076  \\
          & 10 52 35.06 &+57 31 23.7 &17539&221\,$\pm$\,36          & 7.00  &0.075  \\
LOCK850.21& 10 52 56.79 &+57 30 37.9 & 2832&97.9\,$\pm$\,14.1       & 0.57  &{\bf 0.004}  \\
          & 10 52 57.80 &+57 30 35.3 & 2833&124\,$\pm$\,18          &(8.07) &(0.218)\\
LOCK850.22& 10 51 37.09 &+57 33 16.9 & 2895&402\,$\pm$\,21          & 7.41  &{\bf 0.045} \\
          & 10 51 36.68 &+57 33 32.8 & 2896&377\,$\pm$\,20          &(11.8) &(0.127)\\
LOCK850.23& 10 52 12.83 &+57 32 00.5 & 2722&116\,$\pm$\,17          &(9.70) &(0.288)\\
          & 10 52 14.71 &+57 31 54.7 &17516&57.3\,$\pm$\,11.3       & 7.86  &0.213  \\
LOCK850.24& 10 52 00.45 &+57 20 39.7 & 1842&455\,$\pm$\,21          & 2.45  &{\bf 0.007}  \\
LOCK850.26& 10 52 41.13 &+57 23 19.8 &  239&75.9\,$\pm$\,12.7       & 7.92  &0.193  \\
          & 10 52 40.66 &+57 23 09.7 & 5601&195\,$\pm$\,16          & 3.29  &{\bf 0.029}  \\
LOCK850.27& 10 52 03.45 &+57 18 19.3 & 1984&106\,$\pm$\,15          & 5.93  &0.117  \\
          & 10 52 04.77 &+57 18 05.9 & 1986&196\,$\pm$\,13          &(12.3) &(0.247)\\
LOCK850.28& 10 52 57.69 &+57 30 58.6 &13901&252\,$\pm$\,14          &(10.2) &(0.154)\\
LOCK850.29& 10 51 31.65 &+57 20 40.8 &18689&111\,$\pm$\,14          & 7.63  &0.149  \\
LOCK850.30& 10 52 07.68 &+57 19 04.1 & 2004&233\,$\pm$\,19          & 2.63  &{\bf 0.016}  \\
LOCK850.31& 10 52 15.96 &+57 16 19.2 & 3434&467\,$\pm$\,19          & 2.06  &{\bf 0.005}  \\
LOCK850.33& 10 51 55.40 &+57 23 12.9 & 1917&104\,$\pm$\,14          & 4.79  &0.091  \\
LOCK850.34& 10 52 13.66 &+57 33 21.3 & 2932&93.5\,$\pm$\,12.0       & 6.96  &0.153  \\
          & 10 52 14.21 &+57 33 27.9 & 2933&84.9\,$\pm$\,16.7       & 5.70  &0.134  \\
          & 10 52 13.97 &+57 33 32.8 & 2934&128\,$\pm$\,19          & 5.99  &0.101  \\
LOCK850.35& 10 52 46.46 &+57 20 56.8 &  153&51.0\,$\pm$\,12.7       & 3.72  &0.124  \\
          & 10 52 45.94 &+57 20 51.4 &15952&161\,$\pm$\,14          &(9.26) &(0.206)\\
          & 10 52 46.42 &+57 21 06.6 &15953&110\,$\pm$\,38          &(11.1) &(0.346)\\
          & 10 52 46.91 &+57 21 06.1 &15954&108\,$\pm$\,34          &(9.85) &(0.309)\\
          & 10 52 47.94 &+57 21 01.3 &19555&75.0\,$\pm$\,11.4       &(9.71) &(0.393)\\
LOCK850.36& ---		& ---        &  ---&5$\sigma$\,$<$\,60      &---    &---    \\
LOCK850.37& 10 51 24.60 &+57 23 31.0 & 1870&250\,$\pm$\,17          & 5.42  &{\bf 0.047}  \\
          & 10 51 24.27 &+57 23 41.4 &17334&126\,$\pm$\,16          & 6.64  &0.116  \\
LOCK850.38& 10 53 07.06 &+57 24 31.6 & 5682&260\,$\pm$\,16          & 0.41  &{\bf 0.001}  \\
LOCK850.39& ---		& ---	     &  ---&5$\sigma$\,$<$\,60      &---    &---    \\
LOCK850.40& 10 52 01.54 &+57 19 15.9 & 1994&91.9\,$\pm$\,15.0       & 3.84  &0.077  \\
          & 10 52 03.07 &+57 19 23.5 & 1997&85.2\,$\pm$\,14.0       &(11.5) &(0.422)\\
LOCK850.41& 10 52 00.24 &+57 24 21.5 &13508&475\,$\pm$\,37          & 3.71  &{\bf 0.013}  \\
          & 10 51 59.81 &+57 24 25.1 &13509&651\,$\pm$\,46          & 1.56  &{\bf 0.002}  \\
          & 10 51 59.27 &+57 24 13.3 &17394&108\,$\pm$\,15          &(11.4) &(0.358)\\
          & 10 52 00.19 &+57 24 15.3 &17395&212\,$\pm$\,22          &(8.72) &(0.147)\\
LOCK850.43& 10 52 56.64 &+57 23 51.4 & 5780&261\,$\pm$\,24          & 4.30  &{\bf 0.031}  \\
          & 10 52 56.61 &+57 23 58.0 & 5781&456\,$\pm$\,35          & 7.66  &{\bf 0.042}  \\
LOCK850.47& 10 52 34.85 &+57 25 04.6 &17453&107\,$\pm$\,16          &(11.3) &(0.359)\\
\end{tabular}
\end{center}

\noindent
$a$) Used to identify sources in Fig.~A1.\\
$b$) $P$ was calculated using a search radius of 8\,arcsec. For possible
counterparts with 8--15-arcsec offsets, $P$ was calculated using a
search radius of 15\,arcsec --- these values are listed in parentheses.
Reliable identifications ($P\le\rm 0.05$) within 8\,arcsec are listed in {\bf bold}.

\label{24um-lh}
\end{table}

%
%
\setcounter{table}{2}
\begin{table}
\scriptsize
\caption{Cont...}
\vspace{0.2cm}
\begin{center}
\begin{tabular}{lccrccc}
Nickname&\multicolumn{2}{c}{Position at 24\,$\mu$m}&ID&$S_{24\mu\rm m}$&Off &$P$\\
    &$\alpha_{\rm J2000}$&$\delta_{\rm J2000}$&   &/$\mu$Jy        &-set &   \\
    &h m s&$^{\circ}\ '\ ''$                  &   &                &$''$ &   \\
&&&&&&\\
LOCK850.48& 10 52 56.03 &+57 32 42.3 &18826&203\,$\pm$\,17          & 3.90  &{\bf 0.035}  \\
          & 10 52 55.37 &+57 32 46.5 &20105&85.2\,$\pm$\,13.7       & 7.03  &0.165  \\
LOCK850.52& 10 52 46.16 &+57 31 20.2 &18804&561\,$\pm$\,86          & 5.36  &{\bf 0.019}  \\
          & 10 52 45.76 &+57 31 20.6 &20079&310\,$\pm$\,35          & 2.28  &{\bf 0.009}  \\
LOCK850.53& 10 52 40.29 &+57 19 24.4 &13519&168\,$\pm$\,15          & 4.33  &{\bf 0.050}  \\
LOCK850.60& 10 51 43.50 &+57 24 35.8 & 1941&150\,$\pm$\,15          &(10.2) &(0.247)\\
          & 10 51 43.90 &+57 24 43.6 &13512&87.8\,$\pm$\,12.0       & 3.49  &0.070  \\
          & 10 51 43.08 &+57 24 52.2 &13513&82.5\,$\pm$\,13.4       & 7.44  &0.176  \\
          & 10 51 43.81 &+57 24 54.9 &13514&109\,$\pm$\,15          &(9.12) &(0.282)\\
LOCK850.63& 10 51 53.43 &+57 25 06.2 & 1925&53.0\,$\pm$\,13.0       & 4.01  &0.130  \\
          & 10 51 54.27 &+57 25 02.7 & 1931&236\,$\pm$\,17          & 3.78  &{\bf 0.029}  \\
          & 10 51 55.24 &+57 24 59.3 & 1932&79.1\,$\pm$\,12.0       &(12.2) &(0.461)\\
LOCK850.64& 10 52 51.67 &+57 32 48.7 & 2740&88.5\,$\pm$\,12.5       & 6.56  &0.150  \\
          & 10 52 52.57 &+57 32 48.9 & 2741&53.1\,$\pm$\,11.8       &(9.06) &(0.454)\\
          & 10 52 52.32 &+57 32 33.0 &12103&425\,$\pm$\,25          &(10.1) &(0.089)\\
LOCK850.66& 10 51 39.57 &+57 20 27.1 &13365&71.2\,$\pm$\,12.1       &(12.2) &(0.484)\\
LOCK850.67& 10 52 08.07 &+57 23 48.0 & 2044&102\,$\pm$\,14          &(10.4) &(0.340)\\
          & 10 52 08.87 &+57 23 56.3 & 2045&108\,$\pm$\,14          & 1.56  &{\bf 0.017}  \\
LOCK850.70& 10 51 47.88 &+57 30 44.6 & 2571&106\,$\pm$\,12          & 5.53  &0.108  \\
LOCK850.71& 10 52 19.10 &+57 18 57.3 & 3487&181\,$\pm$\,20          & 7.57  &0.100  \\
          & 10 52 19.53 &+57 19 04.8 & 3488&54.3\,$\pm$\,16.0       & 7.46  &0.212  \\
LOCK850.73& 10 51 41.92 &+57 22 18.6 & 1855&278\,$\pm$\,19          & 2.32  &{\bf 0.011}  \\ 
LOCK850.75& 10 53 15.19 &+57 26 45.9 & 5713&147\,$\pm$\,17          & 5.96  &0.089  \\
          & 10 53 15.02 &+57 26 53.2 & 5714&150\,$\pm$\,16          &(10.7) &(0.260)\\
          & 10 53 15.52 &+57 26 37.1 &16059&262\,$\pm$\,18          &(8.99) &(0.124)\\
LOCK850.76& 10 51 49.12 &+57 28 40.1 & 2512&592\,$\pm$\,26          & 5.07  &{\bf 0.016}  \\
LOCK850.77& 10 51 56.99 &+57 22 08.4 & 3602&51.7\,$\pm$\,13.1       & 1.67  &{\bf 0.042}  \\
          & 10 51 57.57 &+57 22 13.4 & 3603&154\,$\pm$\,15          & 5.66  &0.080  \\
          & 10 51 56.23 &+57 22 12.3 & 3608&55.4\,$\pm$\,13.5       & 6.65  &0.199  \\
LOCK850.78& 10 51 43.93 &+57 17 44.9 & 1734&85.6\,$\pm$\,14.7       &(13.0) &(0.462)\\
LOCK850.79& 10 51 51.22 &+57 21 27.8 & 1884&92.8\,$\pm$\,13.1       & 7.16  &0.158  \\
          & 10 51 52.63 &+57 21 24.4 & 1892&292\,$\pm$\,18          & 5.20  &{\bf 0.037}  \\
LOCK850.81& 10 52 31.52 &+57 17 51.6 &17353&3,667\,$\pm$\,51        &(9.59) &(0.007)\\
LOCK850.83& 10 53 07.17 &+57 28 40.0 & 2815&344\,$\pm$\,25          & 6.26  &{\bf 0.041}  \\
LOCK850.87& 10 51 53.36 &+57 17 30.5 & 1975&399\,$\pm$\,22          & 2.92  &{\bf 0.011}  \\
LOCK850.100&10 51 38.76 &+57 15 04.7 & 1623&118\,$\pm$\,13          & 5.65  &0.101  \\
\end{tabular}
\end{center}

\end{table}

%
%
\setcounter{table}{3}
\begin{table}
\scriptsize
\caption{Mid-IR properties of SMGs in the SXDF SHADES Source Catalogue.}
\vspace{0.2cm}
\begin{center}
\begin{tabular}{lccccc}
Nickname&\multicolumn{2}{c}{Position at 24\,$\mu$m}&$S_{24\mu\rm m}^a$&Off&$P^b$\\
    &$\alpha_{\rm J2000}$&$\delta_{\rm J2000}$&/$\mu$Jy           &-set&  \\
    &h m s&$^{\circ}\ '\ ''$                  &                   &$''$&  \\
&&&&&\\
SXDF850.01 &02 17 29.59 &$-$04 59 36.6&485\,$\pm$\,47      &(14.1)&(0.109)\\
SXDF850.02 &02 18 03.54 &$-$04 55 26.9&313\,$\pm$\,47      &0.58  &{\bf 0.001}  \\
SXDF850.04 &02 17 38.69 &$-$05 03 39.2&488\,$\pm$\,47      &2.01  &{\bf 0.005}  \\
SXDF850.05 &02 18 02.83 &$-$05 00 31.0&956\,$\pm$\,47      &1.88  &{\bf 0.002}  \\
SXDF850.06 &02 17 29.77 &$-$05 03 19.6&873\,$\pm$\,47      &7.21  &{\bf 0.017}\\
           &02 17 29.91 &$-$05 03 33.3&179\,$\pm$\,47      &6.82  &0.060\\
           &02 17 30.15 &$-$05 03 24.2&532\,$\pm$\,47      &6.26  &{\bf 0.023}  \\
SXDF850.07 &02 17 38.86 &$-$05 05 29.1&325\,$\pm$\,47      &5.46  &{\bf 0.031}  \\
SXDF850.08 &02 17 43.98 &$-$04 55 52.1&221\,$\pm$\,47      &7.26  &0.056  \\
SXDF850.10 &02 18 25.61 &$-$04 55 59.2&153\,$\pm$\,47      &5.78  &0.057  \\
           &02 18 24.88 &$-$04 56 03.3&177\,$\pm$\,47      &(8.21)&(0.132)  \\
SXDF850.11 &02 17 25.16 &$-$04 59 35.0&195\,$\pm$\,47      &2.52  &{\bf 0.017}  \\
SXDF850.12 &02 17 58.60 &$-$05 05 03.8&397\,$\pm$\,47      &(11.5)&(0.101)\\
SXDF850.14 &02 18 19.58 &$-$05 02 32.2&256\,$\pm$\,47      &(13.0)&(0.161)\\
           &02 18 18.77 &$-$05 02 49.0&240\,$\pm$\,47      &(8.70)&(0.110)  \\
SXDF850.16 &02 18 14.41 &$-$04 57 49.0&247\,$\pm$\,47      &(10.7)&(0.136)\\
SXDF850.17 &02 17 55.11 &$-$04 52 50.5&174\,$\pm$\,47      &(12.5)&(0.195)\\
SXDF850.19 &02 18 27.83 &$-$04 58 36.7&536\,$\pm$\,47      &5.40  &{\bf 0.019}  \\
SXDF850.21 &02 17 42.54 &$-$05 04 25.8&6,844\,$\pm$\,47    &4.37  &{\bf 0.001}  \\
SXDF850.24 &02 17 34.87 &$-$05 04 32.7&381\,$\pm$\,47      &6.64  &{\bf 0.034}  \\
SXDF850.27 &02 18 07.93 &$-$05 01 44.8&334\,$\pm$\,47      &3.83  &{\bf 0.018}  \\
SXDF850.28 &02 18 06.32 &$-$04 59 14.3&176\,$\pm$\,47      &(10.9)&(0.175)\\
           &02 18 06.43 &$-$04 59 20.3&477\,$\pm$\,47      &(10.4)&(0.075)\\
           &02 18 06.87 &$-$04 59 12.4&877\,$\pm$\,47      &4.04  &{\bf 0.007}  \\
SXDF850.29 &02 18 16.49 &$-$04 55 08.2&971\,$\pm$\,47      &3.63  &{\bf 0.005}  \\
SXDF850.30 &02 17 40.00 &$-$05 01 15.1&523\,$\pm$\,47      &4.69  &{\bf 0.016}  \\
SXDF850.31 &02 17 36.75 &$-$04 56 10.4&452\,$\pm$\,47      &(14.6)&(0.120)\\
           &02 17 35.83 &$-$04 55 56.7&594\,$\pm$\,47      &7.10  &{\bf 0.025}  \\
           &02 17 36.37 &$-$04 56 03.4&251\,$\pm$\,47      &6.03  &{\bf 0.043}  \\
SXDF850.32 &02 17 22.58 &$-$05 00 44.4&168\,$\pm$\,47      &7.80  &0.066\\
SXDF850.35 &02 18 00.86 &$-$04 53 06.6&215\,$\pm$\,47      &4.66  &{\bf 0.036}  \\
SXDF850.36 &02 18 31.92	&$-$04 59 59.1&162\,$\pm$\,47      &(13.0)&(0.205)\\
           &02 18 31.95	&$-$04 59 53.2&177\,$\pm$\,47      &7.69  &0.065\\
           &02 18 33.04	&$-$04 59 41.4&181\,$\pm$\,47      &(12.9)&(0.195)\\
           &02 18 31.86	&$-$04 59 37.3&182\,$\pm$\,47      &(11.7)&(0.181)\\
SXDF850.37 &02 17 24.41 &$-$04 58 42.0&183\,$\pm$\,47      &2.14  &{\bf 0.015}  \\
SXDF850.45 &02 18 30.11	&$-$05 05 35.4&157\,$\pm$\,47      &(12.8)&(0.206)\\
SXDF850.47 &02 17 34.37 &$-$04 58 59.9&298\,$\pm$\,47      &7.56  &{\bf 0.048}  \\
           &02 17 33.72 &$-$04 58 58.7&250\,$\pm$\,47      &2.69  &{\bf 0.015}  \\
SXDF850.52 &02 18 05.09 &$-$05 04 52.7&151\,$\pm$\,47      &3.08  &{\bf 0.029}  \\
SXDF850.56 &02 17 51.23	&$-$05 06 30.5&299\,$\pm$\,47      &(8.34)&(0.086)\\
SXDF850.69 &02 17 51.77 &$-$05 02 58.6&157\,$\pm$\,47      &(9.59)&(0.168)\\
           &02 17 51.06 &$-$05 03 02.8&724\,$\pm$\,47      &(13.0)&(0.068)\\
SXDF850.71 &02 18 21.28 &$-$04 58 58.8&404\,$\pm$\,47      &4.47  &{\bf 0.019}  \\
SXDF850.76 &02 17 56.32 &$-$05 06 25.5&183\,$\pm$\,47      &(8.86)&(0.140)  \\
SXDF850.77 &02 17 36.02 &$-$05 04 28.2&726\,$\pm$\,47      &7.32  &{\bf 0.021}\\
           &02 17 36.51 &$-$05 04 25.6&295\,$\pm$\,47      &6.65  &{\bf 0.042}\\
SXDF850.86 &02 18 16.66 &$-$05 04 00.0&208\,$\pm$\,47      &(9.13)&(0.131)\\
SXDF850.88 &02 18 01.54 &$-$05 04 42.1&446\,$\pm$\,47      &(10.4)&(0.079)\\
SXDF850.91 &02 17 34.24	&$-$04 57 14.3&203\,$\pm$\,47      &(12.9)&(0.184)\\
SXDF850.94 &02 17 40.26 &$-$04 58 24.0&187\,$\pm$\,47      &6.83  &0.059\\
           &02 17 39.24 &$-$04 58 13.1&198\,$\pm$\,47      &(13.4)&(0.192)\\
SXDF850.96 &02 18 00.40 &$-$05 02 01.5&478\,$\pm$\,47      &(12.7)&(0.097)\\
SXDF850.119&02 17 56.20 &$-$04 53 02.1&784\,$\pm$\,47      &7.20  &{\bf 0.019}  \\
           &02 17 55.65 &$-$04 52 58.0&202\,$\pm$\,47      &(10.8)&(0.158)  \\
           &02 17 56.24 &$-$04 52 50.9&275\,$\pm$\,47      &4.62  &{\bf 0.029}  \\
\end{tabular}
\end{center}

\noindent
$a$) Objects missing here, but listed in Table~\ref{radio-sxdf}, have upper
limits of 5$\sigma$\,$<$\,235\,$\mu$Jy at 24\,$\mu$m.\\
$b$) $P$ was calculated using a search radius of 8\,arcsec. For possible
counterparts with 8--15-arcsec offsets, $P$ was calculated using a
search radius of 15\,arcsec --- these values are listed in parentheses.
Reliable identifications ($P\le\rm 0.05$) within 8\,arcsec are listed in {\bf bold}.

\label{24um-sxdf}
\end{table}

%
%
\begin{table}
\scriptsize
\caption{Alternative names for the SHADES Source Catalogue.}
\vspace{0.2cm}
\begin{center}
\begin{tabular}{cccccc}
\multicolumn{2}{c}{SHADES}&8-mJy$^a$&MAMBO$^b$ &Bolocam$^c$  &Chapman$^d$\\
LOCK J---&LOCK--  &LE850.--    &LE1200.--     &LE1100.--       &SMM J--   \\
&&&&&\\
105201+572443&850.01&01&005&14&105201.25+572445.7\\
105257+572105&850.02&--&004&01&--                \\
105238+572436&850.03&02&001&08&105238.30+572435.8\\
105204+572658&850.04&14&003&--&--\\
105204+572526&850.06&04&-- &--&--\\
105153+571839&850.08&27&104&--&--\\
105216+572504&850.09&29&042&--&--\\
105227+572513&850.12&16&006&16&105227.58+572512.4\\
105230+572215&850.14&06&010&05&105230.73+572209.5\\
105151+572637&850.16&07&096&--&105151.69+572636.0\\
105158+571800&850.17&03&011&--&105158.02+571800.2\\
105227+572217&850.18&--&009&--&105227.77+572218.2\\
105200+572038&850.24&32&-- &--&--\\
105203+571813&850.27&--&007&04&--\\
105130+572036&850.29&11&-- &--&--\\
105207+571906&850.30&12&-- &--&105207.49+571904.0\\
105155+572311&850.33&18&012&--&105155.47+572312.7\\
105202+571915&850.40&21&-- &--&--\\
105159+572423&850.41&08&014&17&105200.22+572420.2\\
105148+572838&850.76&--&-- &15&--\\
\end{tabular}
\end{center}

\noindent
$a$) Scott et al.\ (2002).\\
$b$) Greve et al.\ (2004); Ivison et al.\ (2005).\\
$c$) Laurent et al.\ (2005).\\
$d$) Chapman et al.\ (2005).

\label{names}
\end{table}

\subsection{Optical imaging}

$R$-band optical imaging for the LH and SXDF were obtained
using the Subaru 8-m telescope. The LH data were taken from
the archive and are described in Ivison et al.\ (2004) and reach a
3$\sigma$ depth of 27.7 mag; similar data for SXDF are described by
Furusawa et al.\ (in preparation), reaching a 3$\sigma$ depth of 27.5
mag (both on the Vega scale, for 2-arcsec-diameter apertures).

\subsection{Near- and mid-IR imaging}

The near- and mid-IR data employed here were obtained using IRAC (at
4.5 and 8\,$\mu$m) and MIPS (at 24\,$\mu$m). The imaging covers the
entire SHADES region of the LH to near-uniform depths of
$\sigma$ = 0.54, 4.4 and 11\,$\mu$Jy at 4.5, 8 and 24\,$\mu$m,
respectively (Egami et al., in preparation), with flux calibration
accurate to $\pm$10 per cent, that is approximately 3$\times$ deeper
at 24\,$\mu$m than the data used by Egami et al.\ (2004), Serjeant et
al.\ (2004) and Ivison et al.\ (2004). In the SXDF, IRAC and MIPS data
are available from the {\em Spitzer} Wide-area InfraRed Extragalactic
(SWIRE -- Lonsdale et al.\ 2003) survey and reach a near-uniform depth
of $\sigma$ = 1.1, 7.5 and 48\,$\mu$Jy at 4.5, 8 and 24\,$\mu$m (Shupe
et al., in preparation). For comparison, the 5-$\sigma$ confusion
limit at 24\,$\mu$m, with 20 beams per source, is around 56\,$\mu$Jy.

\section{Associations between submm galaxies and radio/mid-IR sources}
\label{associations}

Observations in the submm waveband are sensitive to cold dust created
for the most part by supernovae (SNe) and stellar winds, re-radiating
energy absorbed from hot, young stars (Whittet 1992).  The radio
waveband is also sensitive to SNe -- and hence to recent star
formation -- via synchrotron radiation from relativistic
electrons. Near-IR wavelengths probe photospheric emission from stars
and in the mid-IR, at 24\,$\mu$m, we are sensitive to emission from
dust in the circumnuclear torus of AGN and to the warmest dust in
starbursts. The correlation between submm and radio emission from SMGs
is poorer than expected (from local studies -- e.g.\ Dunne et al.\
2000), probably due to a wide range of characteristic dust
temperatures and to the effect of radio-loud AGN; nevertheless,
predicting the rest-frame far-IR properties of SMGs is better
accomplished from the radio end of the SED than from the near- or
mid-IR, adding to the benefit of very high spatial resolution
($\sim$0.1\,arcsec) available at radio wavelengths (Chapman et al.\
2004; Muxlow et al.\ 2005; Biggs et al., in preparation) and making it
the waveband of choice for the identification of counterparts at other
wavelengths and several related objectives.

A radio source peaking at $\ge$\,4\,$\sigma$ with an integrated flux
density in excess of 3\,$\sigma$, in either the high- or
low-resolution images, where $\sigma$ is the noise measured locally,
is considered a {\it robust} detection. In the LH and SXDF,
the surface densities of all radio sources above this threshold are
$\rm 1.9\pm 0.1$\,arcmin$^{-2}$ (Ivison et al.\ 2005) and $\rm 1.6\pm
0.1$\,arcmin$^{-2}$, respectively. Where a robust detection does not
exist, we list those sources peaking at $\ge$\,3\,$\sigma$ with an
integrated flux density in excess of 2\,$\sigma$, these being
considered {\it tentative} detections.  Positions and flux densities
were measured using {\sc jmfit} with multi-component fits: usually a
Gaussian and a surface baseline, with an extra Gaussian component for
close multiple radio sources. To enable us to make appropriate
corrections for bandwidth smearing -- the radio flavour of chromatic
aberration which causes the peak flux density to fall as a function of
distance from the pointing centre -- measurements were made in images
of each pointing rather than in the final, large mosaic. In cases
where sources appeared in more than one 400-arcmin$^2$ pointing,
error-weighted means were obtained.

For each SMG we have searched for potential radio (1.4-GHz)
counterparts inside a positional error circle of radius 8\,arcsec (see
\S4), also listing those within 12.5\,arcsec for completeness. This
relatively large search area ensures that no real associations are
missed. At the extreme depths reached by the radio imaging reported
here, the cumulative surface density of radio sources in the
8-arcsec-radius error circles yields one robust source in every ten
search areas, though not all will be regarded as significant
associations as we shall see shortly.

We have also searched for potential 24-$\mu$m counterparts inside a
positional error circle of radius 8\,arcsec, listing those within
15\,arcsec for completeness (a slightly larger radius than for the
radio counterparts to account for the larger 24-$\mu$m beam).

To quantify the formal significance of each of the potential
submm/radio and submm/mid-IR associations we have used the method of
Downes et al.\ (1986; see also Dunlop et al.\ 1989). This corrects the
raw Poisson probability, $P$, that a radio or 24-$\mu$m source of the
observed flux density could lie at the observed distance from the SMG,
for the number of ways that such an apparently significant association
could have been uncovered by chance.

%
%
\begin{table}
\scriptsize
\caption{Identification summary.}
\vspace{0.2cm}
\begin{center}
\begin{tabular}{lclc}
Nickname        &Robust identification?    &Nickname       &Robust identification?   \\
&&&\\
LOCK850.01 &$\bullet\;\circ$   &SXDF850.01 & $\bullet$         \\
LOCK850.02 &$\bullet \circ\dagger$   &SXDF850.02 & $\bullet\;\circ$   \\
LOCK850.03 &$\bullet \circ\dagger$   &SXDF850.03 & $\bullet$         \\
LOCK850.04 &$\bullet \circ\dagger$   &SXDF850.04 & $\bullet\;\circ$   \\
LOCK850.05 &                  &SXDF850.05 & $\bullet\;\circ$   \\
LOCK850.06 &$\bullet\;\circ$   &SXDF850.06 & $\bullet\;\circ\;\dagger$   \\
LOCK850.07 &$\bullet\;\circ$   &SXDF850.07 & $\bullet\;\circ$   \\
LOCK850.08 &$        \circ$   &SXDF850.08 & $\bullet$         \\
LOCK850.09 &$\bullet\;\circ$   &SXDF850.09 &                   \\
LOCK850.10 &$\bullet      $   &SXDF850.10 & $\bullet$         \\
LOCK850.11 &                  &SXDF850.11 & $\bullet\;\circ$   \\
LOCK850.12 &$\bullet\;\circ$   &SXDF850.12 & $\bullet$         \\
LOCK850.13 &                  &SXDF850.14 & $\bullet$         \\
LOCK850.14 &                  &SXDF850.15 &                   \\
LOCK850.15 &$\bullet \circ\dagger$   &SXDF850.16 & $\bullet$         \\
LOCK850.16 &$\bullet\;\circ$   &SXDF850.17 &                   \\
LOCK850.17 &$\bullet\;\circ$   &SXDF850.18 & $\bullet$         \\
LOCK850.18 &$\bullet      $   &SXDF850.19 & $\bullet\;\circ$   \\
LOCK850.19 &$        \circ$   &SXDF850.20 &                   \\
LOCK850.21 &$        \circ$   &SXDF850.21 & $\bullet\;\circ$   \\
LOCK850.22 &$        \circ$   &SXDF850.22 &                   \\
LOCK850.23 &                  &SXDF850.23 & $\bullet$         \\
LOCK850.24 &$\bullet\;\circ$   &SXDF850.24 & $\bullet \circ \dagger$   \\
LOCK850.26 &$\bullet\;\circ$   &SXDF850.25 &                   \\
LOCK850.27 &                  &SXDF850.27 & $\bullet\;\circ$   \\
LOCK850.28 &                  &SXDF850.28 & $\bullet \circ \dagger$   \\
LOCK850.29 &$\clubsuit    $   &SXDF850.29 & $\bullet\;\circ$   \\
LOCK850.30 &$\bullet\;\circ$   &SXDF850.30 & $\bullet\;\circ$   \\
LOCK850.31 &$\bullet\;\circ$   &SXDF850.31 & $\bullet\;\circ$   \\
LOCK850.33 &$\bullet      $   &SXDF850.32 &                   \\
LOCK850.34 &$\bullet\ddagger$ &SXDF850.35 & $\bullet\;\circ$   \\
LOCK850.35 &                  &SXDF850.36 &                   \\
LOCK850.36 &                  &SXDF850.37 & $\bullet\;\circ$   \\
LOCK850.37 &$\bullet\circ\ddagger$&SXDF850.38 & $\bullet$         \\
LOCK850.38 &$\bullet\;\circ$   &SXDF850.39 &                   \\
LOCK850.39 &                  &SXDF850.40 & $\bullet$         \\
LOCK850.40 &$\bullet      $   &SXDF850.45 &                   \\
LOCK850.41 &$\bullet \circ\dagger$   &SXDF850.47 & $\bullet \circ \dagger$   \\
LOCK850.43 &$        \circ$   &SXDF850.48 &                   \\
LOCK850.47 &                  &SXDF850.49 &                   \\
LOCK850.48 &$        \circ\ddagger$   &SXDF850.50 & $\bullet \ddagger$         \\
LOCK850.52 &$\bullet\;\circ$   &SXDF850.52 & $\bullet \circ \dagger$   \\
LOCK850.53 &$        \circ$   &SXDF850.55 & $ \bullet$        \\
LOCK850.60 &                  &SXDF850.56 &                   \\
LOCK850.63 &$\bullet\;\circ$   &SXDF850.63 &                   \\
LOCK850.64 &                  &SXDF850.65 &                   \\
LOCK850.66 &                  &SXDF850.69 &                   \\
LOCK850.67 &$        \circ$   &SXDF850.70 &                   \\
LOCK850.70 &$\clubsuit    $   &SXDF850.71 & $\circ$           \\
LOCK850.71 &$\bullet      $   &SXDF850.74 & $\bullet$         \\
LOCK850.73 &$\bullet \circ\dagger$   &SXDF850.76 &                   \\
LOCK850.75 &                   &SXDF850.77 & $\bullet\;\circ$   \\
LOCK850.76 &$\bullet\;\circ$   &SXDF850.86 &                   \\
LOCK850.77 &$\bullet\;\circ$   &SXDF850.88 &                   \\
LOCK850.78 &                  &SXDF850.91 &                   \\
LOCK850.79 &$        \circ$   &SXDF850.93 &                   \\
LOCK850.81 &                  &SXDF850.94 &                   \\
LOCK850.83 &$        \circ$   &SXDF850.95 &                   \\
LOCK850.87 &$\bullet\;\circ$   &SXDF850.96 & $\bullet$         \\
LOCK850.100&$\clubsuit     $   &SXDF850.119& $\bullet\;\circ$   \\
\end{tabular}
\end{center}

\noindent
$\bullet$ indicates a robust ($P\le\rm 0.05$) radio identification.\\
$\circ$ indicates a robust identification at 24\,$\mu$m.\\
$\clubsuit$ coincident radio and 24-$\mu$m emission (both $P\le\rm 0.1$) yields
reliable identification.\\
$\dagger$ indicates multiple robust ($P\le\rm 0.05$) identifications.\\
$\ddagger$ close visual inspection of the data reveals more than one good identification.

\label{id-summary}
\end{table}

The positions, flux densities and $P$ values of all LH and SXDF radio
and 24-$\mu$m counterparts are presented in Tables~\ref{radio-lh}
through \ref{24um-sxdf}, adopting those counterparts within 8\,arcsec
with $P\le\rm 0.05$ as robust. $P$ values for counterparts with larger
separations are listed in parenthesis, using search radii of 12.5 or
15\,arcsec at 1.4\,GHz and 24\,$\mu$m, respectively. We have also
searched for cases where {\em coincident} radio and 24-$\mu$m
counterparts within 8\,arcsec have $P_{\rm 1.4GHz}$ and $P_{\rm 24\mu
m}$ $\le$ 0.10, finding three such cases. Figs A1 and A2 contain
25-arcsec\,$\times$\,25-arcsec postage stamp images centred on the LH
and SXDF SMG positions, respectively. Alternative names used for these
SMGs in the literature are listed in Table~\ref{names}.

Our identifications -- based on radio and/or 24-$\mu$m data -- are
summarised in Table~\ref{id-summary}. Clements et al.\ (in
preparation) and Dye et al.\ (in preparation) will present independent
identification analyses in SXDF and LH, respectively, using optical
and near-IR colours which are believed to provide a useful complement
to deep radio imaging (e.g.\ Webb et al.\ 2003b; Pope et al.\ 2005).

Of the 32 identifications made in only one waveband -- equal numbers
in each field -- 21 are radio counterparts, mainly in SXDF. Of these
21 SMGs, only seven have detections at 24\,$\mu$m that have not made
the grade via the $P$ statistic. Of the 11 mid-IR-only
identifications, five have radio counterparts just above our adopted
$P\rm\le 0.05$ threshold.

In total, we find robust counterparts for two thirds (79) of the 120
sources in the SHADES Source Catalogue, entirely consistent with
previous studies (Ivison et al.\ 2002, 2005; Pope et al.\ 2006).

\section{On the uncertainty in SMG positions}

SCUBA-2 will herald a vast increase in the number of catalogued SMGs,
covering tens of square degrees. Radio coverage of such areas at the
depth employed here will not be trivial to acquire, even in the era of
e-VLA and LOFAR. It is interesting, therefore, to speculate about
whether our knowledge of panchromatic SMG properties will progress in
the absence of radio detections (and hence accurate positions and
counterparts at other wavelengths) for the majority of SMGs. Can we
determine the significance of submm detections required to enable
spectroscopic follow-up with modern integral-field unit (IFU)
spectrometers such as KMOS on the 8-m Very Large Telescope (Sharples
et al.\ 2006)?

Submm positions for the SHADES Source Catalogue were deduced by
fitting to the beam pattern in an optimally filtered map (i.e.\ after
smoothing with the beam), then averaging over four independent
reductions of the same raw data (Coppin et al.\ 2006). One reduction
adopted the centre of the nearest 3-arcsec pixel as the position, while
the others used 1-arcsec pixels, so a small rounding error adds to the
uncertainty.  Ignoring this minor effect, the positional uncertainty
should be $\rm \Delta \alpha = \Delta \delta =
0.6\,\theta\,(SNR)^{-1}$ in the limit where centroiding uncertainty
dominates over systematic astrometry errors and for uncorrelated
Gaussian noise. Here, $\Delta \alpha$ and $\Delta \delta$ are the
r.m.s.\ errors in R.A.\ and Dec., respectively, $\theta$ is the {\sc
fwhm} of the submm beam and SNR is the signal-to-noise ratio after
correction for flux boosting (see Appendix B for derivations).

We can use our radio associations, which should provide near-perfect
positions, to check whether the uncertainties in submm position are
consistent with this theoretical expectation, given the size of the
JCMT's beam and the SNR of the 850-$\mu$m sources.

For a Gaussian distribution of errors in R.A.\ and Dec., the
distribution of radial offsets ($r e^{-r^2/2\sigma^2}$) peaks at
$\sigma$ (= $\Delta \alpha = \Delta \delta$). This peak bounds only
39.3 per cent of sources, with 68 per cent of the anticipated radial
offsets lying within 1.51$\sigma$ (close to, but not precisely equal
to $\sqrt{\rm 2}\times\sigma$). 86.5, 95.6 and 98.9 per cent of
offsets are expected to lie within 2\,$\sigma$, 2.5\,$\sigma$ and
3\,$\sigma$, respectively.

\begin{figure}
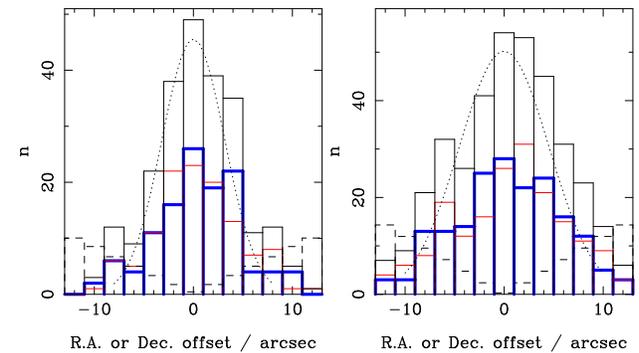

\centerline{\psfig{file=f2a.eps,angle=270,width=1.6in}
\psfig{file=f2b.eps,angle=270,width=1.6in}}
\vspace{-0.5cm}
\noindent{\small\addtolength{\baselineskip}{-3pt}}
\caption{Histograms of positional offsets between the positions of the
SMGs and those of the counterparts ({\em left:} radio; {\em right:}
24\,$\mu$m), in R.A.\ ($\alpha$, thick blue), Dec.\ ($\delta$, red)
and both together (black). The dashed lines show the expected
distribution and absolute level for a randomly distributed population
with the average counts seen in the LH and SXDF images. The
dotted lines show Gaussian fits with $\sigma$ = 3.2\,arcsec which were
constrained to be centred at $\alpha = \delta = \rm 0$ arcsec.}
\label{hist-offsets}
\end{figure}

Fig.~\ref{hist-offsets} shows a histogram of offsets between the
positions of the SMGs and those of all the radio counterparts listed
in Tables~\ref{radio-lh} and \ref{radio-sxdf}. Here, $\alpha$ and
$\delta$ represent the R.A./cos\,$\delta$ and Dec.\ offsets between
submm and radio positions; radial offsets are thus $\sqrt{\alpha^2 +
\delta^2}$. The value of $\sigma$ observed in Fig.~\ref{hist-offsets}
is approximately 3\,arcsec so our adopted search radius of 8\,arcsec
(\S3) corresponds to $\sim$2.5\,$\sigma$ and should thus include
$\sim$95 per cent of all genuine radio identifications; moreover,
since the typical deboosted SNR of the submm sources is $\sim$3
(Coppin et al.\ 2006), the theoretical expectation is also $\sigma
\sim\rm 3$\,arcsec (from equation 2 of Appendix B for $\theta$ =
14.5\,arcsec and $\rm SNR=3$). It is clear, therefore, that the
observed distribution of radial offsets for the radio identifications
is at least comparable with theoretical expectations.

%
%

We can quantify this more precisely in two ways. First, we can use the
distribution of radial offsets for all radio identification
counterparts and attempt to correct statistically for background
contamination: the dashed line in Fig.~\ref{hist-offsets} represents
the distribution and absolute level of a randomly distributed radio
population with the counts seen in the LH radio image
(\S2.1). The number of radio identifications within a 6-arcsec radius
of the submm positions is seen to exceed the random level by almost
two orders of magnitude, which gives us (additional) confidence that
the vast majority of the radio identifications are truly associated
with the SMGs. The finite search radius within which we have hunted
for radio counterparts explains why the observed number of
counterparts falls below that predicted for a random population in the
outermost bins of Fig.~\ref{hist-offsets}. Note that
Fig.~\ref{hist-offsets} uses {\em all} the radio identifications,
rather than just those with the lowest $P$ values, so any bias present
is due only to the finite search radii used to find radio emitters for
this analysis (12.5\,arcsec).

Having corrected the observed distributions for the expected unrelated
`field' radio sources (those in the background and foreground), a
Gaussian fit centred at $\alpha = \delta = \rm 0$\,arcsec, shown in
Fig.~\ref{hist-offsets}, yields a {\sc fwhm} of
7.5\,$\pm$\,0.7\,arcsec (7.4\,$\pm$\,0.6\,arcsec if the centroid is
unconstrained). This translates into $\Delta \alpha = \Delta \delta$ =
{\sc fwhm}/2.354 = 3.2\,arcsec. Our correction for the expected
`field' sources should have dealt with any broadening due to radio
sources unrelated to the SMGs. The median SNR of the radio-detected
sample used in this analysis is 3.0, after correction for
Malmquist-type bias, which implies that $\rm \Delta \alpha=\Delta
\delta =\rm 0.66\,\theta\,(SNR)^{-1}$, adopting $\theta$ =
14.5\,arcsec, i.e.\ 10 per cent higher than expected.

This procedure was replicated for the 24-$\mu$m counterparts listed in
Tables~\ref{24um-lh} and \ref{24um-sxdf}, correcting for blank-field,
background source densities of 4.5 and 1.2\,arcmin$^{-2}$ to limits of
50 and 150\,$\mu$Jy, respectively. The result, shown in the right
panel of Fig.~\ref{hist-offsets}, is a wider distribution, borne out
by the best-fit Gaussian: a {\sc fwhm} of 10.7\,$\pm$\,1.0\,arcsec,
when constrained to be centred at $\alpha = \delta = \rm 0$\,arcsec,
or $\Delta \alpha=\Delta \delta$ = 4.5\,arcsec. The low accuracy of
the 24-$\mu$m positions relative to those determined at 1.4\,GHz can
account for most of the extra width.

\begin{figure}
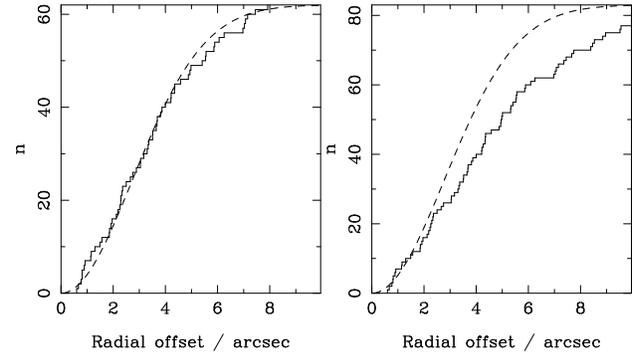

\centerline{\psfig{file=f3a.eps,angle=270,width=1.6in}
\psfig{file=f3b.eps,angle=270,width=1.6in}}
\vspace{-0.5cm}
\noindent{\small\addtolength{\baselineskip}{-3pt}}
\caption{{\em Left:} Cumulative distribution of radial offsets between
the radio and submm positions for the 62 statistically secure ($P
\leq\rm 0.05$) radio identifications. The dashed line shows the
predicted distribution ($1-e^{-r^2/2\sigma^2}$) assuming that the
positional uncertainty in R.A.\ or Dec.\ is given by $\sigma\rm =
0.6\, \theta$/SNR, as discussed in Appendix B, where we have used the
mean SNRs for the sample. A KS test yields a probability of 0.57 that
the data are consistent with the model. {\em Right:} The same
predicted probability distribution (dashed line), this time compared
with the cumulative distribution for all 83 sources with candidate
radio identifications (i.e.\ including those for which $P >\rm 0.05$).
The poor fit in the right-hand plot -- a KS test yields a probability
of 0.0003 that the data are consistent with the model -- demonstrates
the importance of using the $P$ statistic to filter the candidate list
of associations.}
\label{hist-jsd}
\end{figure}

As a second way of quantifying this approach, we can consider only the
subset of `robust' radio identifications ($P \leq\rm 0.05$) on the
basis that this should provide the most secure measure of the true
distribution of uncertainty in the submm positions. The radial offset
distribution for this subset of 62 sources is shown in the left-hand
panel of Fig.~\ref{hist-jsd}, where it is compared with the predicted
cumulative distribution ($1-e^{-r^2/2\sigma^2}$), using $\sigma\rm =
0.6\,\theta\,(SNR)^{-1}$ as discussed in Appendix B. For this
calculation we have adopted $\theta$ = 14.5\,arcsec and SNR = 3.17
(the average SNR for the deboosted 850-$\mu$m flux densities of these
62 sources).  It is clear from this plot that the predicted
distribution is in excellent agreement with that observed for this
secure subset of identified sources; indeed, a Kolmogorov-Smirnov (KS)
test yields a 57-per-cent probability that the data are consistent
with the model. For completeness, the right-hand panel in
Fig.~\ref{hist-jsd} shows the same predicted probability distribution,
this time compared with the cumulative distribution for all 83 sources
with candidate radio identifications, i.e.\ including those for which
$P >\rm 0.05$. The same KS test now yields a probability of less than
0.1 per cent. These plots give confidence that the radial offset
distribution of secure identifications is consistent with that
expected given the JCMT's beam and the deboosted 850-$\mu$m flux
densities of the SHADES sources and that there is no additional
significant source of astrometric error in the submm maps. They also
demonstrate the importance of using the $P$ statistic to filter the
candidate list of associations for robust identifications.

In conclusion, there is no evidence for significant additional sources
of positional error. For an SMG discovered in a submm survey where a
Malmquist-type bias correction has not been applied, we can
parameterise its positional uncertainty as follows. Its position
having been determined after smoothing with the beam, a circle of
radius $\rm 0.91\,\theta\,(SNR_{\rm app}^2 - (2\beta+4))^{-1/2}$, for
power-law counts of the form $N(>f)\propto f^{-\beta}$, has a 68 per
cent chance of containing the submm emitter (where SNR$_{\rm app}$ is
the raw SNR, uncorrected for flux boosting -- see Appendix B), or $\rm
0.91\,\theta\,(SNR)^{-1}$ if a correction {\em has} been
applied. These correspond to conventional 1-$\sigma$ error circles.

%
%

We return now to our initial motivation for this study of positional
uncertainty, namely the feasibility of a spectroscopic redshift
distribution for SMGs based on KMOS near-IR spectroscopy of an
unbiased sample. Such a programme could afford to lose one SMG due to
positional error during a single deployment of the 24 KMOS IFUs. Each
IFU covers 2.8\,$\times$\,2.8-arcsec$^2$ so, leaving room for seeing
effects, we require $\rm 2\sigma\sim 2.5$\,arcsec to ensure that 95.6
per cent of SMGs fall within the central 5\,arcsec$^2$ of each
IFU. Our parameterisation suggests that this level of accuracy
requires an SMG sample cut at $\rm SNR\ge 20$. Adopting the source
counts of Coppin et al.\ (2006), a source density of
$\sim$2200\,deg$^{-2}$ -- sufficient to employ all 24 KMOS IFUs --
would require that we probe the 3-mJy SMG population; this, in turn,
would require that we delve well below the 850-$\mu$m confusion limit
to ensure $\rm SNR\ge 20$, or that we utilise positions determined
using the 450-$\mu$m data that are acquired simultaneously by
SCUBA-2. Optimal exploitation of KMOS may require sharing the IFUs
with other programmes in all but the deepest SCUBA-2 survey fields.

\section{Multiple radio counterparts}
\label{multiple}

%
%
\begin{figure}
\centerline{\psfig{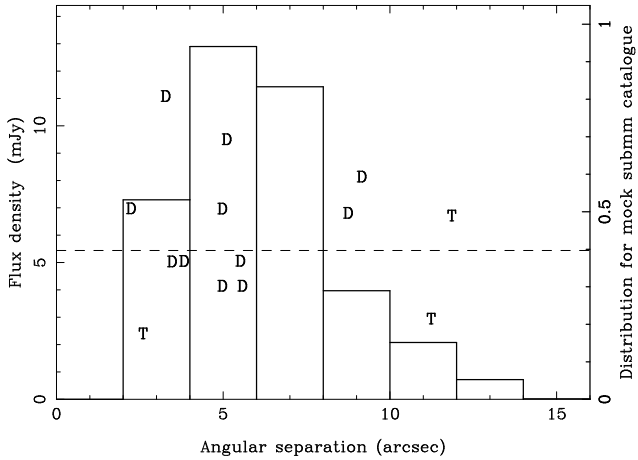}}
\vspace{-0.5cm}
\noindent{\small\addtolength{\baselineskip}{-3pt}}
\caption{Deboosted submm flux density versus the angular separation of
the counterparts for SMGs with more than one robust radio
identification ($P\le\rm 0.05$) within 8\,arcsec of the 850-$\mu$m
position. Points denote radio doubles (D) and triples (T). The average
of the single, unresolved or barely resolved radio counterparts is
represented by a dashed line. The paucity of data at very low and high
separations is due to our finite spatial resolution on the one hand
and to our use of a finite search area and the $P$ statistic on the
other. The histogram shows the distribution of angular separations,
scaled arbitrarily, for multiple identifications found in the
Monte-Carlo simulations described in \S5.}
\label{doubles}
\end{figure}

%
%
\begin{figure}
\centerline{\psfig{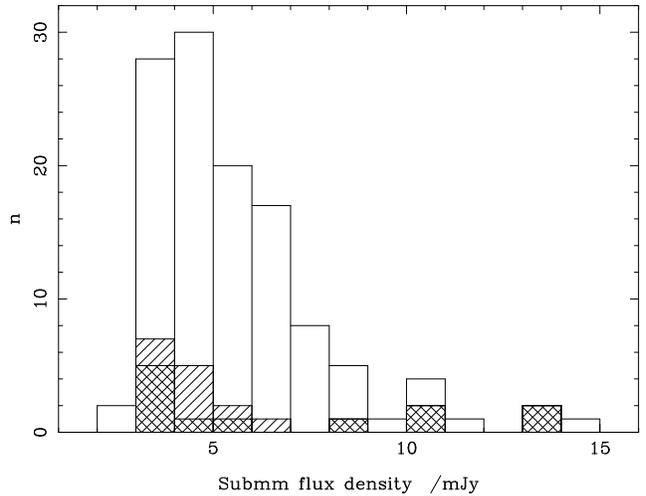}}
\vspace{-0.5cm}
\noindent{\small\addtolength{\baselineskip}{-3pt}}
\caption{Histogram of deboosted submm flux density for the full SHADES
Source Catalogue.  Cross-hatched areas represent the 12 SMGs with two
or more radio components within 8\,arcsec of the 850-$\mu$m position,
associated robustly with the SMG ($P\le\rm 0.05$); single-hatched
areas represent the seven SMGs with multiple, significant 24-$\mu$m
identifications. Five SMGs have multiple, significant radio {\em and}
24-$\mu$m identifications.}
\label{nflux}
\end{figure}

A number of SMGs with more than one robust ($P\le\rm 0.05$) radio
counterpart are apparent in Tables~\ref{radio-lh}--\ref{radio-sxdf}
and Figs~A1--A2: seven in the LH and five in the SXDF. This
tendency for $\sim$10 per cent of SMGs to have multiple radio
identifications was noted previously by Ivison et al.\ (2002) and Pope
et al.\ (2006). The probability of an SMG possessing two statistically
significant radio counterparts was quantified by placing 10$^6$ fake
sources into the real LH and SXDF radio fields and counting the number
of $P<\rm 0.05$ radio counterparts -- a simple Monte-Carlo
approach. This revealed that the calibration of the $P$ statistic is
secure, with $P=\rm 0.05$ yielding 5.05 spurious associations for
every 100 fake SMGs. Multiple robust counterparts are far rarer,
however. For every 100 fake SMGs the simulations suggest that only
0.22 will have more than one secure radio identification by chance, a
figure dominated by doubles, so at first sight the observed tendency
for multiple robust radio counterparts is highly significant.
However, we know that around half (65) of the SHADES SMGs have a {\em
real} association with a radio emitter, or 59 after accounting for the
six spurious identifications we expect (0.05\,$\times$\,120), so
should we be surprised to find a dozen SMGs with multiple radio
identifications? Of the radio-identified SHADES SMGs, 5 per cent will
be spuriously associated with another radio source. We thus expect
three multiple identifications whereas we see a dozen: a significant
difference.

Looking at this another way, the fraction of radio-identified SMGs
with multiple radio counterparts is 18.5\,$\pm$\,5.3 per cent (12/65),
15.4\,$\pm$\,4.9 per cent (10) with separations below 6\,arcsec. How
frequent are such cases amongst the general radio population? The
proportion of radio sources in the SHADES fields with radio companions
within 4, 6, 8 and 10\,arcsec are (cumulatively) 1.2\,$\pm$\,0.3,
3.9\,$\pm$\,0.5, 7.1\,$\pm$\,0.6 and 10.3\,$\pm$\,0.7 per cent
(Poisson uncertainties). The number of SMGs with separations below
10\,arcsec, and particularly below 6\,arcsec, is thus
significant. Interestingly, bright SMGs make up one in seven of all
radio multiples with separations below 6\,arcsec.

What causes this multiplicity? At least three mechanisms could be
responsible: AGN-driven jets; physical interactions; and
confusion.

Discriminating between these mechanisms is extremely difficult. The
first -- jets -- could be revealed via their morphology or their radio
spectral index, but to date neither property has been probed for a
significant sample. The spectroscopic evidence required to reveal the
second possibility -- a physical association -- is available only
rarely in the SHADES fields, although a number of linked, multiple
systems with few-arcsec separations and near-identical redshifts have
been documented elsewhere (Ivison et al.\ 1998, 2000; Ledlow et al.\
2002; Neri et al.\ 2003; Smail et al.\ 2003a; Chapman et al.\ 2005;
Tacconi et al.\ 2006) which leaves little doubt that many SMGs with
multiple radio identifications are interaction-driven starbursts with
separations of ten (or a few tens) of kpc.

%
%
\begin{figure*}
\centerline{\psfig{file=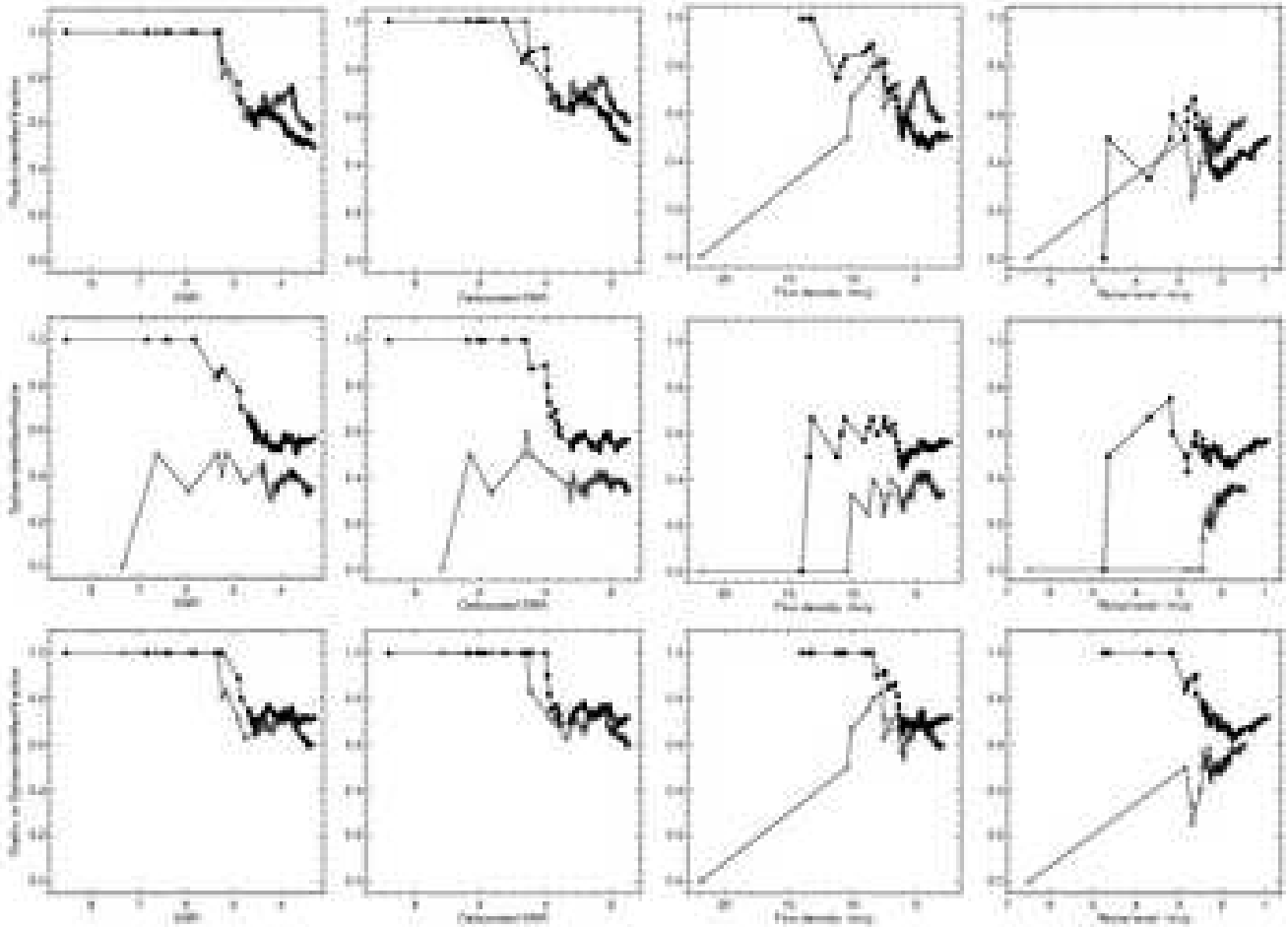,angle=0,width=7in}}
\vspace{-0.3cm}
\noindent{\small\addtolength{\baselineskip}{-3pt}}
\caption{{\it Top row:} from left to right, plots of the cumulative
radio-identified fraction for the LH SMG sample (filled
circles) and the SXDF SMG sample (open circles) against submm SNR {\em
before} flux deboosting, submm SNR {\em after} deboosting, 850-$\mu$m
flux density {\em after} deboosting and 850-$\mu$m noise level. {\it
Middle row:} the same plots, but for source identification at
24\,$\mu$m. {\it Bottom row:} the same plots, allowing for
identifications at 24\,$\mu$m, 1.4\,GHz, or coincident weak emission
at both as summarised in Table~\ref{id-summary}.}
\label{trends}
\end{figure*}

Fig.~\ref{doubles} shows a plot of submm flux density versus angular
separation for those SHADES SMGs with more than one radio counterpart
and we see no contradiction of the previous trend: two thirds of the
multiple identifications have separations of 2--6\,arcsec. However,
our data and our approach bias us against finding systems with smaller
and larger separations, as can be seen by the distribution of
separations found for fake SMGs with multiple radio counterparts
during our Monte-Carlo simulations (Fig.~\ref{doubles}).
High-resolution radio imaging from MERLIN has provided examples of
multiple, discrete radio sources separated by 0.2--2\,arcsec (Chapman
et al.\ 2004; Biggs et al., in preparation), though they are rare.

The size of the SHADES survey provides a unique opportunity to probe
the third mechanism -- confusion. The steepness of the submm counts
may yield examples where two or more faint, unrelated SMGs share a
sightline and thus conspire to create a seemingly bright SMG.  There
is approximately one SMG in the $\rm 2<S_{\rm 850\mu m}<4$\,mJy flux
density range for every 4.3\,arcmin$^2$ of sky, according to the
differential counts presented by Coppin et al.\ (2006). We thus expect
$\rm 185\pm 50$ such sources in the SHADES fields. The probability of
a 2--4-mJy SMG lying within 7\,arcsec of another source is $\sim$1 per
cent, so we could expect to see two of these amalgamated sources at
flux densities between 4 and 8\,mJy in the SHADES sample. This flux
density range accounts for 62.5 per cent of the full sample, so we
might expect around three such sources in total (perhaps rather more
if we included amalgamations of far more common, fainter sources). Of
these three, two should have a real radio identification; one may have
several. The difficulty we face in exploring this small subset of
amalgamated sources is in knowing which of the SHADES SMGs they
are. One prediction might be that they are expected to have fainter
counterparts at other wavelengths, but even this may be premature
(Serjeant et al.\ 2007). We must content ourselves with the knowledge
that they should be revealed via SCUBA-2 450-$\mu$m imaging in the
near future.

Without spectroscopic data we cannot determine whether physical
interactions or confusion make up the majority of the SMGs with
multiple identifications, let alone whether bright SMGs are special
cases where two massive components are merging, as suggested by Smail
et al.\ (2003b). The median deboosted submm flux density of the SHADES
Source Catalogue is 5.0\,mJy; the error-weighted mean 850-$\mu$m flux
density of SMGs with more than one radio counterpart is $\rm 5.8\pm
0.4$\,mJy; that for a comparison sample, the 48 SMGs with a single
$P\le\rm 0.05$ radio counterpart, is $\rm 5.4\pm 0.2$\,mJy, so the
simplest approach yields no evidence of a difference between SMGs with
single and multiple identifications. Fig.~\ref{nflux} shows the
distribution of deboosted submm flux density for the whole SHADES
Source Catalogue and for those sources with multiple robust
counterparts at 1.4\,GHz or 24\,$\mu$m. Taking the median SHADES flux
density as our threshold, eight multiple identifications lie above and
11 lie below, respectively (six apiece using only the radio). However,
as our flux density threshold rises to 10\,mJy so the fraction of
sources with multiple identifications rises from 15/111 to 4/9 (or
8/111 to 4/9 using only the radio); even ignoring the high probability
that one of the remaining five bright sources may be spurious
(SXDF850.45) and that another has several possible counterparts
(Lock850.34 -- Table~\ref{id-summary}), this is a significant
trend. It is plausible that these sources are examples of confusion
(i.e.\ amalgamated sources) but we note that the physically linked
systems reported to date are often similarly bright.

We conclude that the incidence of very high flux density and
counterpart multiplicity are weakly linked and that the case for a
preferred separation between multiple counterparts is plausible but
not proven. In particular, we note that almost half of the brightest
nine SMGs -- all $>$10\,mJy -- have multiple radio counterparts and
that all have separations in the range 2--6\,arcsec, or 20--70\,kpc at
their likely redshifts and at an inclination of 45$^{\circ}$ to the
sky, perhaps enabling efficient gas fueling for central starbursts or
AGN via overlapping galactic disks --- see the qualitative discussion
and illustrations (particularly Figs~11--13) in the merger simulations
of Springel, Di Matteo, \& Hernquist (2005) where a particularly
intense burst of activity occurs on first passage for systems that
lack prominent bulges, with galaxy separations of $\sim$30\,kpc for
the subsequent few tens of Myr.

\section{Radio and mid-IR identification trends and submm sample refinement}

Following Ivison et al.\ (2002), we seek to exploit the clear
prediction that spurious SMGs will lack radio or mid-IR
counterparts. Genuine sources can, of course, evade radio or mid-IR
detection -- because they lie at extreme redshift, for example (see
Ivison et al.\ 2005) -- but general trends in the identification rate
may be evident. In this section we therefore explore what can be
learned about the SMGs without counterparts.

Fig.~\ref{trends} shows the cumulative identification rate for SMGs in
the LH and SXDF fields as a function of submm SNR (before
and after flux deboosting), deboosted submm flux density and submm
flux uncertainty.

\subsection{Radio trends}

Looking at the radio identification trends as functions of submm flux
density and noise, we see the recovery rate tailing off at the
faintest flux density limits ($<$5\,mJy) in SXDF, whereas the rate is
remarkably flat for fainter flux densities in the LH field. Both
fields show improving identification rates as the submm noise
declines, despite the deboosting procedures outlined in Paper {\sc ii}
-- a worrying trend, though we should bear in mind that searching for
identifications within a fixed radius must act as a bias against
low-SNR sources. For the highest values of submm flux density and
noise we see similarities with trends discussed by Ivison et al.\
(2002) for the 8-mJy Survey, i.e.\ the brightest source in each field
lies in a region with high noise, and neither has a robust radio
counterpart.

The SXDF radio identification rate versus raw submm SNR shows a steep
decline below an SNR of 4; after flux deboosting this effect is
mitigated somewhat, with matching trends in the SXDF and LH fields. It
is noteworthy that the overall radio recovery rate in SXDF is over 10
per cent higher than in the LH field, despite the shallower depth of
the SXDF radio imaging. We attribute this to three effects, each of
which we believe contributes to the unexpectedly low LH identification
rate: first, the LH radio image is a single pointing, designed
originally to identify SMGs in the small 8-mJy Survey field (cf.\ a
mosaic of three in SXDF), so the pernicious effect of bandwidth
smearing will be evident for a significantly larger fraction of the
SHADES field in LH than in SXDF; second, although it is clearly useful
to work with the best possible radio data, deep imaging inevitably
yields more faint, unrelated, background sources, causing $P$ values
for relatively bright counterparts to rise relative to those
calculated for a lower source density; third, it is possible (though
it has yet to be shown unambiguously -- Ivison et al.\ 2002; Chapman
et al.\ 2004; Muxlow et al.\ 2005) that a significant fraction of the
emission in some SMGs is resolved by high-resolution radio data. That
these effects are significant, collectively, is demonstrated by the
significantly higher SMG identification rate in the shallower, lower
resolution SXDF data; in addition, seven LH SMGs (LOCK850.10, .34,
.37, .38, .40, .77 and .100) are detected robustly only in the
noisier, low-resolution radio image, though we note that in several
cases the 4.2-arcsec {\sc fwhm} image alone does not allow us to
differentiate between plausible spectroscopic targets. There are
several lessons here: ensure interferometric data contain an adequate
fraction of short spacings -- a synthesised beam with 1.5--2\,arcsec
{\sc fwhm} provides a good compromise for identification of
FIR-luminous galaxies; where necessary, i.e.\ when the area of
interest is similar to that of the radio interferometer's primary beam
and the spectral resolution is poor ($\delta\lambda/\lambda <\rm
1000$), obtain data in a compact mosaic of pointings rather than a
single, deep pointing.

\subsection{Mid-IR trends}

The trend of overall recovery rate is reversed in the mid-IR, the LH
yielding a 20 per cent higher identification rate than the SXDF.  The
reason is obvious, however: it is due to the substantial extra depth
of the LH {\em Spitzer} 24-$\mu$m data ($\sigma\rm = 11$ versus
47\,$\mu$Jy). Only one SMG is identified solely on the basis of its
mid-IR emission in SXDF compared with ten in the LH.  For both fields
the decline at low deboosted SNR is less marked than the radio
trend. Against submm flux density and noise, the 24-$\mu$m
identification trends for both fields match those at radio wavelengths
(with the aforementioned 20 per cent offset for the SXDF sources); the
very brightest sources again lack robust counterparts.

\subsection{Overall trends}

The lower row of plots in Fig.~\ref{trends} show the {\em overall}
identification trends -- the fraction of sources identified at
1.4\,GHz and/or 24\,$\mu$m, including the three cases mentioned in
\S\ref{associations} where weak radio and 24-$\mu$m counterparts are
coincident (one of which is the brightest LH source, LOCK850.34).

The identification trends are similar for the two SHADES fields:
identification is essentially complete above a deboosted submm SNR of
$\sim$4 with an abrupt step down to 60--70 per cent thereafter; also,
success rates improve as the submm noise declines. The SXDF
identification rate tails off below a deboosted submm SNR of 2.5 and
at submm flux densities below 5\,mJy. This may be due to the limited
depth of the SXDF radio and 24-$\mu$m imaging rather than any
deficiency of the SXDF catalogue, but we note that it is a strong
tendency.

Summarising these plots, the best available complementary data in the
LH -- equivalent to those available in the Great Observatories Origins
Deep Survey (GOODS) northern field -- allows us to identify robustly
over two thirds of SMGs to current submm detection limits.  The
observed trends in identification rate give no strong rationale for
rejecting any sources from the parent SHADES Source Catalogue,
although a slight question mark is thrown over some of the lowest SNR
sources.

%
%
\begin{figure}
\centerline{\psfig{file=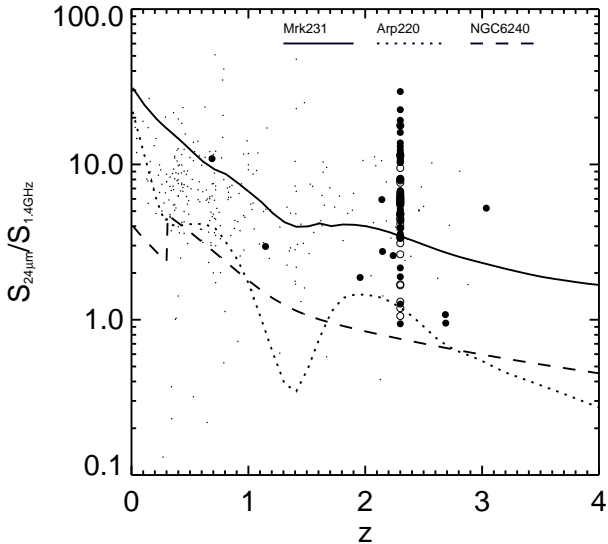,angle=0,width=3.3in}}
\vspace{-0.5cm}
\noindent{\small\addtolength{\baselineskip}{-3pt}}
\caption{Ratio of $S_{\rm 24\mu m}/S_{\rm 1.4GHz}$ as a function of
redshift, $z$, for SHADES sources with robust counterparts (filled
circles: LH; empty circles: SXDF). Those without spectroscopic
redshifts -- the majority -- are plotted arbitrarily at $z=\rm
2.3$. The tracks of Arp\,220, Mrk\,231 and NGC\,6240 are shown
together with a sample of faint radio sources in SXDF (small dots --
Ibar et al., in preparation).}
\label{radio-24um}
\end{figure}

%
%
\begin{figure}
\centerline{\psfig{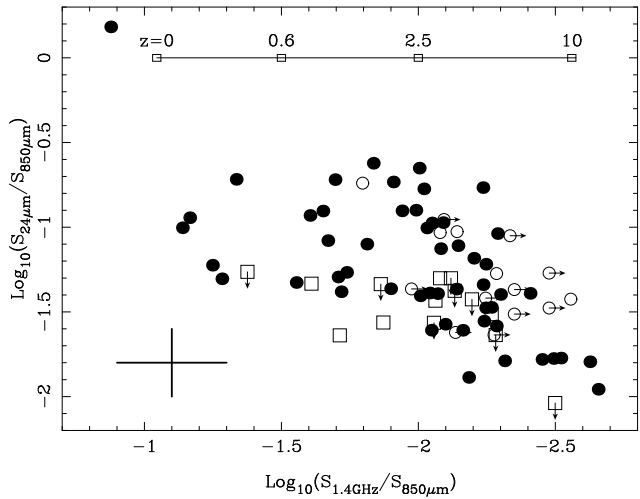}}
\vspace{-0.5cm}
\noindent{\small\addtolength{\baselineskip}{-3pt}}
\caption{Log$_{10}\,S_{\rm 24\mu m}/S_{\rm 850\mu m}$ versus
  log$_{10}\,S_{\rm 1.4GHz}/S_{\rm 850\mu m}$ for SHADES SMGs with
  both mid-IR and radio identifications (filled circles), with only
  radio identifications (squares) and with only mid-IR identifications
  (open circles). A representative error bar is shown, lower left. The
  redshift parameterisation of Chapman et al.\ (2005) is shown as a
  horizontal bar at log$_{10}\,S_{\rm 24\mu m}/S_{\rm 850\mu m}\rm =0$
  (see \S7.2).}
\label{s24s850-s1400s850}
\end{figure}

\section{Constraints from spectral indices}

\subsection{\em S$_{\rm\bf 24\mu m}$/S$_{\rm\bf 1.4GHz}$}

Since the spectral slopes at 24\,$\mu$m and 1.4\,GHz are similar, it
may prove instructive to examine the behaviour of $S_{\rm 24\mu
m}/S_{\rm 1.4GHz}$ as a function of redshift, as shown in
Fig.~\ref{radio-24um}. We expect this plot to betray AGN contributions
to the radio flux density in so-called `radio-excess AGN' (Drake et
al.\ 2003; Donley et al.\ 2005) or, conversely, `mid-IR-excess AGN'
which have QSO-heated dust but little or no AGN-related emission in
the radio. For star-forming galaxies this ratio is tightly constrained
out to $z=\rm 1$ (Appleton et al.\ 2004). Galaxies with low values of
$S_{\rm 24\mu m}/S_{\rm 1.4GHz}$, i.e.\ those with strong radio with
respect to 24-$\mu$m emission, are unlikely to be dominated by star
formation.

The SHADES SMGs share approximately the same distribution of $S_{\rm
24\mu m}/S_{\rm 1.4GHz}$ values as the other radio sources in SXDF
(Ibar et al., in preparation). Fig.~\ref{radio-24um} shows the
redshift tracks of Arp\,220, NGC\,6240 and Mrk\,231 -- archetypal
ultraluminous IR galaxies with increasing degrees of AGN
contribution. Measured values of $S_{\rm 24\mu m}/S_{\rm 1.4GHz}$ for
the SHADES SMGs are consistent with any of these SEDs but Mrk\,231 is
the preferred template, implying an AGN contribution to the mid-IR
luminosity. Only at $z\rm <1$ could the most extreme SMG be classified
confidently as having a radio excess.

\subsection{\em S$_{\rm\bf 850\mu m}$/S$_{\rm\bf 1.4GHz}$}

Hughes et al.\ (1998) and Carilli \& Yun (1999) pointed out the value
of $S_{\rm 850\mu m}/S_{\rm 1.4GHz}$ as an indicator of redshift for
SMGs, at least for $z\rm< 3$. Smail et al.\ (2000) and Ivison et al.\
(2002) were the first to employ the technique for significant samples
of SMGs, finding median redshifts, $z\rm\gs 2$.

Chapman et al.\ (2005) found that the relation showed a large
dispersion for their sample of radio-identified SMGs with
spectroscopic redshifts, indicative of a range of SEDs. They noted
that a purely submm-selected sample should show an even wider range of
$S_{\rm 850\mu m}/S_{\rm 1.4GHz}$ than their radio-identified SMGs,
since the need for an accurate radio position biases the sample in
redshift and temperature.

The surprisingly flat trend identified by Chapman et al., uncorrected
for a probable redshift-dependent $\sim$0.3\,dex shift attributable to
their radio selection criteria, was parameterised as $S_{\rm 850\mu
m}/S_{\rm 1.4GHz}$ = 11.1 + 35.2\,$z$. This parameterisation was not
intended as a careful photometric redshift technique -- the r.m.s.\
scatter in redshift is $\sim$1, after all -- but likely remains the
best way to estimate the median redshift of radio-identified SMG
samples. Applying this to our sample of 65 SMGs with robust radio
counterparts yields a median redshift of 2.8, with an interquartile
range of 1.3--3.8, somewhat higher and broader than the spectroscopic
redshift distribution reported by Chapman et al.\ (median $z=\rm 2.2$,
interquartile 1.7--2.8, before their small correction for the radio
selection function). The Chapman et al.\ parameterisation is not
appropriate for SMGs without radio identifications, but for the entire
SHADES sample (adopting the limits in
Tables~\ref{radio-lh}--\ref{radio-sxdf} for those lacking formal
detections) it indicates a median redshift of 3.3.

The difference between the distribution reported here and that of
Chapman et al.\ (2005) is quite marked, but can be explained by a
variety of effects: spectroscopic bias; field-to-field variations;
strong clustering of the SMG population (Blain et al.\ 2004); our
adoption of deboosted flux densities for all SHADES sources (a large
proportion of the Chapman et al.\ sample is likely to have suffered a
submm flux density boost of one form or another); and, not least, the
difficulty of measuring accurate and consistent radio flux densities
using data with different $uv$ coverage.

Fig.~\ref{s24s850-s1400s850} shows a log--log plot of $S_{\rm 24\mu
m}/S_{\rm 850\mu m}$ versus $S_{\rm 1.4GHz}/S_{\rm 850\mu m}$ for
SHADES SMGs, with different symbols representing identifications made
in different wavebands (radio plus mid-IR; mid-IR only; radio
only). As we have discussed, $S_{\rm 1.4GHz}/S_{\rm 850\mu m}$ is
sensitive to redshift (and temperature) and the Chapman et al.\
parameterisation is shown as a horizontal bar. It is apparent that
24-$\mu$m flux density is correlated significantly with redshift, as
expected for the $K$ correction at that wavelength. The SMG with
$S_{\rm 1.4GHz}/S_{\rm 850\mu m}>\rm 0.1$ is SXDF850.21, the most
obvious example of a local galaxy in the sample ($z=\rm 0.044$,
Simpson et al.\ 2006; see Appendix A, Fig.~A2).

%
%
\begin{figure*}
\centerline{\psfig{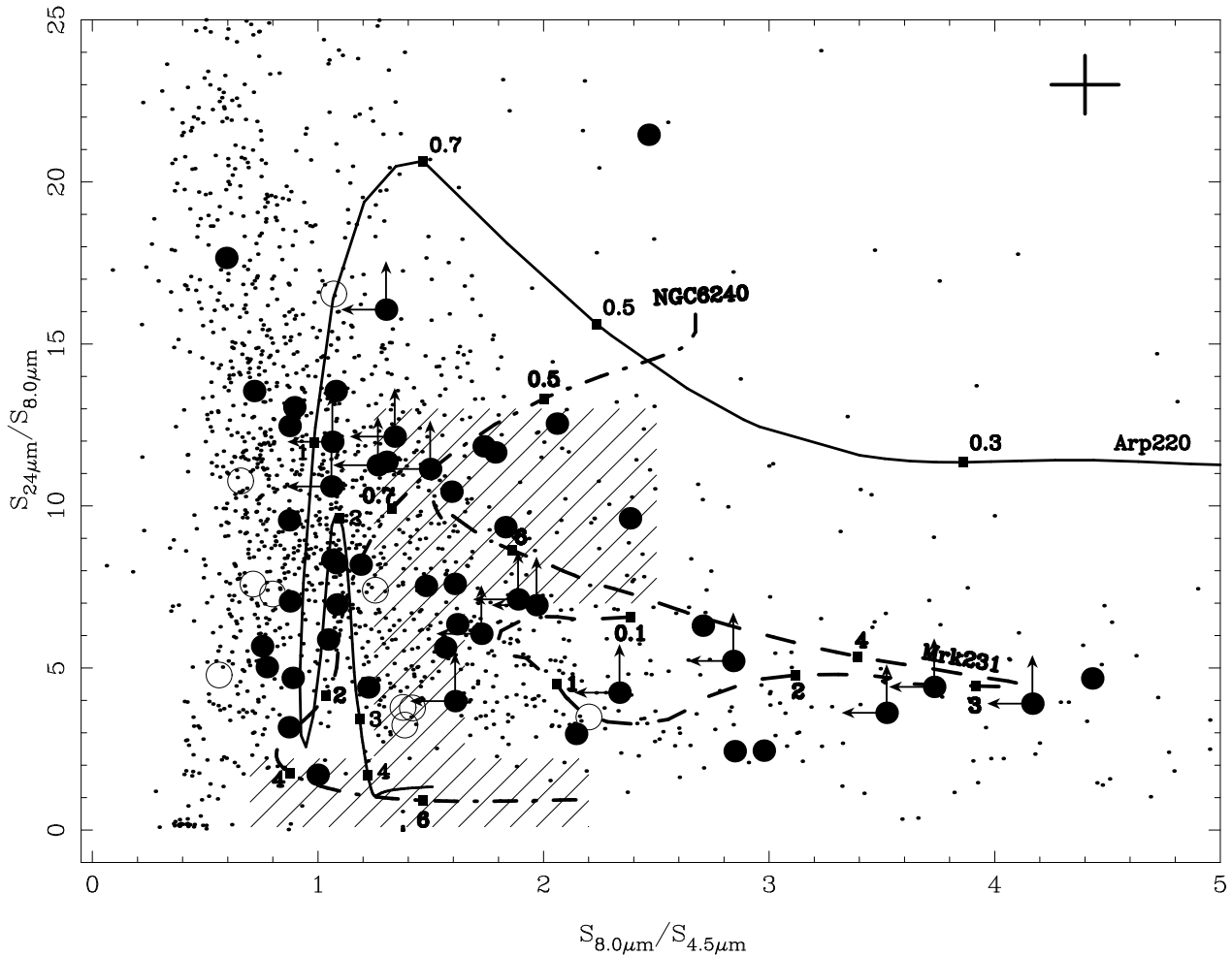}}
\vspace{-0.5cm}
\noindent{\small\addtolength{\baselineskip}{-3pt}}
\caption{$S_{24\mu \rm m}/S_{8\mu \rm m}$ versus $S_{8\mu \rm
m}/S_{4.5\mu \rm m}$ colour-colour diagram for a faint
24-$\mu$m-selected sample of 4,457 galaxies (dots) with 4.5- and
8.0-$\mu$m detections. Robustly identified SMGs are plotted
as filled circles and less securely identified SMGs as open circles. A
representative error bar is shown, top right. The redshift tracks of
Arp\,220, Mrk\,231 and NGC\,6240 are shown (adopting SEDs from
Rigopoulou et al.\ 1999 and Spoon et al.\ 2004), with squares to
indicate redshift. Hatched regions indicate regions of colour-colour
space where one might expect to find the most distant SMGs.}
\label{colour-colour}
\end{figure*}

\section{The diagnostic power of mid-IR colour}

Ivison et al.\ (2004) used a colour-colour plot to exploit the strong
diagnostic potential of the mid-IR for discriminating between galaxies
dominated by starbursts and AGN. Key spectral indices for
high-redshift galaxies are available between 3.6 and 24\,$\mu$m since
the rest-frame $\sim$3--10\,$\mu$m slope for starbursts is steeper
than for AGN, with a flatter region between 1 and 3\,$\mu$m, whereas
AGN exhibit power-law spectra covering rest-frame
$\sim$0.2--10\,$\mu$m (e.g.\ Mrk\,231).

Fig.~\ref{colour-colour} shows $S_{24\mu \rm m}/S_{8\mu \rm m}$ versus
$S_{8\mu \rm m}/S_{4.5\mu \rm m}$. We expect the low-$S_{8\mu \rm
m}/S_{4.5\mu \rm m}$ portion -- the left side -- to be occupied by
$z\gs\rm 0.7$ starbursts, represented here by the redshift track of
Arp\,220. High-redshift starbursts are expected in the lower left
region of Fig.~\ref{colour-colour}, but spectral features in
Arp\,220's SED yield several kinks which limit the diagnostic power of
the plot; power-law AGN, represented in Fig.~\ref{colour-colour} by
Mrk\,231, track left-to-right with increasing redshift across the
lower third of the plot, returning to the left only $z\gs\rm 4$. The
redshift track of NGC\,6240 -- a classical Compton-thick AGN
displaying mid-IR PAH features indistinguishable from those of a
starburst galaxy -- overlaps significantly with the colour-colour
space occupied by Arp\,220, at $z\sim\rm 0.4$ and at much higher
redshifts, but most of the confusing overlap occurs where we expect
NGC\,6240-type SEDs at $z\sim\rm 0.6$ and Mrk\,231-like SEDs at $z>\rm
6$.

Do SMGs stand out from a 24-$\mu$m-selected {\em Spitzer} sample in
colour-colour space? Fig.~\ref{colour-colour} shows an independent
galaxy sample selected at 24\,$\mu$m in the LH, at depths
commensurate with our {\em Spitzer} identifications, and we can see
that the data are clustered along the track occupied by Arp\,220-like
SEDs for $z\ge\rm 0.7$, with a significant number of sources along the
track defined by a Mrk\,231-like SED. SMGs are similarly positioned
and do not stand out clearly from 24-$\mu$m-selected
galaxies. However, the hatched areas of Fig.~\ref{colour-colour} --
those colour combinations where we might expect to find SMGs with the
highest redshifts ($z\gs\rm 4$) -- are well populated with SMGs. The
fraction of SMGs in these regions is significantly larger than for the
control sample: we find only 14 per cent of the 4,457 mid-IR-selected
galaxies in the hatched regions. Based on the Chapman et al.\
parameterisation of $S_{\rm 850\mu m}/S_{\rm 1.4GHz}$, their median
redshift is higher than that of the radio-detected fraction of SHADES,
3.2 versus 2.8, although we note that some of the best $z\ls\rm 1$
candidates also fall in these regions, e.g.\ SXDF850.52. Nevertheless,
it seems sensible that any search for a high-redshift population of
SMGs should base its target selection on a combination of the $S_{\rm
850\mu m}/S_{\rm 1.4GHz}$, $S_{\rm 1200\mu m}/S_{\rm 850\mu m}$ (Eales
et al.\ 2003; Greve et al.\ 2004), $S_{24\mu \rm m}/S_{8\mu \rm m}$
and $S_{8\mu \rm m}/S_{4.5\mu \rm m}$ colours.

\section{Concluding remarks}

We have determined the most likely radio and/or mid-IR
identifications, and hence accurate positions, for the SHADES Source
Catalogue presented by Coppin et al.\ (2006). We have identified
robust counterparts to over two thirds of this sample (54 and 46 per
cent at 1.4\,GHz and 24\,$\mu$m, respectively), presenting optical,
24-$\mu$m and radio images of each SMG.

Employing the submm/radio flux density ratio as an indicator of
redshift, guided by the Chapman et al.\ (2005) parameterisation, we
find a median redshift of 2.8 for the radio-identified sample,
somewhat higher than the spectroscopic median.

We present a diagnostic colour-colour plot, based on {\em Spitzer}
data, in which we identify regions commensurate with SMGs at very high
redshift.

We further exploit our identifications to show that:
\begin{itemize}
\item observed trends in identification rate give no strong rationale
for pruning the parent SHADES sample (cf.\ Ivison et al.\ 2002);
\item uncertainties in submm position are consistent with theoretical
expectations, with no evidence for significant additional sources of
positional error;
\item significantly more SMGs have multiple robust counterparts than would
be expected by chance, indicative of physical associations. These
multiple systems are most common amongst the brightest SMGs and are
typically separated by 2--6\,arcsec, $\sim$15--50/sin $i$\,kpc at
$z\sim\rm 2$, consistent with early bursts seen in merger simulations.
\end{itemize}

\section*{Acknowledgements}

AWB acknowledges support from the Alfred P.\ Sloan Foundation and the
Research Corporation. IS acknowledges support from the Royal
Society. CS and SR acknowledge financial support from the PPARC. IA and DHH
acknowledge support from CONACYT grants 39548-F and 39953-F.

\setlength{\bibhang}{2em}

\newpage

\appendix

\section{POSTAGE STAMP IMAGES}

This section presents 25\,$\times$\,25-arcsec postage stamp images of
each SMG in the SHADES Source Catalogue as well as a description of
the most unusual examples.

Figs~A1 and A2 show greyscale $R$-band optical data in the
left-hand panels, where available, and greyscale 24-$\mu$m data in the
right-hand panels. Superimposed on the $R$-band images are
high-resolution (1.3\,arcsec {\sc fwhm} for the LH, 1.7\,arcsec {\sc
fwhm} for SXDF) radio contours, plotted at $-$3, 3, 4 ... 10, 20
... 100 $\times$ $\sigma$, where $\sigma$ was measured in source-free
regions around each SMG and is quoted in the lower-right corner of
each image in units of $\mu$Jy\,beam$^{-1}$. Superimposed on the
24-$\mu$m data are low-resolution (4.2\,arcsec, {\sc fwhm}) radio
contours, plotted at $-$3, 3, 4 ... 10, 20 ... 100 $\times$ $\sigma$,
where $\sigma$ was measured in source-free regions around each SMG and
is again quoted in the lower-right corner of each image. Broken
crosses mark the positions of all 24-$\mu$m sources brighter than
150\,$\mu$Jy found within 15\,arcsec of SMG positions in SXDF -- their
positions are listed in Table~\ref{24um-sxdf}. The large central
circles indicate 2\,$\sigma$ positional uncertainties where $\sigma =
0.6\,\theta$/SNR and deboosted SNR values have been adopted (Coppin et
al.\ 2006). As shown in \S4, there is an 86.5 per cent probability
that these circles contain the source of submm emission. For
counterpart identification we simply use a radius of 8\,arcsec (or
12.5\,arcsec for the radio, 15\,arcsec at 24\,$\mu$m, to be more
complete).

Solid boxes indicate robust identifications, where $P\le\rm 0.05$
based on the radio or 24-$\mu$m counts, or a combination of the
two. Dashed boxes indicate tentative associations.

\subsection*{Cases worthy of comment}

Some of the SMGs present unusual combinations of observed
characteristics and we comment on them here.

\noindent
{\bf LOCK850.06:} Betrayed at both 24\,$\mu$m and 1.4\,GHz, but
invisible optically.

\noindent
{\bf LOCK850.07:} As LOCK850.06, though with an optical counterpart
within 1\,arcsec; possibly typical of the composite blue-red
pairs noted by Ivison et al.\ (2002).

\noindent
{\bf LOCK850.08:} An optical counterpart likely lies behind the
diffraction spike. An ideal target for adaptive-optics-
(AO-) assisted studies, exploiting the bright star to the north.

\noindent
{\bf LOCK850.11:} This apparently obvious 24-$\mu$m identification
just fails to qualify as a `robust' counterpart because it comprises
two fainter sources. We view these as likely counterparts. They are
coincident with a disturbed optical galaxy which should be targeted
spectroscopically.

\noindent
{\bf LOCK850.14:} The nearest radio emitter does not qualify as a
robust identification but has an excellent spectroscopic redshift in
the catalogue of Chapman et al.\ (2005).

\noindent
{\bf LOCK850.15:} A complex system with as many as three plausible
identifications, suggestive of a colossal merger.

\noindent
{\bf LOCK850.16:} Described in detail by Ivison et al.\ (2002, 2005).

\noindent
{\bf LOCK850.18:} An obvious -- though faint -- radio
identification, yet there is no sign of 24-$\mu$m or optical emission.

\noindent
{\bf LOCK850.19:} A straightforward 24-$\mu$m identification with
support from faint radio emission.

\noindent
{\bf LOCK850.21:} A solid 24-$\mu$m identification; 24-$\mu$m and
distorted optical emission to the south-east may be related
physically.

\noindent
{\bf LOCK850.23:} Faint 24-$\mu$m and radio emission point to a faint
optical counterpart (circled in Fig.~A1); well worth targeting
spectroscopically, though not formally a robust identification.

\noindent
{\bf LOCK850.29:} Faint radio and 24-$\mu$m emission yield a formal
identification; the double optical galaxy seems to be offset to the
north east and yet it resembles many SMGs; it should be targeted
spectroscopically.

\noindent
{\bf LOCK850.30:} A multiple radio identification. The weakest radio
component remains stubbornly above $P=\rm 0.05$; the brightest radio
emitter was reported by Ivison et al.\ (2002) to have an inverted
radio spectrum (see Bertoldi et al.\ 2000 for other examples of this
phenomenon). The 24-$\mu$m emission appears to lie between the radio
components. In one obvious interpretation the radio emission may
emanate from lobes powered by a central, black hole- and star-forming
galaxy.

\noindent
{\bf LOCK850.34:} A multitude of multiple counterparts. An opportunity
for detailed study of a potentially complex, interacting system.

\noindent
{\bf LOCK850.37:} Robust but distinct identifications at 24\,$\mu$m
and 1.4\,GHz. Challenging, optically.

\noindent
{\bf LOCK850.48:} A seemingly straightforward identification, yet
a potentially complex system.

\noindent
{\bf LOCK850.52:} An extended counterpart at 24\,$\mu$m, barely
visible in the high-resolution radio image and yet obvious and
extended in the lower-resolution map; extra resolution available in
the LH has clearly hindered the identification process. The optical
counterpart must be part of an extensive system, presumably largely
obscured.

\noindent
{\bf LOCK850.53:} A typical counterpart consisting of two optical
galaxies, betrayed by their 24-$\mu$m emission.

\noindent
{\bf LOCK850.60:} Several plausible identifications at 24\,$\mu$m, the
closest of which just fails to qualify as a robust counterpart.

\noindent
{\bf LOCK850.63:} Another plausible AO target.

\noindent
{\bf LOCK850.67:} Optically faint SMG, blank at 1.4\,GHz, given away
by its 24-$\mu$m emission.

\noindent
{\bf LOCK850.70:} A classic optical pair betrayed at 24\,$\mu$m and by
weak radio emission.

\noindent
{\bf LOCK850.77:} As LOCK850.34: a pair of pairs.

\noindent
{\bf LOCK850.79:} Another SMG with several plausible identifications,
though only one of these is statistically robust.

\noindent
{\bf LOCK850.87:} Optically invisible, yet bright at 24\,$\mu$m and
1.4\,GHz.

\noindent
{\bf SXDF850.01:} Optically invisible, yet bright at 1.4\,GHz.

\noindent
{\bf SXDF850.02:} The radio morphology resembles the base of a wide-angle
tail radio galaxy.

\noindent
{\bf SXDF850.03:} The radio emission is apparently associated with a
bright, nearby galaxy, though the alignment is poor and lensing of a
background SMG must be a possibility.

\noindent
{\bf SXDF850.05:} Seemingly a multi-component merger; sufficiently
bright at 24\,$\mu$m and 1.4\,GHz to suggest it lies at relatively low
redshift.

\noindent
{\bf SXDF850.06:} An immensely complex region with at least three
radio-detected components. The brightest 24-$\mu$m identification is
coincident with the radio source most distant from the SMG centroid.

\noindent
{\bf SXDF850.07:} An optically faint SMG in a complex region, betrayed
by its 24-$\mu$m and 1.4-GHz emission.

\noindent
{\bf SXDF850.08:} A robust radio identification, offset by several
arcsec from a plausible 24-$\mu$m counterpart.

\noindent
{\bf SXDF850.10:} It is plausible that the submm emission emanates
from between the hotspots of a lobe-dominated radio galaxy.

\noindent
{\bf SXDF850.11:} An excellent, clearly identified target for
AO-assisted study, exploiting the nearby star.

\noindent
{\bf SXDF850.12:} An distorted optical counterpart lies beneath very
faint radio emission close to the SMG centroid.

\noindent
{\bf SXDF850.14:} Near-coincident, faint 24-$\mu$m and 1.4-GHz
emission, though it would be tempting to target the distorted optical
galaxy north of the {\em Spitzer} emission for spectroscopy.

\noindent
{\bf SXDF850.16:} Faint radio emission is offset from a seemingly
distorted optical counterpart by $\sim$1\,arcsec.

\noindent
{\bf SXDF850.21:} A local galaxy lies close to this SMG -- VLA0077 in
the catalogue of Simpson et al.\ (2006), at $z=\rm 0.044$; this must
be viewed as the most likely identification -- a rare example of a
nearby galaxy in a blank-sky submm survey.

\noindent
{\bf SXDF850.23:} As SXDF850.16.

\noindent
{\bf SXDF850.24:} Two robust radio identifications, one
near-coincident with faint 24-$\mu$m emission.

\noindent
{\bf SXDF850.28:} An immensely complex region with at least three
radio-detected components, each with different 24-$\mu$m properties.

\noindent
{\bf SXDF850.29:} A bright radio identification -- VLA0225 in the
catalogue of Simpson et al.\ (2006) -- offset significantly from the
centroid of a bright $z=\rm 0.264$ optical galaxy. The correct
identification becomes obvious in the near-IR (Clements et al., in
preparation).

\noindent
{\bf SXDF850.30:} This SMG is betrayed by 24-$\mu$m and 1.4-GHz
emission; a nearby optical galaxy may be the unobscured component of a
larger system.

\noindent
{\bf SXDF850.31:} Two robust 24-$\mu$m identifications, one coincident
with radio emission, both with bright optical counterparts.

\noindent
{\bf SXDF850.37:} Optically faint SMG with near-coincident 24-$\mu$m
and 1.4-GHz emission.

\noindent
{\bf SXDF850.47:} A complex region with three radio-detected
components, each with near-coincident 24-$\mu$m emission.

\noindent
{\bf SXDF850.52:} Two robust radio identifications with very different
optical properties, one bright, one invisible; the brightest of the
optical galaxies is not well aligned with its radio emission.

\noindent
{\bf SXDF850.77:} A complex SMG with two radio emitters, neither of
which is aligned well with the two 24-$\mu$m emitters in the region.

\noindent
{\bf SXDF850.119:} Two plausible identifications, each with very
different optical properties -- one bright and presumably relatively
local; the other optically invisible, likely at high redshift.

%
%
\setcounter{figure}{0}
\begin{figure*}
\centerline{\psfig{file=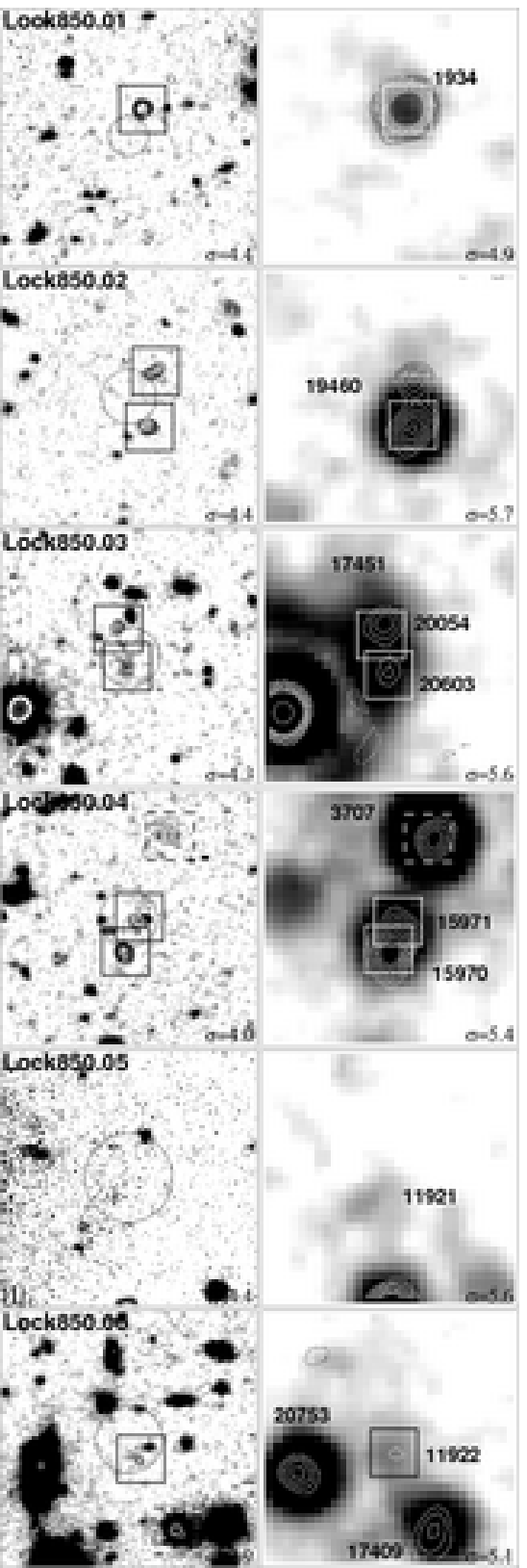,angle=0,width=3.in}
\hspace*{2mm}
\psfig{file=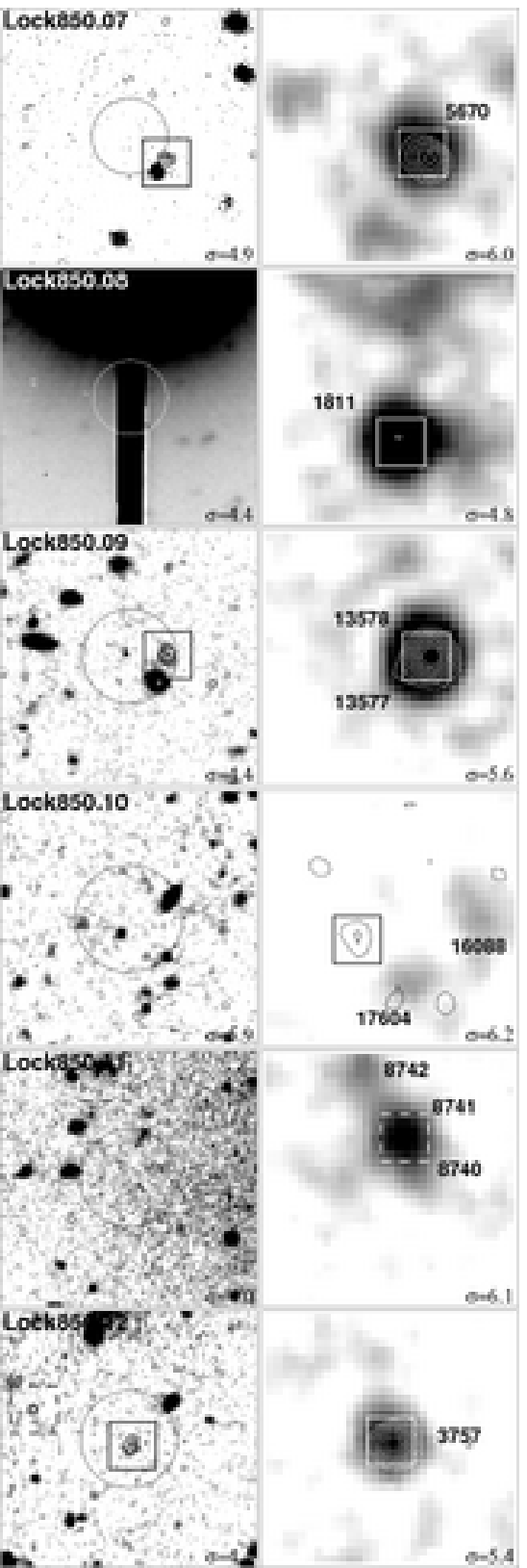,angle=0,width=3.in}}
\vspace{-0.3cm}
\noindent{\small\addtolength{\baselineskip}{-3pt}}
\caption{25\,$\times$\,25-arcsec postage stamp images of each SMG in
the LH SHADES Source Catalogue. Greyscale $R$-band and 24-$\mu$m data
are shown in the left- and right-hand panels, respectively,
superimposed with radio contours. Circles indicate 2\,$\sigma$
positional uncertainties.  Solid boxes indicate robust
identifications, where $P\le\rm 0.05$ based on the radio or 24-$\mu$m
counts, or a combination of the two. Dashed boxes indicate tentative
associations.}
\end{figure*}

\setcounter{figure}{0}
\begin{figure*}
\centerline{\psfig{file=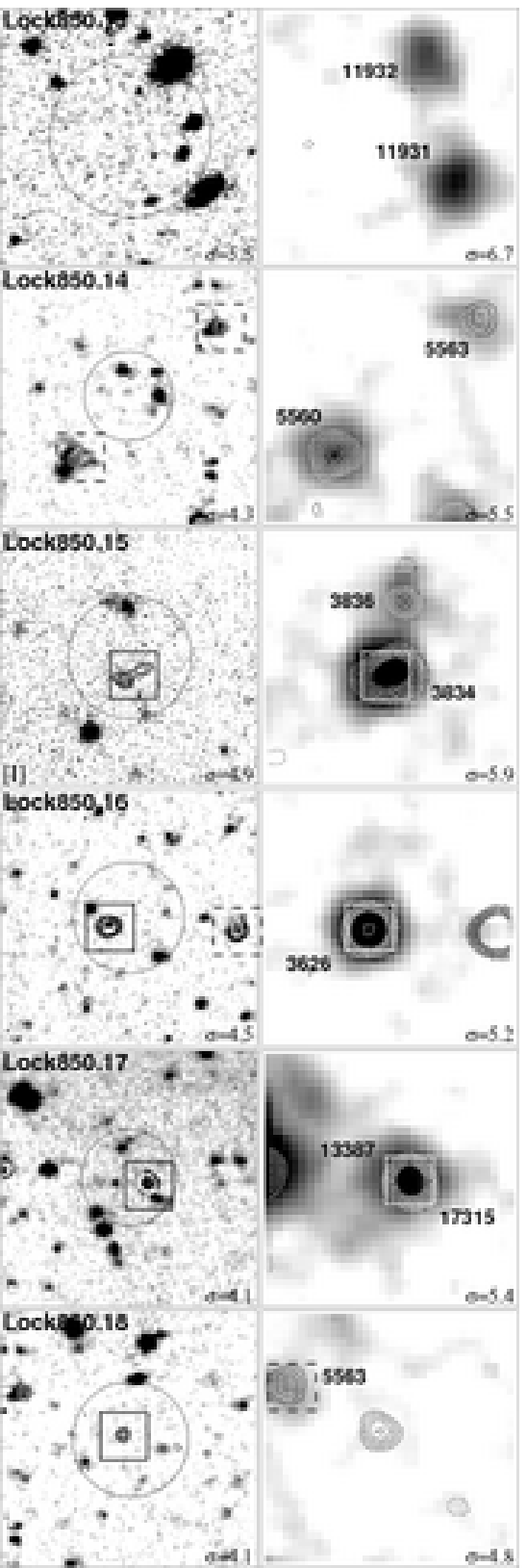,angle=0,width=3.in}
\hspace*{2mm}
\psfig{file=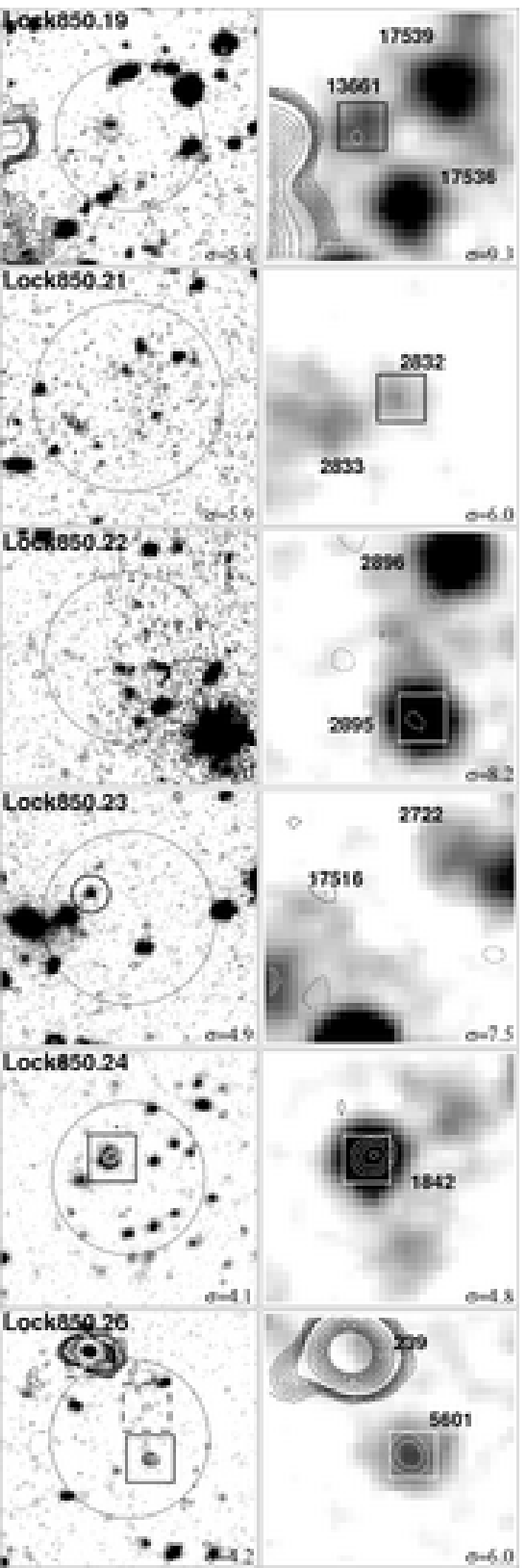,angle=0,width=3.in}}
\vspace{-0.3cm}
\noindent{\small\addtolength{\baselineskip}{-3pt}}
\caption{Cont...}
\end{figure*}

\setcounter{figure}{0}
\begin{figure*}
\centerline{\psfig{file=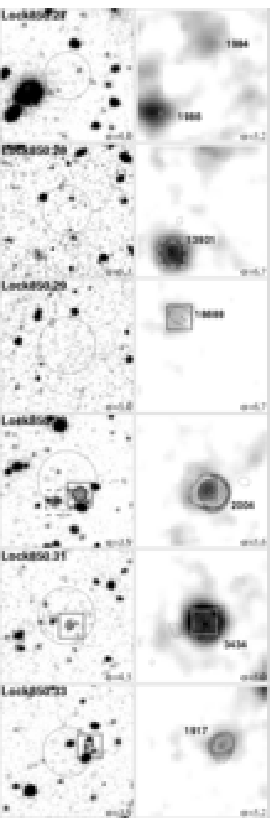,angle=0,width=3.in}
\hspace*{2mm}
\psfig{file=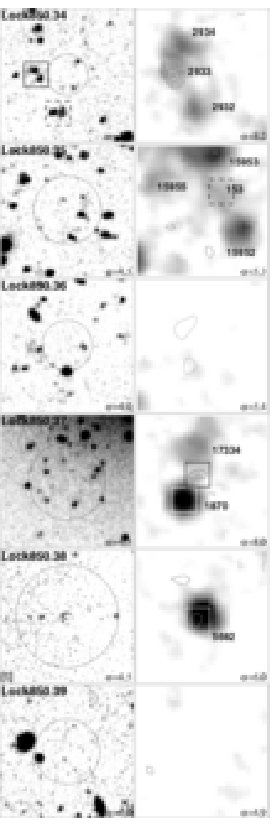,angle=0,width=3.in}}
\vspace{-0.3cm}
\noindent{\small\addtolength{\baselineskip}{-3pt}}
\caption{Cont...}
\end{figure*}

\setcounter{figure}{0}
\begin{figure*}
\centerline{\psfig{file=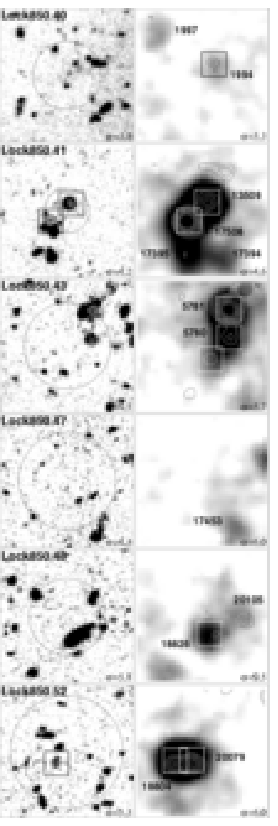,angle=0,width=3.in}
\hspace*{2mm}
\psfig{file=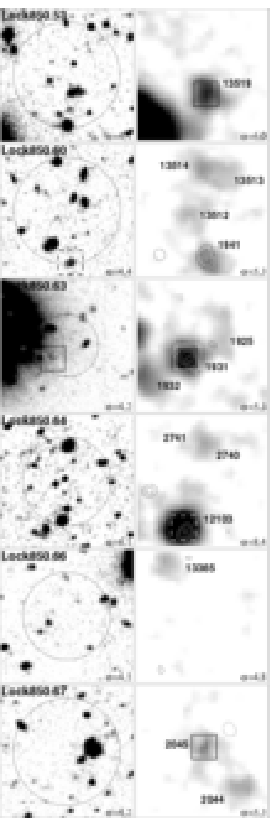,angle=0,width=3.in}}
\vspace{-0.3cm}
\noindent{\small\addtolength{\baselineskip}{-3pt}}
\caption{Cont...}
\end{figure*}

\setcounter{figure}{0}
\begin{figure*}
\centerline{\psfig{file=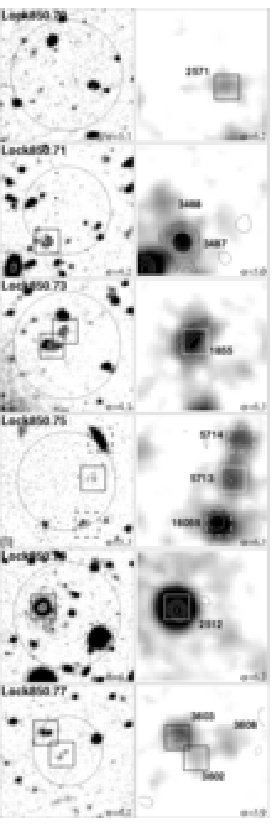,angle=0,width=3.in}
\hspace*{2mm}
\psfig{file=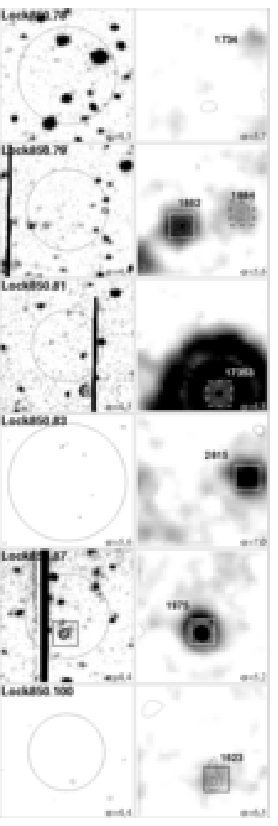,angle=0,width=3.in}}
\vspace{-0.3cm}
\noindent{\small\addtolength{\baselineskip}{-3pt}}
\caption{Cont...}
\end{figure*}

\setcounter{figure}{1}
\begin{figure*}
\centerline{\psfig{file=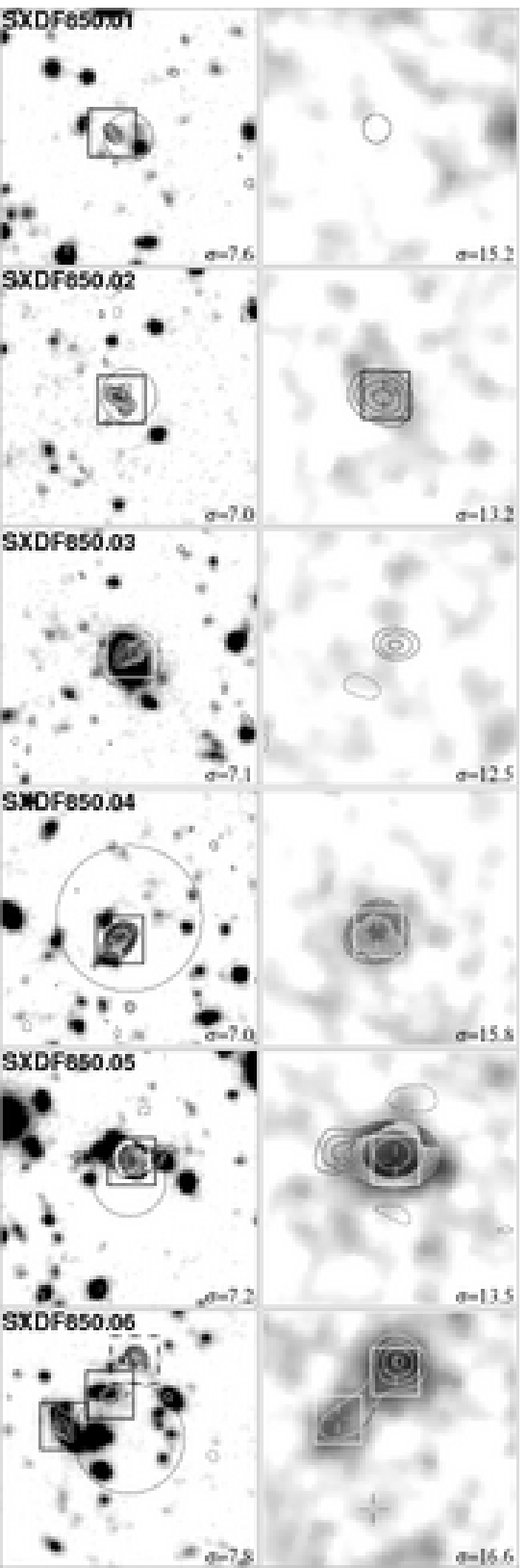,angle=0,width=3.in}
\hspace*{2mm}
\psfig{file=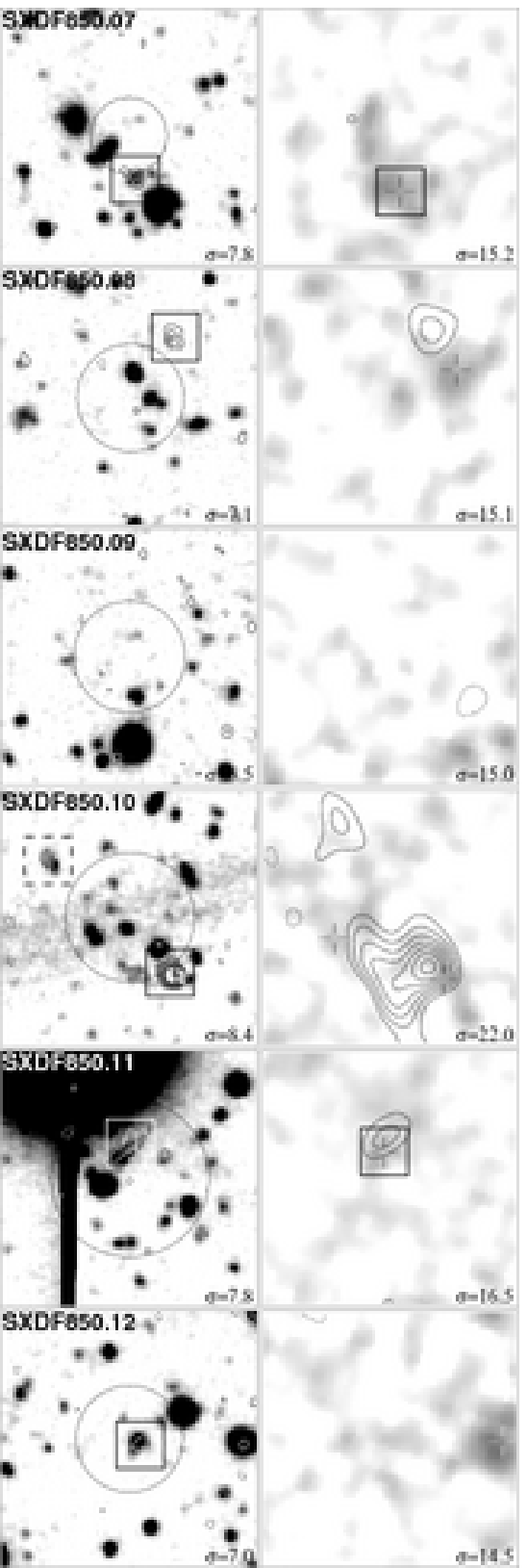,angle=0,width=3.in}}
\vspace{-0.3cm}
\noindent{\small\addtolength{\baselineskip}{-3pt}}
\caption{25\,$\times$\,25-arcsec postage stamp images of each SMG in
the SXDF SHADES Source Catalogue. Greyscale $R$-band and 24-$\mu$m
data are shown in the left- and right-hand panels, respectively,
superimposed with radio contours. Circles indicate 2\,$\sigma$
positional uncertainties. Broken crosses mark 24-$\mu$m sources
brighter than 150\,$\mu$Jy within 15\,arcsec of SMG positions --
their positions are listed in Table~\ref{24um-sxdf}. Solid boxes
indicate robust identifications, where $P\le\rm 0.05$ based on the
radio or 24-$\mu$m counts, or a combination of the two. Dashed boxes
indicate tentative associations.}
\end{figure*}

\setcounter{figure}{1}
\begin{figure*}
\centerline{\psfig{file=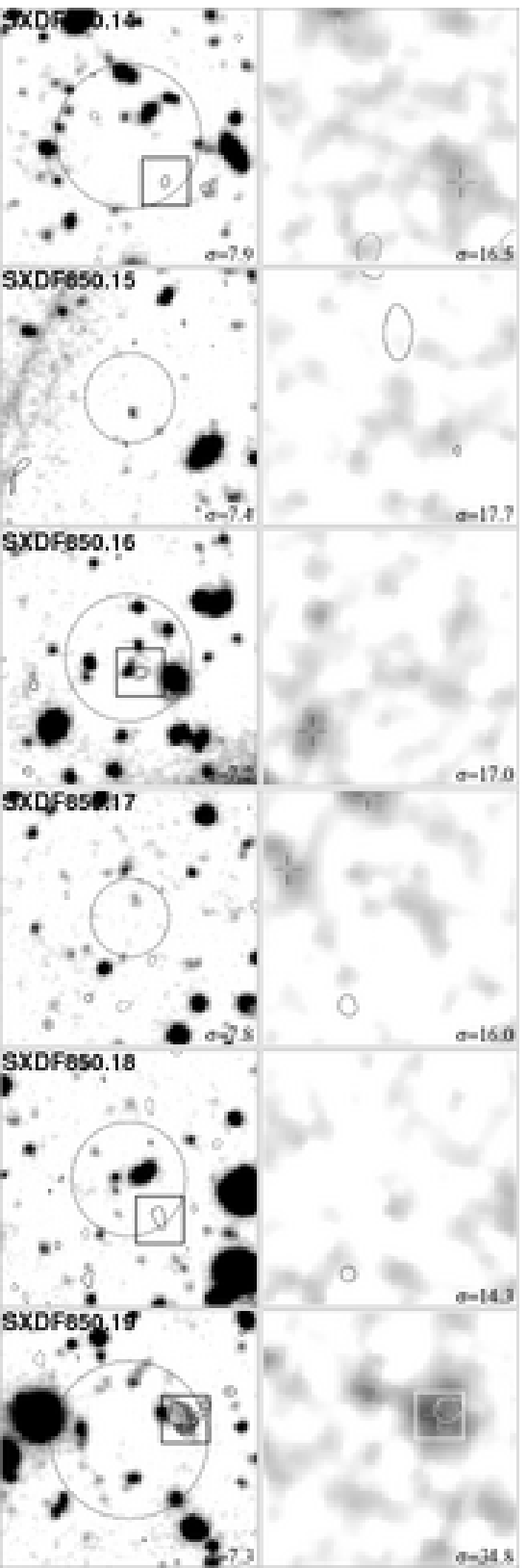,angle=0,width=3.in}
\hspace*{2mm}
\psfig{file=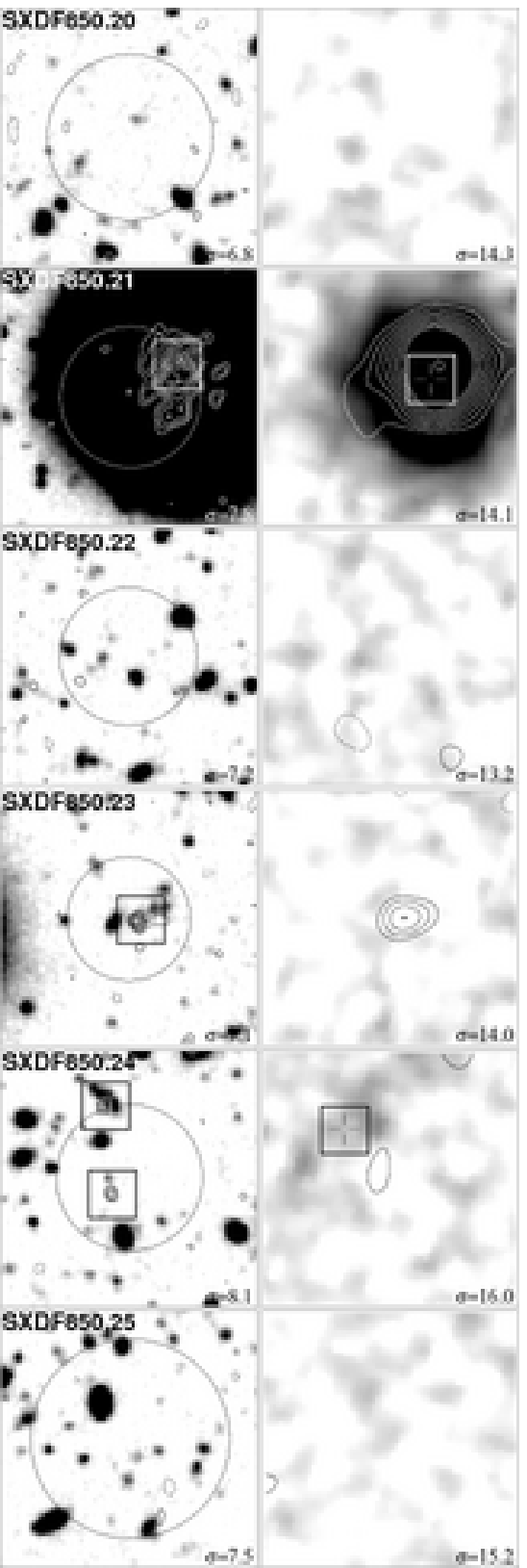,angle=0,width=3.in}}
\vspace{-0.3cm}
\noindent{\small\addtolength{\baselineskip}{-3pt}}
\caption{Cont...}
\end{figure*}

\setcounter{figure}{1}
\begin{figure*}
\centerline{\psfig{file=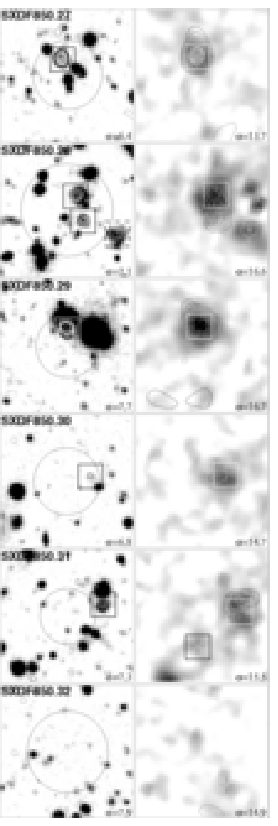,angle=0,width=3.in}
\hspace*{2mm}
\psfig{file=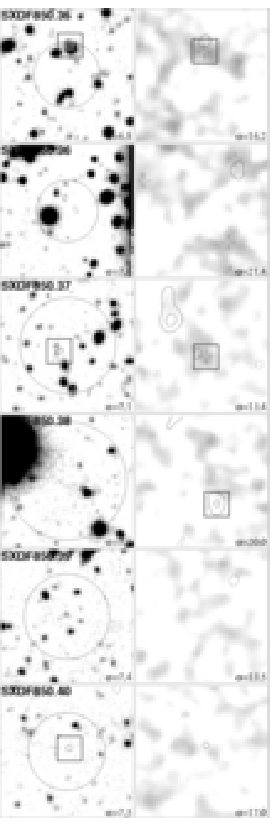,angle=0,width=3.in}}
\vspace{-0.3cm}
\noindent{\small\addtolength{\baselineskip}{-3pt}}
\caption{Cont...}
\end{figure*}

\setcounter{figure}{1}
\begin{figure*}
\centerline{\psfig{file=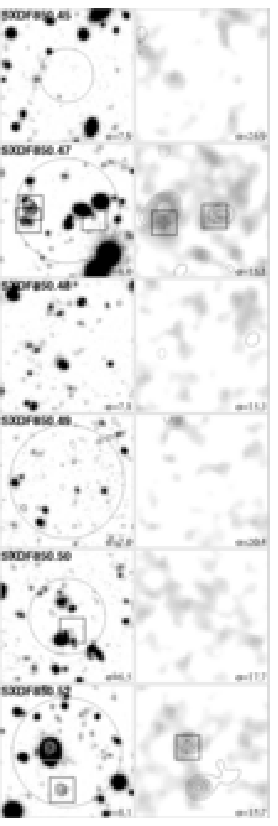,angle=0,width=3.in}
\hspace*{2mm}
\psfig{file=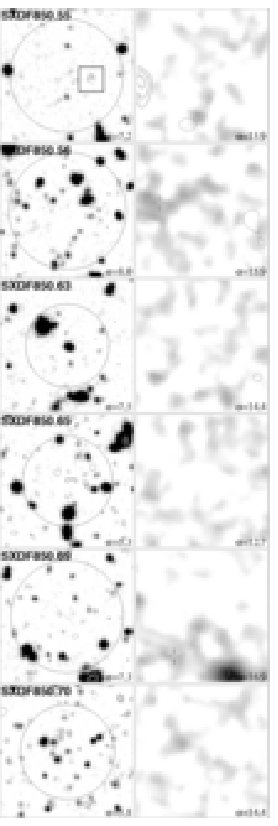,angle=0,width=3.in}}
\vspace{-0.3cm}
\noindent{\small\addtolength{\baselineskip}{-3pt}}
\caption{Cont...}
\end{figure*}

\setcounter{figure}{1}
\begin{figure*}
\centerline{\psfig{file=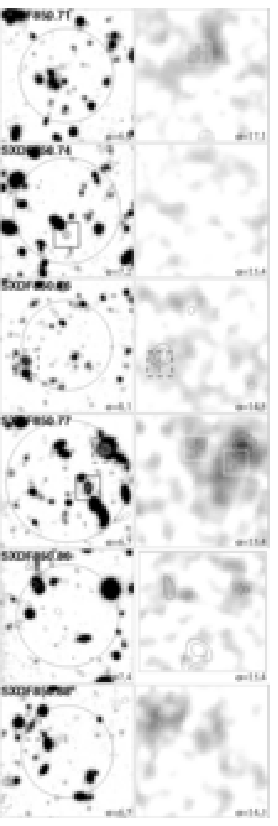,angle=0,width=3.in}
\hspace*{2mm}
\psfig{file=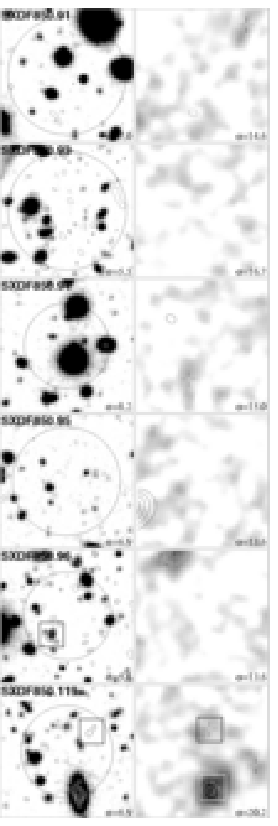,angle=0,width=3.in}}
\vspace{-0.3cm}
\noindent{\small\addtolength{\baselineskip}{-3pt}}
\caption{Cont...}
\end{figure*}

\newpage

\section{POSITION AND FLUX ERRORS}

\subsection*{Uncorrelated noise}

Much of the theory needed for an understanding of SCUBA position
errors can be found in Condon (1997), which treats the general case of
fitting a Gaussian ellipsoid to map data, for which there are six free
parameters: source coordinates, total flux, two principal axes and a
position angle. For the present application, we generally prefer to
assume that SCUBA sources will not be resolved by the beam, although
resolved or blended sources are certainly known (Ivison et al.\ 2000;
Stevens et al.\ 2003; Pope et al.\ 2005). The map should therefore
consist of a scaled and shifted replica of the beam, plus noise. This
leaves just three free parameters.

We follow Condon and assume that the beam is a single 2D Gaussian with
an r.m.s.\ `width' $\sigma$ ($\simeq {\rm FWHM}/2.354$) in each
coordinate. Let the coordinates of the centroid be $(\alpha,\delta)$
and assume that the map is digitised on a (fine) grid where the pixel
spacing is $h$ and the noise value at each pixel is an independent
zero-mean Gaussian deviate with r.m.s.\ value, $\mu$; the units of
$\mu$ are those of surface brightness.  The peak value of the fitted
profile is $A$; strictly, this is a surface brightness value and the
total integrated flux density will be $S = 2\pi \sigma^2 A$. However,
normally the factor $2\pi \sigma^2$ will be absorbed into map units of
mJy\,beam$^{-1}$ or equivalent, so that $A$ has the numerical value of
the flux density of a fitted unresolved source.  With this notation,
Condon's solution for the r.m.s.\ errors ($\Delta$) on the
three-parameter fit is
\begin{equation}
\eqalign{
\Delta A &= \sqrt{1\over \pi}\, {h \over \sigma}\, \mu \cr
\Delta \alpha= \Delta \delta &= \sqrt{2\over \pi}\, {\mu \over A}\, h. \cr
}
\end{equation}
For a practical formula, it makes sense to combine these by defining
the flux signal-to-noise ratio: ${\rm SNR} = A/\Delta A$:
\begin{equation}
\Delta \alpha= \Delta \delta = \sqrt{2}\, ({\rm SNR})^{-1} \, \sigma
\simeq 0.6\, ({\rm SNR})^{-1} \, {\rm FWHM}.
\end{equation}
This is independent of $h$, as makes intuitive sense (although the
derivation assumes $h\ll\sigma$). Note that Condon quotes a larger
error in $A$ for the 6-parameter case: this appears to be an error,
but is in any case irrelevant for the present purpose.

\subsection*{Correlated noise and optimal filtering}

A more serious problem with this result is that often the noise is not
independent from pixel to pixel. This may be inherent in the data
(e.g.\ interferometry maps, where the noise has the same coherence
structure as a point source), or may be a result of smoothing the
map. Smoothing may arise either via some form of `drizzling' in the
data reduction software, or can be an explicit convolution. The prime
example of the latter is `optimal filtering' in which the map is
convolved again with the beam in an attempt to improve the visibility
of true sources in comparison with the noise. This was the strategy
used by Scott et al.\ (2002) to identify candidate sources for
detailed fitting to the unsmoothed data. In this case there is no
fitting of the position of the source: the position is taken as the
location of a peak in the filtered map.

A slightly more general problem is now to consider a `source' in the
form of a Gaussian of height $B$ and width $\sigma_{\rm s}$ superimposed on
a correlated noise field $n(\alpha)$ produced by smoothing white
noise with a Gaussian of width $\sigma$. The resulting noise field
will have an r.m.s.\ value $\epsilon$ and we are interested in both
${\rm SNR}=B/\epsilon$ and the error in the position of the peak. The
latter can be solved by considering a Taylor expansion of the signal
$S$ around the peak (in one coordinate, $x$, and assuming the true source to
be centred at $x=0$):
\begin{equation}
S(\alpha) \simeq B(1-x^2/2\sigma_{\rm s}^2) + n + n'x + n'' x^2/2,
\end{equation}
where $n'$ denotes $dn/dx$, etc.  For large $B$, the last term is
negligible, so the apparent position of the peak is just
$x=\sigma_{\rm s}^2 n'/B$. The r.m.s.\ positional errors
in each coordinate are then
\begin{equation}
\Delta \alpha= \Delta \delta = {\sigma_{\rm s}^2 \over B}\, \left\langle (n')^2 \right\rangle^{1/2}.
\end{equation}
The r.m.s.\ value of the gradient in Gaussian-filtered white noise is
straightforward to evaluate 
(e.g. using equations 16.41 and 16.42 of Peacock 1999):
\begin{equation}
\left\langle (n')^2 \right\rangle = {\left\langle n^2 \right\rangle \over 2 \sigma^2} = {\epsilon^2 \over 2 \sigma^2}.
\end{equation}
In terms of ${\rm SNR}=B/\epsilon$, this gives
\begin{equation}
\Delta \alpha= \Delta \delta = 2^{-1/2} \, {\sigma_{\rm s}\over \sigma}\, ({\rm SNR})^{-1} \, \sigma_{\rm s}.
\end{equation}

The appropriate value of $\sigma_{\rm s}$ depends on the application. For
interferometry data, $\sigma_{\rm s}=\sigma$, so we have
\begin{equation}
\Delta \alpha= \Delta \delta = 2^{-1/2} \, ({\rm SNR})^{-1} \, \sigma
\simeq 0.3\, ({\rm SNR})^{-1} \, {\rm FWHM}.
\end{equation}
For the case of optimal filtering, the source is broadened so that
$\sigma_{\rm s}=\sqrt{2}\, \sigma$, yielding
\begin{equation}
\Delta \alpha= \Delta \delta = \sqrt{2}\, ({\rm SNR})^{-1} \, \sigma
\simeq 0.6\, ({\rm SNR})^{-1} \, {\rm FWHM}.
\end{equation}
This is of the identical form to the result for the gridded data.
However, the definitions of ${\rm SNR}$ are different in the two cases;
to finish, we need to prove that they are, in practice, identical.

First, suppose we allow ourselves any filtering scale, $\sigma_{\rm f}$.
The filtered source width satisfies $\sigma_{\rm s}^2 = \sigma^2 +
\sigma_{\rm f}^2$ and flux conservation gives $B=A(\sigma/\sigma_{\rm s})^2$. The
r.m.s.\ of the filtered white noise can be worked out most simply by
Fourier transforming the original noise field, multiplying by the
transform of a Gaussian filter and squaring to get the new noise
power spectrum, which is then integrated to get the new noise
variance. The unfiltered noise variance is derived by considering
a constant power spectrum over the Nyquist range of wavenumbers
between $-\pi/h$ and $+\pi/h$.
The filtered result can then be expressed as
\begin{equation}
\epsilon = {h\over \sqrt{4\pi}\,\sigma_{\rm f}}\, \mu
\end{equation}
(provided $\sigma_{\rm f} \gg h$), so the SNR of the filtered peak is
\begin{equation}
{\rm SNR}_{\rm peak} = {\sqrt{4\pi}\, A \sigma^2 \sigma_{\rm f}
\over
\mu h \sigma_{\rm s}^2}.
\end{equation}
This has a maximum at $\sigma_{\rm f}=\sigma$, verifying the optimal
filter result and giving
\begin{equation}
{\rm SNR}_{\rm peak} = {\sqrt{\pi}\, A \sigma
\over
\mu h},
\end{equation}
which is identical to Condon's result (eqn 1). 
Thus, we have verified that optimal filtering
returns the same SNR as direct fitting to the pixel data, and shown
that it also yields identical positional errors.

\subsection*{Correction for flux boosting}

It is well known that a flux-limited sample selected in the presence
of noisy fluxes suffers two related effects: too many sources are
found (Eddington bias) and the selected sources have their fluxes
systematically over-estimated. This is sometimes loosely called Malmquist bias
although, strictly speaking, Malmquist bias is the effect on the mean flux of
a distribution  due to the imposition of a flux limit.  A Malmquist bias
persists even without noise. A more prosaic term
for the latter effect is `flux boosting'; in practice the observed SNR
values for SCUBA sources will thus be too high. The standard form for
the Malmquist correction (see e.g. \S3.6.1 of Binney \& Merrifield 1998)
in magnitude units is
\begin{equation}
\Delta m = - \sigma^2\, {d\ln(dN/dm) \over dm},
\end{equation}
where $dN/dm$ is the differential number counts and here $\sigma$
means the r.m.s.\ magnitude error. We shall assume power-law counts
with $N(>f)\propto f^{-\beta}$, so that $\Delta m = - 0.4\beta \ln 10\,
\sigma^2$, and the apparent SNR from the Malmquist formula is
\begin{equation}
{\rm SNR}_{\rm app} = {\rm SNR} \, \exp(\beta/{\rm SNR}^2).
\end{equation}

However, the Binney \& Merrifield formula does not apply in this case, because
the measurements are subject to flux errors, rather than the magnitude
errors assumed in their approach.
It is straightforward to derive the appropriate correction by taking
a Bayesian approach, as has also been followed in Coppin et al.\ (2005).
If the apparent flux is $f_a$, we want to know
the conditional distribution of the corresponding true flux, $f$, which is
\begin{equation}
P(f | f_a) \propto P(f) \, P(f_a|f).
\end{equation}
The prior, $P(f)$ is just the (power-law) number counts, and
$P(f_a|f)$ is just the Gaussian error distribution $\propto \exp[-(f_a-f)^2/2]$
(we implicitly set the r.m.s.\ noise equal to unity, so as to work in SNR units).
This equation has the drawback that $P(f | f_a)$ diverges at $f=\rm 0$, reflecting the
fact that the confusion limit has not been allowed for, but there is a
well-defined maximum is the conditional distribution, and we take this as
the best estimate of $f$ given $f_a$. This is easily shown to be
\begin{equation}
f = f_a/2 + \sqrt{f_a^2/4 - (\beta+1)}.
\end{equation}

Before adopting this as a correction for flux boosting, however, there is
one further correction to consider, which increases the size of the
effect. This arises because we have assumed implicitly that the location
of the source is known, so that the apparent flux is the true flux plus
a noise term. But we have shown above that the existence of a noise field
inevitably introduces position errors, so that we are never measuring
exactly at the true position of the source.
The effect of position errors on the apparent flux is easy to analyse,
following our earlier formulae. The variation in signal with
one coordinate, $x$, around a peak is approximately
\begin{equation}
S(x) \simeq f(1-x^2/2\sigma_{\rm s}^2) + n + n'x,
\end{equation}
and we have already shown the effect of the noise gradient $n'$ in
perturbing the position of the peak. But it also perturbs the
{\it height\/} of the peak, which is the apparent flux:
\begin{equation}
f_a = f + n + (n'\sigma_{\rm s})^2/2f.
\end{equation}
For Gaussian noise, the gradient $n'$ is independent of the
amplitude of the noise, $n$, so there is an additional boost
of the flux -- which is largest for those sources with the
largest positional errors. In terms of the offset in one coordinate,
$\Delta x$, the flux boost is
\begin{equation}
\Delta f / f = (\Delta x)^2/2\sigma_s^2.
\end{equation}
There is an independent effect
from each coordinate, so that the expected size of the boost
from gradients is
\begin{equation}
\langle \Delta f \rangle = \langle (n')^2\rangle \sigma_{\rm s}^2/f
= \epsilon^2\sigma_{\rm s}^2/2f\sigma^2,
\end{equation}
using our previous expression for the r.m.s.\ gradient.
We will ignore the dispersion in this correction, since it is
usually much smaller than the dispersion in $n$.
Since the noise field and the noise gradient are independent,
we can correct for them in turn. If we take out previous deboosted
estimate, $f$, the correction for gradient bias to yield the
final estimate of the true flux, $f_t$, is
\begin{equation}
f_t = f/2 + \sqrt{f^2/4 -1}
\end{equation}
(where we have assumed optimal filtering, so $\sigma_{\rm s}^2=2\sigma^2$).

Combining these two steps yields a cumbersome expression for the
true SNR in terms of the apparent SNR, and we advocate the following
convenient approximation as suitable for use when the apparent SNR 
exceeds 3:
\begin{equation}
{\rm SNR} = \sqrt{ {\rm SNR}_{\rm app}^2 - (2\beta+4) }.
\end{equation}
Our final suggested formula for the expected position errors is thus
\begin{equation}
\Delta \alpha= \Delta \delta =
0.6\, [{\rm SNR}_{\rm app}^2 - (2\beta+4)]^{-1/2} \, {\rm FWHM}.
\end{equation}

\subsection*{Strategy for optimal source reliability}

It may seem self-evident that optimal filtering as discussed above is also the best
strategy for source detection (neglecting confusion) -- but this is not so obvious.
Optimal filtering gives the most accurate measurement of the flux for a given source.
For detection, we want to minimise the probability of noise alone yielding a
spurious source of the observed height. If we smooth an image with a filter that is
broader than optimal, the apparent SNR goes down -- but nevertheless the expected
number of noise peaks on the image of this new height may go down, just because of
the larger coherence length in the new noise field.

This all works out quite simply for Gaussian filtering and a Gaussian source:
the apparent SNR ($\equiv \chi$) is
\begin{equation}
 \chi = {\rm SNR} \;  2 \xi / (1+\xi^2),
\end{equation}
where $\xi = \theta_{\rm filter}/\theta_{\rm beam}$ and SNR means the standard
optimally-filtered value. The number density of peaks with height above
$\chi$ is proportional to $N = \xi^{-2} \chi \exp(-\chi^2 / 2)$ (for $\chi\gs\rm 3$;
see Bond \& Efstathiou 1987). So, we need to vary $\xi$ to
minimise $N$. As a function of the optimally-filtered SNR, the numerical value of the
required $\xi$ can be approximated empirically by 
\begin{equation}
 \xi \simeq 1 + 2/{\rm SNR}^2.
\end{equation}
Thus, for our typical 4-$\sigma$ threshold, we should in principle filter with something
about 15 per cent broader than the beam to give us the best chance that the
sources are real. This is not a big effect and we have chosen to ignore it,
but it is an interesting point of principle.

\end{document}